\documentclass[aps,epsf,amsmath,superscriptaddress,citeautoscript,showpacs,floatfix,twocolumn,nofootinbib,prd]{revtex4-2}
\usepackage[T1]{fontenc}
\usepackage{txfonts}

\usepackage[dvipsnames]{xcolor}
\definecolor{red}{rgb}{0.9, 0,0}
\definecolor{cerulean}{rgb}{0., 0.42,0.9}
\definecolor{navy}{rgb}{0.05, 0.05,0.8}

\usepackage{setspace}
\usepackage[colorlinks]{hyperref}
\hypersetup{
    colorlinks = true,
    citecolor  = red,
	linkcolor  = navy
}

\usepackage{multirow}
\usepackage{slashed}
\usepackage{graphics}
\usepackage{amsmath}
\usepackage{amssymb}
\usepackage{latexsym}
\usepackage{mathtools}
\usepackage{dsfont}
\usepackage{cleveref}
\usepackage{hyperref}
\usepackage{amsfonts}
\usepackage{enumitem}
\usepackage{xstring} 
\usepackage{xspace} 
\usepackage[normalem]{ulem}
\usepackage{fontawesome}
\usepackage{braket}
\usepackage{mathrsfs}
\usepackage{float}

\newcommand{\be}{\begin{equation}}
\newcommand{\ee}{\end{equation}}
\newcommand{\bs}{\begin{split}}
\newcommand{\es}{\end{split}}
\newcommand{\R}[1]{\textcolor{red}{#1}}

\newcommand{\simgt}{\lower.5ex\hbox{$\; \buildrel > \over \sim \;$}}
\newcommand{\simlt}{\lower.5ex\hbox{$\; \buildrel < \over \sim \;$}}

\newcommand{\comment}[1]{}

\usepackage{subfiles} 

\begin{document}

\title{The Role of Quantum Measurements when Testing the Quantum Nature of Gravity}

\author{Daisuke Miki}
\email{miki.daisuke@phys.kyushu-u.ac.jp}
\affiliation{Department of Physics, Kyushu University, 744 Motooka, Nishi-Ku, Fukuoka 819-0395, Japan}
\author{Youka Kaku}
\email{kaku.yuka.g4@s.mail.nagoya-u.ac.jp}
\affiliation{Department of Physics, Graduate School of Science, Nagoya University, Chikusa, Nagoya 464-8602, Japan}
\author{Yubao Liu}
\email{lybphy@hust.edu.cn}
\affiliation{National Gravitation Laboratory, Hubei Key Laboratory of Gravitation and Quantum Physics, School of Physics, Huazhong University of Science and Technology, Wuhan, 430074, China}
\author{Yiqiu Ma}
\email{myqphy@hust.edu.cn}
\affiliation{National Gravitation Laboratory, Hubei Key Laboratory of Gravitation and Quantum Physics, School of Physics, Huazhong University of Science and Technology, Wuhan, 430074, China}
\author{Yanbei Chen}
\email{yanbei@caltech.edu}
\affiliation{Burke Institute for Theoretical Physics, California Institute of Technology, Pasadena, California 91125, USA}

\date{\today}

\begin{abstract}
In order to test the quantum nature of gravity, it is essential to explore the construction of classical gravity theories that are as consistent with experiments as possible.  In particular, the classical gravity field must receive input regarding matter distribution.  Previously, such input has been constructed by taking expectation values of the matter density operator on the quantum state, or by using the outcomes of all measurements being performed on the quantum system --- or by using information obtained by auxiliary observers (like those that lead to the CSL and Diosi-Penrose collapses) that continuously monitor the quantum dynamics.  We propose a framework that unifies these models, and argue that the Causal Conditional Formulation of Schr\"odinger-Newton (CCSN) theory, which takes classical inputs only from experimental and environmental channels --- without auxiliary observers --- is a minimum model within this framework.  Since CCSN can be viewed as a quantum feedback control scheme, it can be made explicitly causal and free from pathologies that previously plagued Schr\"odinger-Newton (SN) theories. Since classical information from measurement results are used to generate classical gravity, CCSN can mimic quantum gravity better than one would naively expect for a classical theory --- making it more subtle to perform tests of the quantum nature of gravity.  We predict experimental signatures of CCSN in two concrete scenarios: (i) a single test mass continuously monitored by light, and (ii) two objects interacting via mutual gravity, each monitored separately. In case (i), we show that the mass-concentration effect of self classical gravity still makes CCSN much easier to test than testing the establishment of mutual entanglement, yet the signatures are more subtle than previously thought for classical gravity  theories.  Using time-delayed measurements and non-stationary measurements, which  delay or suspend the flow of classical information into classical gravity, one can make CCSN more detectable. In case (ii), we show that mutual gravity generated by CCSN can lead to correlations that largely mimic signatures of quantum entanglement in steady-state measurements. Rigorous protocols that rule out LOCC channels, which are experimentally more challenging than simply testing steady-state entanglement, must be applied in order to completely rule out CCSN. 
\end{abstract}

\maketitle

%

\singlespacing 
\section{Introduction}
\label{sec:intro}
The reconciliation between quantum mechanics and general relativity has been a long-standing problem in physics, which most believe should be solved by creating a consistent theory of quantum gravity.  However, from an empirical point of view, it is still a valid question to ask whether gravity should be quantized~\cite{carlip2008quantum,carney2019tabletop}.  With the progress of experimental physics, a series of work has proposed testing the quantum nature of gravity, for example, (i) testing nonlinearities that arise due to semiclassical gravity~\cite{Yang13,giulini2011gravitationally,giulini2014centre,helou2017measurable,helou2017extensions,scully2022semiclassical,helou2019testing,Yubao,Liu2024,howl2021non}, (ii) testing whether the mutual gravity between objects can be implemented via (a simple) classical channel~\cite{kafri2014classical,kafri2015bounds}, or (iii) whether gravity can be used to establish quantum entanglement between two objects~\cite{bose2017spin,marletto2017gravitationally,Miao}. This has been generalized to the consideration of whether gravity can be realized by general LOCC channels~\cite{lami2024testing}.


{
\renewcommand{\baselinestretch}{0.9}
\begin{table*}
\begin{tabular}{c|c|c|c|c|c}
Class & Model  &  \begin{tabular}{c} 
Auxiliary \\ Observers \\ 
Introduced?
\end{tabular}  & \begin{tabular}{c} 
Auxiliary \\ Outcomes 
used \\ to Generate $\phi$?
\end{tabular}   & \begin{tabular}{c} 
Experimental \\ Measurement  Outcomes 
 \\used  to 
Generate $\phi$?
\end{tabular}   & Features \\
\hline 
\hline 
\multirow{2}{*}{\rotatebox[origin=c]{90}{
\begin{tabular}{c} Collapse\\ Models
\end{tabular}}}&
Diosi-Penrose~\cite{diosi1987universal,penrose1996gravity} & 
\begin{tabular}{c}
Measure $\mathbf{g}$\\
everywhere
\end{tabular} &  No & No
& \multirow{2}{*}{\begin{tabular}{c}
Gravity \\ not implemented
\end{tabular}}
\\
\cline{2-5}
& CSL~\cite{ghirardi1986unified,bassi2023collapse} & 
\begin{tabular}{c}
Measure Smeared\\
Matter Distribution 
\end{tabular}
&  No & No &  \\
\hline \hline
{\multirow{3}{*}{\rotatebox[origin=c]{90}{
\begin{tabular}{c}
Schr\"odinger-Newton
\end{tabular}
}}}&
\begin{tabular}{c}
Pre-Selection~\cite{Yang13,helou2017measurable}\\
S-N
\end{tabular} & No & No & No  & \begin{tabular}{c} Violates  \\ Page-Geilker\end{tabular}
\\
\cline{2-6}
& \begin{tabular}{c}
Post-Selection\\
S-N~\cite{helou2017measurable}
\end{tabular} & No & No & Yes  & 
\begin{tabular}{c}
Future measurement \\ choices influence \\ past.
\end{tabular}
\\
\cline{2-6}
& \begin{tabular}{c}
Causal-\\
Conditional\\
S-N~\cite{helou2017extensions,scully2022semiclassical,Yubao,Liu2024}
\end{tabular} & No& No & \begin{tabular}{c}
Obtain conditional \\ expectation of positions  \\ then generate gravity \\ via classical feedback \end{tabular} & Preserves causality
\\
\hline
\hline 
{\multirow{4}{*}{\rotatebox[origin=c]{90}{
\begin{tabular}{c}
Classical Gravity with \\ Auxiliary Observers
\end{tabular}
}}}&
\begin{tabular}{c}
N-H extension of \\
S-N~\cite{nimmrichter2015stochastic}
\end{tabular} & \begin{tabular}{c} Measure $\mathbf{g}$ \\ everywhere \end{tabular} & Yes & No  & 
\begin{tabular}{c} Classical gravity \\ via  Diosi-Penrose \\ measurements
\end{tabular}
\\
\cline{2-6}
& \begin{tabular}{c}
KTM\\
Model ~\cite{kafri2014classical,kafri2015bounds}
\end{tabular} & \begin{tabular}{c}
Measure position\\ of each mass
\end{tabular} & \begin{tabular}{c} 
Uses instant \\  outputs of
 \\  position channels\end{tabular}& No  & 
\begin{tabular}{c}
Instant outputs \\ are very noisy
\end{tabular}
\\
\cline{2-6}
& \begin{tabular}{c}
Oppenheim's\\
 Model~\cite{oppenheim2023postquantum}
\end{tabular} & Yes & Yes & No & \begin{tabular}{c}
More general \\
and includes \\
NH and KTM 
\end{tabular} \\
\cline {2-6}
\noalign{\vskip\doublerulesep
         \vskip-\arrayrulewidth} 
\cline {2-6}
&   \begin{tabular}{c} Unified \\ model \end{tabular} &
\begin{tabular}{c}
Measure position\\ of each mass
\end{tabular}
& 
Yes 
& Yes
& 
\begin{tabular}{c}
Can incorporate \\
all above models
\end{tabular}
\\
\hline
\hline 
\end{tabular} 
\caption{A summary of collapse, Schr\"odinger-Newton, and  Classical Gravity Models which rely on auxiliary observers. We propose a unified model in which classical gravity depends on the outcomes of auxiliary observers as well as the results of experiments performed by the experimentalist.\label{tab:models} }
\end{table*}
}

In order to generate gravity in a manner that recovers classical laws of gravity, we need a classical gravitational potential $\phi$ in the Schr\"odinger equation that will depends on a {\it classical matter distribution}. For $n$ quantum objects that mutually only interact via gravity, we can write a so-called Schr\"odinger-Newton (SN) Equation for their joint wavefunction,
\begin{equation}
    i\partial_t \psi(t,\mathbf{x}_1,\ldots,\mathbf{x}_n)= \sum_j \left[\hat {H}_j  +  m_j \phi(t,\mathbf{x}_j)\right] \psi(t,\mathbf{x}_1,\ldots,\mathbf{x}_n)\,.  
\end{equation}
with $\phi(t,\mathbf{x}_j)$ the value of the classical gravitational potential at the position of object $j$.  Other interaction terms can be added as potentials. 

How can we obtain a classical matter distribution from a quantum state? At first sight, we might use the expectation value of the matter density operator. On which quantum state? As we perform an actual experiment on our quantum system, we can (i) use the ``many-world'' quantum state which contains all possible measurement outcomes, or (ii) the particular {\it conditional quantum state} that can be constructed from the measurement outcome perceived by the experimentalist --- or (iii) a quantum state that can be constructed form information collected by an array of {\it auxiliary observers} that continuously monitors all quantum dynamics that goes on within the entire universe. 

\begin{figure*}
    \includegraphics[width=0.90\textwidth]{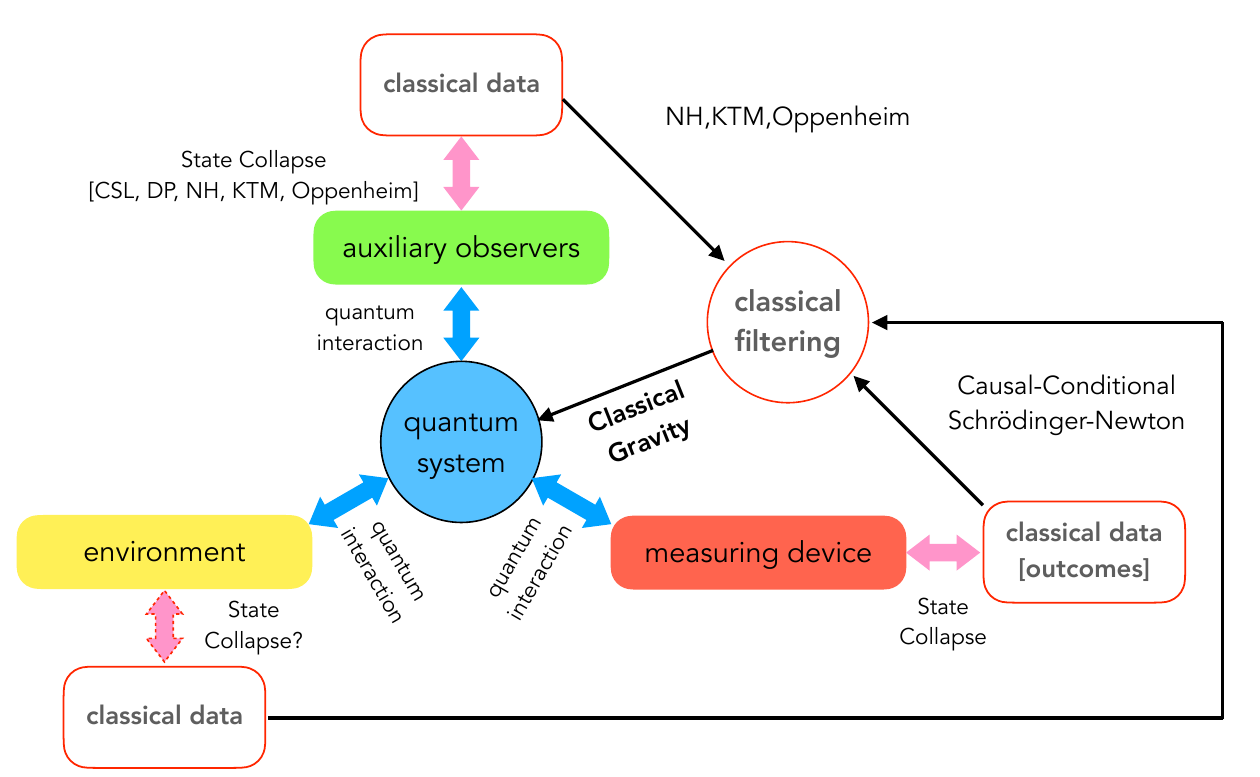}
        \caption{ Schematic plot of how classical gravity can be implemented by collecting classical information from auxiliary observers, measurement results, and possibly the environment. Piecing together the three effects, we arrive at a uniform model of classical gravity.  See Figure~\ref{fig:collapse} for a further model in which auxiliary observers measure the environment, quantum system, and measuring devices.
\label{fig:scheme}}
\end{figure*}

As Page and Geilker have argued, choice (i), although the most convenient, blatantly violates our common experience~\cite{page1981indirect}. As we shall discuss below, different formulations of classical gravity differs in the way they obtain classical information, and how information is {\it processed} to generate gravity.  In particular, (ii) has been adopted by the Causal Conditional Formulation of Schr\"odinger-Newton (CCSN) theory~\cite{helou2017extensions,scully2022semiclassical,helou2019testing,Yubao,Liu2024}, while (iii) can effectively be viewed as the foundation of the so-called Collapse models~\cite{diosi1987universal,penrose1996gravity,ghirardi1986unified,bassi2023collapse}, and has been applied by the Nimmrichter-Hornberger Stochastic  Extension of Schr\"odinger-Newton~\cite{nimmrichter2015stochastic}, the Kafri-Taylor-Milburn model~\cite{kafri2014classical,kafri2015bounds}, and the post-Quantum model constructed by Oppenheim and collaborators~\cite{oppenheim2023postquantum}. 
We will explain how these theories differ in the way they obtain and process classical information about the quantum system, and how they generate classical gravity.  Based on this understanding, we will propose a unified framework that can incorporate all these models (see Figure~\ref{fig:scheme}).

At this time, we would like to highlight the role of the {\it measuring device}, which in standard quantum mechanics only provides a set of measurement-result eigenspaces to project into, and then triggers the {\it Born's rule}. Now that we need to collect all classical information from our system, we will have to carefully consider how that takes place.  Let us consider the effects of the auxiliary observers in the collapse models. In a typical quantum mechanics experiment consisting a microscopic quantum system and a measuring device, these auxiliary observers will only collapse the state of the measuring device -- the {\it only} macroscopic object in the experiment, thereby recovering standard quantum mechanics. In this case, the information the auxiliary observers collect is identical to the information the experimentalist is obtaining. For experiments involving macroscopic quantum systems under quantum measurement, the auxiliary observers will also collapse the quantum state of the macroscopic quantum system, leading to measurable deviations from quantum mechanics~\cite{bassi2023collapse}. So far, no such deviations have been observed~\cite{helou2017lisa,carlesso2016experimental,vinante2017improved,ddonadi2021underground}, indicating that even from the most macroscopic quantum system prepared so far, the collection of quantum information by collapse models -- if indeed taking place --- is still confined to the macroscopic measuring devices.  Based on this discussion, we conclude that, even if the source of classical information for gravity is obtained by auxiliary observers, the main path is still through the experimentalist's information (see Figure~\ref{fig:collapse}).  In this way, CCSN is a minimum model for classical gravity.


After proposing the  framework, and arguing that CCSN is a minimum theory, we shall study several concrete experimental protocols for studying the  self CCSN gravity of a single test mass continuously monitored by light, as well as the mutual gravity between two objects, each monitored separately. These scenarios have been studied by Liu et al.~\cite{Yubao,Liu2024}, showing the ``surprising result'' that CCSN, although classical in nature, can in many cases predict almost the same phenomenology as quantum gravity. Discussions in this paper {explain} the origin 
{of} this result: experiments performed to search for quantum correlations to be established by quantum gravity can pass their classical information to classical gravity, which can generate correlations that can mimic quantum gravity to some extent. 

Since this paper views CCSN as a feedback control scheme, it will adopt the corresponding tools from quantum measurement theory~\cite{wiseman2009quantum}.  In particular, we propose to use the Wiener-filtering approach~\cite{Ebhardt09,chen2013macroscopic}, which can provide simple analytical results for linear systems at Gaussian states.  We will use this approach to analyze two concrete experimental schemes: (i) delayed measurements, and (ii) non-stationary experiments in which measurements are turned off for a while for the classical gravity field to not receive information from the measurement process. It turns out that these non-stationary schemes can circumvent the problem found in Ref.~\cite{Yubao,Liu2024}, and once more reveal distinct signatures of CCSN. 

This paper will be organized as follows. In Sec.~\ref{sec:framework}, we shall describe the general framework of classical gravity models. In Sec.~\ref{sec:single}, we will make predictions to experimental signatures of the causal conditional formulation of Schr\"odginer Newton theory when a single test mass is monitored, quantifying the observability of Schr\"odinger-Newton signatures.  In Sec.~\ref{sec:Mutual}, we discuss signatures in mutual gravity experiments, showing the existence of correlations that mimic quantum entanglement which originate from classical interactions that depends on the settings of measurement devices.  In Sec.~\ref{sec:conclusions}, we summarize our main conclusions. 

\section{Collapse Models and Quantum Matter Sourced Classical Gravity: A General Framework}
\label{sec:framework}

In this section, we shall start by reviewing Schr\"odinger-Newton theories~\cite{Yang13,giulini2011gravitationally,giulini2014centre,helou2017measurable,helou2017extensions,scully2022semiclassical,howl2021non,Yubao,Liu2024}.  Starting as a nonlinear model of quantum mechanics, the Schr\"odinger-Newton theory will be recast into its Causal Conditional formulation that does not violate causality by construction~\cite{helou2017extensions,scully2022semiclassical,Yubao,Liu2024}.  We will next review collapse models~\cite{diosi1987universal,penrose1996gravity,ghirardi1986unified,bassi2023collapse} and connect them with classical gravity models~\cite{nimmrichter2015stochastic,kafri2014classical,kafri2015bounds,oppenheim2023postquantum}.  We then introduce a unified framework that incorporates all these models, and finally pointed out a direction of using generalised measurement scheme for highlighting the signature of the experiment on testing these classical gravity model.

\subsection{Overview}

{We start Table~\ref{tab:models} with the Diosi-Penrose~\cite{diosi1987universal,penrose1996gravity} and  CSL~\cite{ghirardi1986unified,bassi2023collapse} collapse models on the first two rows.  In these models, wavefunctions that contain macroscopically distinct superositions are spontaneously collapsed --- leading to a single classical reality.  Such collapses may be driven by additional --- yet unknown --- physics.  However, mathematically, these models can be viewed from {\it within quantum mechanics} using a much more mundane perspective.  Master equations that describe the evolutions of quantum systems in these models are identical to ones that introduce auxiliary observers that measure the matter distribution in space.  The CSL and Diosi-Penrose models correspond to two different prescriptions for correlations between these measurements.  In this way, these auxiliary observers do extract information about matter distributions in the universe, although in the original collapse models, such information was not used to create a classical gravity field --- until later~\cite{nimmrichter2015stochastic}.}


SN theories are listed in rows 3--5 of Table~\ref{tab:models}. In the pre-section model, the evolution of the {\it initial wavefunction} of the system is used to compute the {\it expected} matter distribution at subsequent times, thereby violating the Geilker-Page argument against nonlinear quantum mechanics.  In the post-selection model\,\cite{Helou17}, the {\it final wavefunction}, consistent with the measurement results, is used to generate the expected matter distribution.  This potentially violates causality since the evolution of a system at time $t$ may depend on choices of which measurements are made at $t'>t$, although it is not clear whether such dependence can actually cause superluminal signaling. 

Over the past decades~\cite{polchinski1991weinberg,simon2001no,diosi2025causality}, nonlinearity of the SN theory has been connected to the violation of causality, which can be viewed as pathological. However, causality violation can be cured by the Causal-Conditional formulation of Schr\"odinger-Newton (CCSN)~\cite{helou2017extensions,helou2019testing,scully2022semiclassical}. In CCSN, all measurement results are fed back in a causal way to generate the gravitational potential. In fact, this formulation is mathematically equivalent to the description of a feedback control system within {\it linear quantum mechanics}.   

 Here we do note that Newtonian gravity, being a non-relativistic theory, can lead to apparent superluminal signal propagation as part of its prediction error, and causality will be restored by implementing the analogy of the 
 Li\'enard-Wiechert potential for linearized gravity.  However, the superluminal signal propagation enabled by the pre-selection and post-selection models are more severe and cannot be restored this way.

The SN theories use the measurement data from the experimental device to generate classical gravity. In contrast, another class of models uses the auxilliary outcomes --- measurement data collected by auxiliary observers introduced by collapse models --- to produce classical gravity. They are listed on rows 6--8 of Table~\ref{tab:models}. For example, the Nimmrichter-Hornberger extension of Schr\"odinger-Newton uses the measurement results of the Diosi-Penrose observers~\cite{nimmrichter2015stochastic}.  The Kafri-Taylor-Milburn  model~\cite{kafri2014classical,kafri2015bounds} has position measurements with variable strengths on individual objects generating gravity, using the results themselves as classical positions of test masses to drive the gravity field.  Oppenheim's post-Quantum model is more general, allowing the measurement of a general set of quantum observables, the performance of general classical filtering (thereby incorporating classical dynamics), and the generation of a classical gravity field~\cite{oppenheim2023postquantum,layton2024healthier}.  

We can use Figure~\ref{fig:scheme} to summarize the above discussion. In this figure a quantum system interacts with a measuring device (essential for any experimental setup), the environment (inevitable for realistic experiments), and auxiliary observers (introduced by some models).  The interaction between the system, the device, and the environment are all quantum.  The device and the auxiliary observers are measured projectively, extracting classical information about the system, and causing back action.  In particular, back action from the auxiliary observers will be seen as noise arising from Collapse Models. Whether the environment is measured or not does not affect any predictions in standard quantum mechanics~\footnote{Even though, as a computational/conceptual tool, one can imagine measuring the environment in different ways, which correspond to different {\it unravelings} of environmental decoherence.}, yet it can affect predictions of classical-gravity models, as discussed by Ref.~\cite{Yubao,Liu2024}. Since it appears optional whether the environment is measured or not, we have drawn the dashed line in Figure~\ref{fig:scheme}. In the original models of Nimmrichter-Hornberger~\cite{nimmrichter2015stochastic}, Kafri-Taylor-Milburn~\cite{kafri2014classical,kafri2015bounds} and Oppenheim and collaborators~\cite{oppenheim2023postquantum,layton2024healthier}, only classical information from auxiliary observers was used to generate classical gravity. In CCSN, information from the measurement device and environment is used to generate classical gravity. Putting these into the same figure leads us to propose a unified model in which {information} from auxiliary observers, measurement devices, and the environments can be combined to generate classical gravity.




In the following, we shall first review the steps toward building the CCSN model, and then discuss the combination of CCSN and auxiliary observers to form a unified model.  As we finally return to the conceptual foundation of the unified model, we will argue that CCSN is a minimum model for classical gravity.

\subsection{Nonlinear Quantum Mechanics Versus Information Obtained from Measurements}
\label{sec:nonlinear_QM_vs_measurement}

\subsubsection{Schr\"odinger-Newton Theory: Nonlinear Quantum Mechanics}

In the absence of measurement processes carried out by the experimentalist, there exist two ways in which gravity can depend on matter distribution.  In the first approach, the Newtonian gravitational potential  $\phi$ is co-evolved with the wavefunction of matter $\psi$, which, in the Newtonian limit, is given by 
\begin{equation}
    \phi(t,x)= \sum_j\int\frac{Gm_j}{|x-x_j|}|\psi(t,x_1,\ldots,x_j)|^2 dx_1\ldots dx_n\,.
\end{equation}
Here for simplicity we have restricted to one-dimensional motion.  In Dirac bracket notation, and in the limit the object's position uncertainties are much less than their separations, we can write 
\begin{align}
\label{SN1}
    id|\psi(t)\rangle &= \sum_j\left[\hat H_j + m_j\phi(t, \hat{x}_j)\right]|\psi(t)\rangle dt,  \\
    \label{SN2}
    \phi(t,{x}) & =- \sum_k \big\langle\psi(t)\big|\frac{G m_k}{|{x} -  \hat{{x}}_k|}\big|\psi(t)\big\rangle.\\
    \label{SN3}
\langle \hat{{x}}_j\rangle &=\langle \psi(t) | \hat{{x}}_j|\psi(t)\rangle.
\end{align}
As we insert Eqs.~\eqref{SN2} and \eqref{SN3} into \eqref{SN1}, the resulting equation is called the Schr\"odinger-Newton (SN) equation. Here we will need to pay special attention to the self-interaction term $k=j$ in Eq.~\eqref{SN3} when evaluating $\phi(t,x)$ at $\hat x_j$.  
 
 In the situation that each object moves around a zero-point position $x_j^{(0)}$, 
 \begin{equation}
     \hat{x}_j ={x}_j^{(0)} +{\hat{X}}_j,
 \end{equation}
 up to quadratic order, after adding constant forces acting on each mass to overcome mutual Newtonian gravity from the other masses and removing potential-energy constant terms, 
we obtain
\begin{align}
     \phi(t,\hat x_j) 
     =&-\frac{1}{2}\omega_{\rm SN}^2 (\hat X_j-\langle \hat X_j\rangle)^2 
     - \frac{1}{2}\sum_{k\neq j}\omega_{jk}^2(\hat X_j- \langle \hat X_k\rangle)^2
     \label{SN:phi_j}
 \end{align}
The first term in the above potential corresponds to the self-gravity of each object, 
and  is unique to a classical theory of gravity. {The quantity $\omega_{\rm SN}$ is given by the oscillation frequency of the object's center-of-mass, when the entire object moves inside the classical gravitational potential created by the expectation value of its own microscopic matter distribution.  This expectation value is taken over the center-of-mass quantum state and the thermal fluctuations of internal motions. 
For a test mass that consists of a single type of atom with mass $m_{\rm atom}$, which oscillates around lattice sites with zero-point position uncertainty $x_{\rm int}$, we have
\begin{equation}
\label{eq:omegaSN}
    \omega_{\rm SN}=\sqrt{{Gm_{\rm atom} }/{(6\sqrt{\pi}x_{\rm int}^3)}}.
\end{equation}}
Although $\omega_{\rm SN}$  depends on the material from which the mass is made, we shall adopt the same value for all masses for simplicity.  In order for this quadratic approximation to be valid for this self-gravity term, the uncertainty of $\Delta \hat X_j$ should be less than the motion of the mass's nuclei around their equilibrium positions.
The subsequent terms in Eq.~\eqref{SN:phi_j} arise from the mutual interaction between the masses, with
\begin{equation}
    \label{eq:omegajk}
    \omega^2_{jk} = 2Gm_k/d_{jk}^3\,,\quad    d_{jk} =|x^{(0)}_k-x^{(0)}_j|\,,\quad  j \neq k\,.
\end{equation} 
Note that in general $\omega_{jk}\neq \omega_{kj}$, although $m_j \omega_{jk} = m_k\omega_{kj}$.  This quadratic approximation is valid for the mutual-gravity term as long as $\Delta \hat X_j$ is much less than the size of the objects and their separations.  

Comparing Eqs.~\eqref{eq:omegaSN} and \eqref{eq:omegajk}, we can see that the self-gravity frequency $\omega_{\rm SN}$ is much greater than the mutual gravity frequencies $\omega_{jk}$.  In practice, the difference can be two orders of magnitude~\cite{Yang13}.  This is because $\omega_{\rm SN}$ arises from an effective matter density of $\sim m_{\rm atom}/x_{\rm zp}^3$ around the lattice sites, which is much greater than the mean matter density of typical materials, which bounds $m_k/d_{jk}^3$ and $m_k/d_{jk}^3$.  This will eventually make the self-classical gravity a lot more detectable than the mutual gravity between objects, making it a promising experimental stepping stone toward {eventually testing quantum nature of gravity}.

 Even though Eqs.~\eqref{SN1} and \eqref{SN:phi_j} form a well-posed system, it is non-trivial to incorporate quantum measurement results, since measurement involves the nonlinear collapse of the wavefunction. Nevertheless, since all terms in the Hamiltonian are up to quadratic order in position and moment, one can develop a state-dependent Heisenberg Picture treatment for this system~\cite{helou2017measurable}, with linear Heisenberg Equations. Two most straightforward ways of implementing quantum measurement in that framework will be to interprete $\langle \hat X_j\rangle$ as expectation value on the initial state prepared for the measurement process, or the final state that corresponds to the measurement outcome.   The former corresponds to the pre-selection model (row 3 of Table~\ref{tab:models}), while the latter corresponds to the post-selection model (row 4 of Table~\ref{tab:models}). The nonlinearity of Eq.~\eqref{SN:phi_j} means that the two approaches will lead to different predictions, as shown by Ref.~\cite{helou2017measurable}.   
 
 For two objects, the pre-selection model leads to the same violation as the Page-Geilker argument. Suppose an object splits into two macroscopically distinct positions  $x_1$ and $x_2$ according to the result of a quantum measurement with two equally likely outcomes, not accounting for the measurement result will lead to a mass distribution with 50\% support at $x_1$ and $x_2$ each, contradicting common experimence.  The post-selection model, while circumventing the Page-Geilker argument by being faithful to measurement outcomes, leads to a violation of causality, since evolutions of the masses will depend on choices of measurements made in the future. 

\subsubsection{Causal Conditional Schr\"odinger-Newton: Linearity and Causality Restored}

In order to incorporate measurement results while protecting causality, one can use a {\it causal-conditional} formulation of Schr\"odinger-Newton  (row 5 of Table~\ref{tab:models}), in which the matter distribution employed to generate gravity at any point in spacetime is obtained from the conditional distribution of matter gathering all measurement results within the past light cone of the point~\cite{helou2017extensions,helou2019testing,scully2022semiclassical}. In the Schr\"odinger Picture, this seems to be the most natural mathematical formulation~\cite{Yubao}. We shall remain in the Newtonian domain by not considering the signal propagation time between different sites, yet we will allow a time delay in performing each quantum measurement, which will allow us to illustrate the causal structure of the theory to some extent. 

Suppose the measurement on each mass is performed on a continuous variable $\hat Q_j$ (which could be $\hat X_j$ itself in the simplest case), we can write the following set of Stochastic Differential Equations:
 \begin{align}
\label{eqmeasure1new}
id|\psi\rangle 
=
\sum_j&\bigg\{ \hat H_j dt\nonumber\\
& + 
\frac{m_j}{2}
\Big[\omega_{\rm SN}^2(\hat X_j - X_{j}^c)^2+\sum_k \omega_{jk}^2(\hat X_j -X_k^c )^2\Big]dt  \nonumber\\
& +{\epsilon_j^Q(\hat {Q}_j - \langle \hat{Q}_j\rangle)} {dW}_j^Q/{\sqrt{2}} 
\nonumber\\
&-i{(\epsilon^Q_j)^2(\hat{Q}_j -\langle \hat Q_j\rangle)^2} dt/{4}\bigg\} |\psi\rangle
\\
\langle \hat{Q}_j\rangle &=\langle \psi| \hat{Q}_j|\psi\rangle  \,,\quad 
 d{z}_j = \langle \hat{{Q}}_j\rangle+{{dW}^Q_j}/({\sqrt{2}\epsilon_j^Q})
 \label{eq:cond}
\\
\label{eqmeasure5new}
X_j^c(t)& =E\left[\hat X_j(t)\big|
\big\{z_k(t'):t'<t
\big\}\right]
\end{align}
Here, Eq.~\eqref{eqmeasure1new} is a Stochastic Schr\"odinger Equation that governs the evolution of the {\it conditional} quantum state $|\psi\rangle$ of the optomechanical system (including masses and optical modes of cavities): the first line is the Hamiltonian of mass $j$; the second line includes self gravity and mutual gravity terms, with $X_j^c$ the {\it conditional expectation value} of the position of mass $j$;
the third line is the stochastic term arising from measurement of $\hat Q_j$, with $\langle \hat{Q}_j\rangle$ the conditional expectation value of the measured quantity; the last line is the Ito term that enforces the normalization of the quantum state. Here the quantity $\epsilon_j^Q$ is the strength at which mass $j$ is being measured by the experimentalist. In Eq.~\eqref{eq:cond}, the conditional expectation  $ \langle \hat Q_j\rangle $ is obtained by taking expectation value of the operator $\hat Q_j$ on the conditional state $|\psi\rangle$, while the measurement outcome $z_j$ is written as a sum of the conditional expectation $\langle \hat Q_j\rangle$ and a stochastic contribution.  In Eq.~\eqref{eqmeasure5new}, the conditional expectation $X_j^c$ of the position of mass $j$ is obtained by computing the expectation value of the position operator $\hat X_j$ over all measurement outcomes of each mass $k$ up to time $t$. In particular, if the experimentalist performs measurement on $\hat X_k$ with a delay time $\tau_k$, one can define $\hat Q_k(t) = \hat X_k(t-\tau_k)$.  This delay {\it prevents classical gravity from promptly receiving information regarding the positions of the masses, and will lead to more significant observational signatures. }

We emphasize that, since the conditional expectation $X_j^c$ is constructed by classical filtering of measurement results $z_k$, which are {\it classical}, they act as the classical position of mass $j$ according to which classical gravity is generated.  We can also see explicitly that the stochastic differential equations \eqref{eqmeasure1new}--\eqref{eqmeasure5new} are identical to those that describe a feedback control system within {\it linear quantum mechanics} --- with causality of the feedback system enforced by only allowing time delays instead of time advances. 

At this stage, one might want to consider the retarded gravitational potential by using the appropriately time-delayed conditional expectation values of position for $X_j^c$.  This will definitely lead to a completely causal theory, although such a naive theory will only work for a scalar version of gravity, which does not have the correct near-zone behavior.  In order to obtain the correct near-zone behavior, one will need to consider the tensorial nature of gravity, and to account for the gravitating effects of forces that drive the masses to move.~\footnote{For the gravitating effect of stresses, and the corresponding gauge dependence, see Ref.~\cite{pang2018quantum}.} We shall not carry out such a process in this paper, except to have faith that causality will be restored in the relativistic version of the theory. On the practical side, the near-zone correction to scalar gravity actually reduces the effect of retardation: in the near zone, tensorial gravity is better approximated by the instantanous Newtonian gravity. More specifically, the correction is of the order  
\begin{equation}
    (\omega L/c)^5 = 4\times10^{-39}\,\left(\frac{\omega}{2\pi \times 1\,{\rm Hz}}\right)^5 \left(\frac{L}{1\,{\rm m}}\right)^5
\end{equation}
    where $\omega$ is the frequency of operation and $L$ is the size of the system. We also note that for macroscopic quantum gravity experiments one usually operates at lower frequencies and shorter distances.

\subsection{Introduction of Auxiliary Measurements and a Unified Model}
\label{sec:auxunify}

As a second approach, the Newtonian gravity potential $\phi$ can be constructed relying on data obtained from {\it auxiliary} observers that continuously monitor the position of all masses at all locations.  One explicit construction was from Kafri, Taylor and Milburn (KTM) ~\cite{kafri2014classical,kafri2015bounds}. 

Since these additional measurement processes will act back onto the masses and cause continuous stochastic localization, we can related the KTM model to the previously studied CSL~\cite{ghirardi1986unified,bassi2023collapse} and  the Diosi-Penrose~\cite{diosi1987universal,penrose1996gravity} collapse models.  Even though those models were not constructed to address classical gravity, we can certainly view the stochastic collapses as arising from continuous measurements of matter density across space. The difference between CSL and Diosi-Penrose lies in the particular linear combinations of matter densities at different spatial locations that are independently monitored. 
Mathematically, this means we unravel the CSL and the Diosi-Penrose master equations to Stochastic Schr\"odinger Equations with stochastic measurement outcomes.  This point of view was indeed adopted by Nimmricheter and Hornberger (NH), who further  argued that the Diosi-Penrose model should be used to provide information to Schr\"odinger-Newton~\cite{nimmrichter2015stochastic}.

\begin{figure*}
    \includegraphics[width=0.65\textwidth]{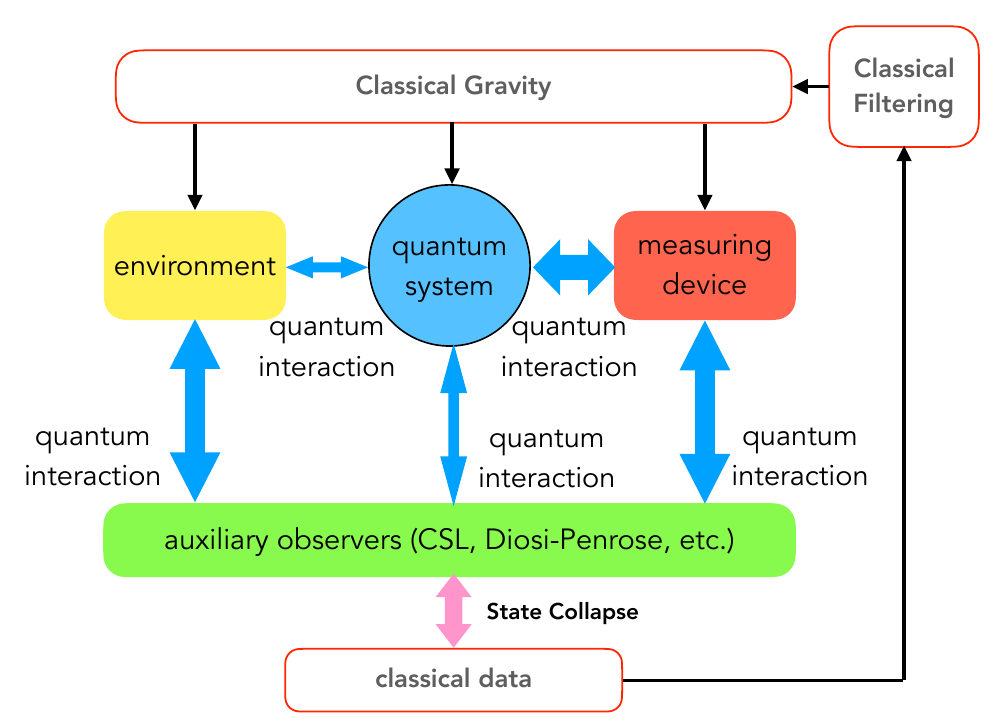}
    \caption{Auxiliary observers in traditional collapse models were introduced to interact strongly with the measuring device, and weakly with microscopic quantum systems, so that the collapses produce the same phenomenology as standard quantum mechanics. No collapses have been discovered so far in macroscopic quantum mechanics experiments.   If we would still like to believe that collapse produce the single-world reality, then in a macroscopic quantum mechanics experiment with low-level of decoherence, the classical reality is still produced by the auxiliary observers's interaction with the measuring device.  In this case, classical gravity should be sourced by the experimentalist's measurement results, as prescribed by Causal Conditional Schr\"odinger-Newton theory.  \label{fig:collapse}}
\end{figure*}

The subsequent proposal by Oppenheim on the post-Quantum gravity model follows a similar strategy, but provides a more general framework~\cite{oppenheim2023postquantum,layton2024healthier}: the classical dynamics extracts information from the quantum dynamics via measurement processes, performs classical filtering, and then act back to the quantum system as adjustments to classical parameters of its Hamiltonian.  Even though measurements on the quantum system can be general, in the situation with only a finite number of masses, and when light travel time between the masses are negligible, we can always attribute the measurements as being performed on the positions of individual test masses, with strengths that are free parameters of the theory. Let us jointly write down the Stochastic Schr\"odinger Equation that incorporates these auxiliary observers: 
\begin{align}
    \label{eq:unified1}
    id|\psi\rangle 
    =
    \sum_j&\bigg\{ \hat H_j dt\nonumber\\
    & + 
    \frac{m_j}{2}
    \Big[\omega_{\rm SN}^2(\hat X_j - Y_{j})^2+\sum_k \omega_{jk}^2(\hat X_j -Y_k )^2\Big] dt  \nonumber\\
    & +{\epsilon_j^Q(\hat {Q}_j - \langle \hat{Q}_j\rangle)} \,{dW}_j^Q /\sqrt{2}
    \nonumber\\
    &-i(\epsilon^Q_j)^2(\hat{Q}_j -\langle \hat Q_j\rangle)^2 dt/4 \nonumber\\
    &+{\epsilon_j^G(\hat {X}_j - \langle \hat{X}_j\rangle)} {dW}_j^G /{\sqrt{2}}\nonumber\\
    & -i{(\epsilon^G_j)^2(\hat{X}_j -\langle \hat X_j\rangle)^2} dt/4\bigg\} |\psi\rangle 
\end{align}
\begin{align}
     \langle \hat{Q}_j\rangle &=\langle \psi| \hat{Q}_j|\psi\rangle  \,,\quad 
  d{z}_j = \langle \hat{{Q}}_j\rangle+{{dW}^Q_j}/({\sqrt{2}\epsilon_j^Q})
      \label{unified:cond1}
     \\
     \langle \hat{X}_j\rangle &=\langle \psi | \hat{X}_j|\psi\rangle  \,,\quad 
     d{z}_j^G = \langle \hat{X}_j\rangle+{dW^G_j}/({\sqrt{2}\epsilon_j^G})
     \label{unified:cond2}
    \end{align}
In Eq.~\eqref{eq:unified1}, we have added two more terms to Eq.~\eqref{eqmeasure1new}, which arise from the measurement-induced back action and decoherence from the auxiliary observers, with $\epsilon_j^G$ a tunable, model-dependent measurement strength on mass $j$.  {Furthermore, for the Wiener increments  $dW_j^G$, they can satisfy a generalized Ito rule of
\begin{equation}
\label{eq:corrdw}
     dW_j^G dW_k^G = c_{jk}^G dt \,,
\end{equation}
with $c_{jk}^G$ quantifying possible correlations between measurements.  Here we have left $Y_j$, the classical position of mass $j$, undefined, since that definition will differ model by model, as we shall describe below. } 

The KTM model~\cite{kafri2014classical,kafri2015bounds} proposes the classical position of test mass $k$ as the instantaneous result of measurement, 
\begin{equation}
    Y_{kj}^{\rm KTM}  = z_k^G\,,
\end{equation}
and with all measurements independently performed.  Since the instantaneous result is not filtered, gravitational force tends to be more noisy in this model, compared with others. As one varies $\epsilon_j^G$, one can trade-off between a more accurate position measurement but a strong back-action, or a less accurate position measurement but a weaker back-action.  This trade-off is also considered within the Oppenheim model, for example by Carney and Matsumura in gravitational scattering~\cite{carney2024classical}. 

The NH model~\cite{nimmrichter2015stochastic} proposes to use auxiliary observers measure the operators that corresponds to the Newtonian gravity field, leading to a Diosi-Penrose-type decoherence, and with conditional expectation extracted from the auxiliary observers: 
\begin{equation}
    Y_{j}^{\rm NH}  = E\left[\hat X_j(t') \big|\{z_k^G(t'): t'<t\}\right]\,.
\end{equation}
{
In a more recent work by Kryhin and Sudhir~\cite{kryhin2025distinguishable}, the Oppenheim model were recasted into a similar form to NH, emphasizing the correlated nature of Diosi-Penrose decoherence on different objects, such as in Eq.~\eqref{eq:corrdw}.  }

{We must point out that in the NH, KTM and the Oppenheim models, {\it information collected by measurements made by the experimentalist is not incorporated when gravity is generated} --- even though in well-performed experiments {\it the main source of classical information actually comes from the experimentalists's measurement.}  We hereby propose that when a high quantum-efficiency experiment is performed on a well-isolated, gravitationally interacting quantum system,  {\it the results of the experimentalist's measurement must be incorprate into classical gravity.}  The most natural way of doing so is given by }
\begin{equation}
    Y_{kj}^{\rm unified}  = E\left[\hat X_k(t) \big| \{ z_l^G(t'),z_l^Q(t'):t'<t\}\right]\,,
\end{equation}
We note that the Oppenheim model is actually general enough to include this model, although the experimentalists's results were not explicitly considered previously. We note that this unified model reduces to the CCSN theory when the auxiliary observers's strengths of measurement becomes very weak, with $\epsilon_j^G \rightarrow 0$.    In this paper, we shall only study the phenomenology for $\epsilon_j^G =0$, and leave the more general study to a future paper.  

Let us now argue that $\epsilon_j^G\rightarrow 0$ is a well-motivated choice, especially as we remind ourselves of the original motivation of the collapse models.   
We should recall that the collapse models were introduced to {\it create} the single-world reality that we experience by collapsing the quantum state of the entire universe --- but mostly the macroscopic measuring devices --- to unique ones that are consistent with the measurement results. In this way, we should modify Figure~\ref{fig:scheme} into Figure~\ref{fig:collapse}, where the auxiliary observers interact with the quantum system, the measuring device, as well as the environment --- and then the states of the auxiliary observers get collapsed to create the classical reality.   In the literature, if microscopic quantum experiments were considered, the auxiliary observers were always {\it postulated} to work to produce the standard quantum measurement phenomenology.  Only for macroscopic quantum systems were the auxiliary observers' actions explicitly considered as a weak continuous measurement on macroscopic test masses. The fact that the collapses were not yet discovered shows that if the collapse models actually work, then their effects on all existing quantum systems were small, yet their effects on all quantum measurement devices are still strong enough to collapse them into a single reality.  In this case, since the single classical reality emerges during the collapse of the measuring devices, classical gravity should be sourced by the measurement results of the experimentalist, as prescribed by the Schr\"odinger-Newton theory. 

\subsection{Effect on the experimental phenomenology}

One direct phenomenological consequence of (classical-information-driven) classical gravity is that the experimental signature for testing these gravity models will be dependent on how information is extracted --- and as we have argued, the setup of the experimeter's measurement plays a crucial role. 

As discussed in Ref.~\cite{helou2017measurable,Yubao}, the pre-selection,  post-selection, and causal-conditional SN theories will have very different experimental signatures. Furthermore, in the experiment aiming to distinguish classical gravity from quantum gravity that involves continuous quantum measurement, classical information obtained by {gravity} from measurements results  --- which ``follows'' quantum-state collapses, as formulated by CCSN, can diminish the stronger signatures in the pre/post-selection SN model. Moreover, for the experiment that targets at probing gravity-induced-entanglement,  measurement-sourced classical gravity can also generate {\it apparent entanglement} as false alarm signals\,\cite{Liu2024}. It is important to note that such apparent entanglement does not violate the LOCC condition set forth by Ref.~\cite{lami2024testing}, hence {underlining} the importance {of} developing such rigorous bounds.  Yet in practice the existence of such apparent entanglement will make experimental signatures of classical gravity more elusive.

The feedback process, as discussed by Refs.~\cite{Yubao,Liu2024} has the feature that the measurement data till time $t'<t$ is used to generate the classical gravity at $t$. With the unified framework, we realize that with more flexible measurement schemes, classical gravity can be generated by the data of a delayed measurement, which points out a direction to circumvent the difficulties discussed in\,\cite{Yubao,Liu2024}. In defining a measurement process, not only what physical quantity we measure is important, but also when we make the measurement. This is because the state evolution in our unified framework is nonlinear in a way that the measurement-sourced gravity will affect the dynamical evolution of the quantum state. With a delayed measurement, the classical gravity does not follow the wave function collapse which allows the manifestation of the SN feature. In the following sections, we shall use an example optomechanical protocol to discuss the test of the quantum nature of gravity using non-stationary measurement schemes.

\section{A Single Macroscopic Object in its Own Classical Gravity Potential}
\label{sec:single}

In this section, we consider a simplest optomechanical device, a single test mass in a harmonic potential, acting as the movable end mirror of a bad cavity (i.e., one whose linewidth is much broader than detection bandwidth). The cavity is pumped by a single carrier field on resonance, and with the cavity output field detected by a homodyne detector. When analyzing this system using CCSN, Ref.~\cite{Yubao} showed, by solving Stochastic Schr\"odinger Equations that CCSN becomes hard to distinguish from quantum gravity when masses are under continuous measurement. Here, we apply the formalism of Wiener filters, once more quantify the size of CCSN signature for continuous measurement, and {further} consider the effect of delayed and non-stationary experiments. 

\subsection{The Desirable Signatures of Single-Object Schr\"odinger-Newton and the Role of Measurements}
The optomechanical Hamiltonian of our system, in Schr\"odinger-Newton theory can be written as
\begin{align}
    \hat{H}&=
    \frac{1}{2M}\hat{p}^2+\frac{1}{2}M\omega_m^2\hat{x}^2
    -\hbar\alpha\hat{x}\hat{a}_1
    +\frac{1}{2}M\omega_{\rm SN}^2(\hat{x}-\langle \hat x\rangle)^2,
\end{align}
where $\hat{x}$ and $\hat{p}$ are the mechanical coordinate and momentum of the object, satisfying $[\hat{x},\hat{p}]=i\hbar$, $M$ and $\omega_m$ are object's mass and mechanical resonant frequency within the potential well.  Operators $\hat{a}_1$ and $\hat{a}_2$ are the amplitude and phase quadratures of the incoming optical field, satisfying $[\hat{a}_1(t),\hat{a}_2(t')]=i\delta(t-t')$. The quantity 
\begin{equation}
    \alpha=
    \sqrt{8P_{\rm cav}\omega_0/(\kappa^2 L^2\hbar)}
\end{equation}
is the optomechanical coupling, written in terms of the intracavity optical power $P_{\rm cav}$, the cavity length $L$, the cavity bandwidth $\kappa$, and the carrier frequency $\omega_0$. Here we have assumed a {``bad cavity limit"}, with bandwidth $\kappa$ much greater than other frequency scales in the problem. 

The quantity $\omega_{\rm SN}$ in the fourth term on the right-hand side is the SN frequency defined in Eq.~\eqref{eq:omegaSN}, 
and {$\langle \hat x\rangle$} is the {\it expectation value} of the mass's position on the system's quantum state.   As discussed in Sec.~\ref{sec:nonlinear_QM_vs_measurement}, this term arises from nonlinear nature of classical gravity, and only takes this quadratic form when range of motion of the mass is much less compared to the zero-point uncertainty $x_{\rm zp}$ of its atoms around their equilibrium positions.

Suppose we start the mass at an initial Gaussian wavefunction with $\langle \hat x \rangle =0$, and do not consider the subsequent results of the continuous measurement, then the expectation value $x_c$ can always be set to zero. In this case, the total potential energy of the mass is given by $M\omega_Q^2 \hat x^2/2$ with 
\begin{equation}\label{eq:selfgrav_effective_freq}
    \omega_Q^2 =\omega_m ^2 +\omega_{\rm SN}^2,
\end{equation}
In this way, the uncertainty ellipse for $(\hat x, \hat p)$ will rotate in phase space with frequency {$\omega_Q$}~\footnote{With components of the $(\hat x,\hat p)$ covariance matrix oscillating at  $2\omega_Q$.}, which differs from $\omega_m$.  If we were to continue with this prescription, we will have
\begin{align}
    \dot{\hat x}=\hat p/M\,, \quad    \dot{\hat p}=-M\omega_Q^2 \hat x + \hbar \alpha \hat a_1\,.
\end{align}
Adding the Heisenerg equations for the out-going field
\begin{align}
    \hat b_1 = \hat a_1\,,\quad 
    \hat b_2 = \hat a_2 +\alpha \hat x,
\end{align}
we will obtain
\begin{equation}
    \hat b_2 = \hat{a}_2+{\hbar\alpha^2 \hat{a}_1 }/{[M(\omega_Q^2-\omega^2)]}
\end{equation}
and its spectral density is
\begin{align}
    S_{b_2b_2}&=1+
    \alpha^4\hbar^2/[{M^2(\omega^2-\omega_Q^2)^2}],
\end{align}
where we have defined the (single-sided and symmetrized) cross spectral density using the convention of 
\begin{equation}
2\pi\delta(\omega-\omega')S_{AB}(\omega)=\braket{\hat{A}(\omega)\hat{B}^\dagger(\omega')+\hat{B}^\dagger(\omega')\hat{A}(\omega)}\,,   
\end{equation}
In this model the quantum radiation pressure noise of a measurement will show up at $\omega_Q$, instead of $\omega_m$, providing a distinct experimental signature. For the standard quantum mechanics $\omega_{\rm SN}=0$, the radiation pressure noise will peak at $\omega_m$.  This signature, if real, is of great experimental significance.  First of all, this shift in oscillation frequency is a distinct signature in dynamics, instead of an additional noise.  Furthermore, the magnitude of the shift is related to $\omega_{\rm SN}$, which is a time scale much faster than the typical gravitational time scale of a uniform object, thanks to the significant concentration of mass around lattice sites. 

Unfortunately, however, as we measure $\hat b_2$, we will inevitably modify the quantum state of the system, and this modification must be reflected back in {$\langle \hat x \rangle$} --- in order to {avoid} the Page-Geilker argument.   According to discussions in Section~\ref{sec:nonlinear_QM_vs_measurement}, in the causal-conditional formulation, we take  \R{$\langle \hat x \rangle$} as the {\it conditional} expectation value of $\hat x$~\cite{helou2019testing,scully2022semiclassical}, {denoted as $x_c$}.  This has been computed by Liu et al., showing that the radiation pressure noise will now peak at $\omega_m$~\cite{Yubao}.

In the rest of this section, after briefly describing the theoretical foundations for analyzing CCSN (in terms of feedback), we shall consider time delayed measurements, as well as non-stationary measurements.  By not letting gravity  receive classical information during certain time intervals, {we} will eventually be able to recover the pre-selection {signatures} to some extent. 

\subsection{Schr\"odinger-Newton with Bad Cavity and Delayed measurement}
\label{sec:self_gravity_SN_badcavity}
Let us now introduce the use of Heisenberg Equations and Wiener filtering to CCSN, and apply them to a time-delayed measurement scheme for our single-object optomechanical system. The time delay is added with the aim of preventing classical gravity from following the quantum collapse, hence leading to more distinct experimental signatures. 

\subsubsection{Heisenberg Equations}

The Heisenberg Equations of Motion for the position and momentum of a mechanical oscillator with mass $M$ and eigenfrequency $\omega_m$, under continuous measurement, within CCSN  is given by 
\begin{equation}
    \dot{\hat{x}}=
    {\hat{p}}/{M},\; 
    \dot{\hat{p}}=
    -M\omega_Q^2\hat{x}-2\gamma\hat{p}+f_{\rm th}+\hbar\alpha\hat{a}_1+M\omega_{\rm SN}^2x_{\rm c}\,,\end{equation}
where  $\gamma$ and $f_{\rm th}$ are the mechanical dissipation and the fluctuating thermal force whose spectrum can be obtained through the dissipation fluctuation theorem. Using the Fourier transformation convention of  
\begin{equation}
    F(\omega)=\int_{-\infty}^\infty dte^{i\omega t}\tilde{F}(t) \,,
\end{equation}
 we have
\begin{align}
    \hat{x}(\omega)&=\chi_Q(\omega)[\hbar\alpha\hat{a}_1(\omega)+M\omega_{\rm SN}^2x_{\rm c}(\omega)+f_{\rm th}(\omega)],
    \label{xfd}
\end{align}
where 
\begin{equation}
\chi_Q(\omega)=[{M(\omega_Q^2-2i\gamma\omega-\omega^2)}]^{-1}
\end{equation}
 is the susceptibility of mechanical mirror --- with the SN-modified frequency $\omega_Q$.
The mechanical position is affected by quantum radiation pressure, classical gravity, and classical thermal noise.
Then, we can decompose $\hat{x}$ into a purely quantum part and a classical part as
\begin{align}
\label{eqxHeis}
    &\hat{x}=\hat x_Q+x_{\rm cl},\; \hat x_Q = \hbar\alpha\chi_Q\hat{a}_1,\;
    x_{\rm cl}=\chi_Q(M\omega_{\rm SN}^2x_{\rm c}+f_{\rm th}).
\end{align}
Here we have used a simplified notation where multiplication of $\chi_Q$ really indicates a convolution:
\begin{equation}
    A = \chi_Q B \Leftrightarrow A(t) = \int_{-\infty}^t dt'\chi_Q(t-t')B(t')\,.
\end{equation}
Under the bad cavity condition, the output amplitude quadrature $\hat{b}_1$ and phase quadrature $\hat{b}_2$ (relative to the carrier field) are derived by the input-output relation as
\begin{align}
    \hat{b}_1=\hat{a}_1,\quad
    \hat{b}_2=\hat{a}_2+\alpha\hat{x}.
\end{align}
The general output quadrature is given by 
\begin{equation}
\hat{b}_{\zeta}=\hat{b}_1\cos\zeta+\hat{b}_2\sin\zeta\,,
\end{equation} 
with $\zeta$ sometimes referred to as the homodyne angle because a homodyne detection with a local-oscillator phase $\zeta$ can be used to detect $b_\zeta $ .  Using Eq.~\eqref{xfd}, we can decompose $\hat b_\zeta$ into quantum and classical parts, 
\begin{align}
    \hat{b}_{\zeta}
    &=\hat{b}_{\zeta{\rm Q}}+\alpha x_{\rm cl}\sin\zeta,\\
    \hat{b}_{\zeta{\rm Q}}&=(\cos\zeta+\hbar\alpha^2\chi_Q\sin\zeta)\hat{a}_1+\hat{a}_2\sin\zeta,
    \label{bzetaq}
\end{align}
where $\hat{b}_{\zeta{\rm Q}}$ is a quantum part of $\hat{b}_\zeta$.

In the above decomposition of operators into quantum and classical parts, we have assume that, during the measurement, the environment's quantum state has also been collapsed so that the value of the thermal force has become {\it classical}~\footnote{This cannot be strictly true since $f_{\rm th}$ here is responsible for driving the quantum zero-point fluctuation of the oscillator.  However, this can be approximately true when $k_BT \gg \hbar\omega_m$. }. In this way, $f_{\rm th}$, even though a {\it random process} for the experimentalist, is treated as {\it known} when gravity is to be generated.   Note that this is {not the only } possible theoretical prescription.  One can also choose to  treat $f_{\rm th}$ as quantum, which will lead to different experimental signatures, as discussed in\,\cite{Liu2024}.

\begin{figure}
    \includegraphics[width=0.30\textwidth]{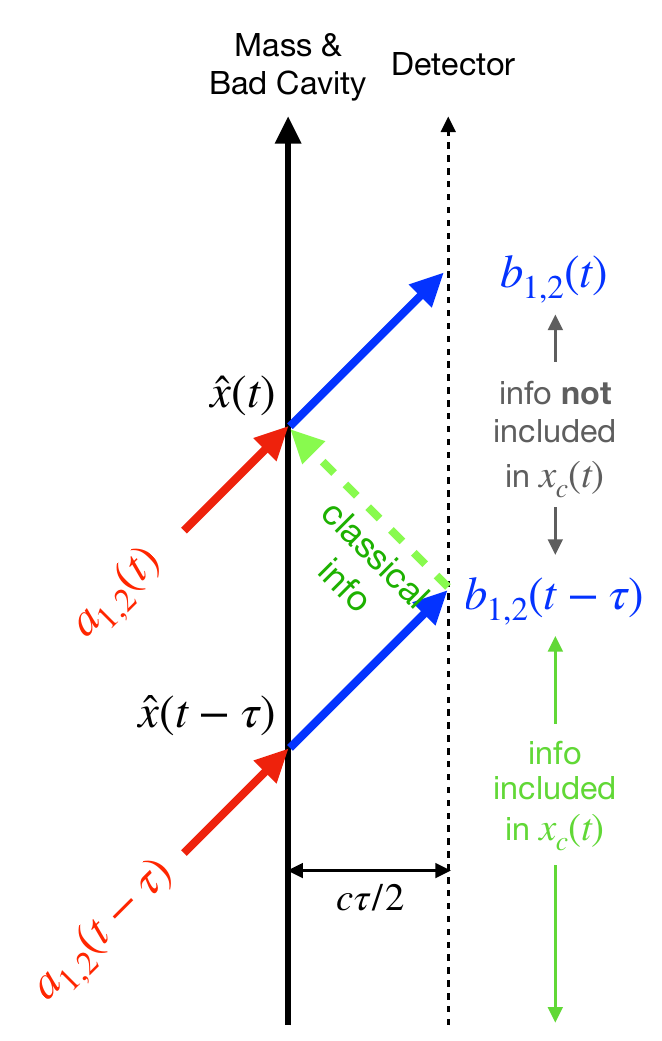}
    \caption{Spacetime diagrams showing a delayed measurement for a single mass inside a bad cavity. For incoming and out-going fields, we label their time coordinates using their arrival and departure time from the mass-bad-cavity system.  In this experiment, a delay of $\tau$ is created by placing the detector $c\tau/2$ away from the cavity.}
    \label{fig:diagram}
\end{figure}

\subsubsection{Determining the central position $x_c$ of the object's self-gravity potential}

\begin{figure}
    \includegraphics[width=0.5\textwidth]{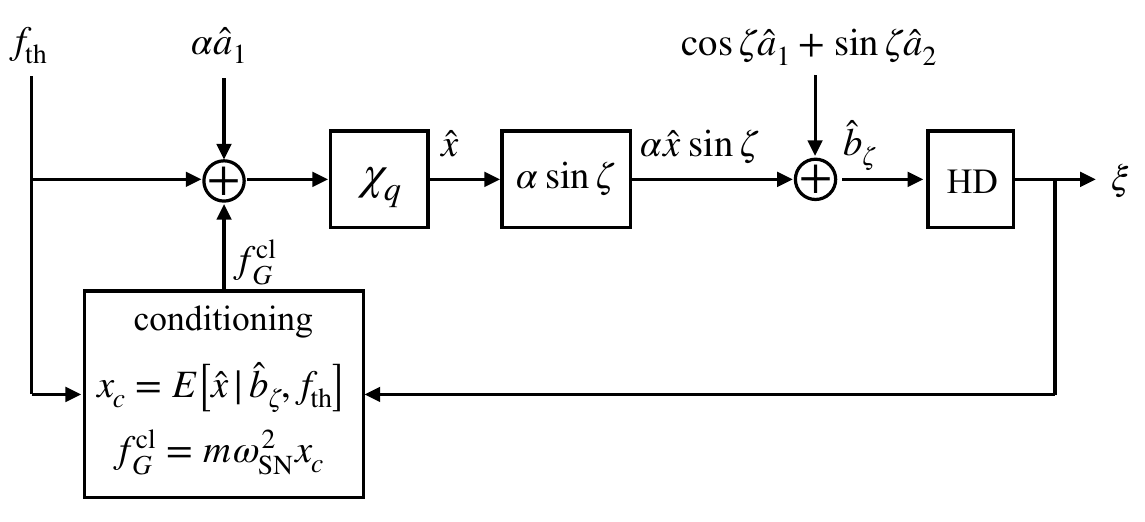}
    \caption{Block diagram illustrating the Heisenberg-picture equations of motion of a single test mass in the Causal Conditional Schr\"odinger-Newton (CCSN) theory.  The displacement $\hat x$ responds to quantum radiation pressure force $\alpha \hat a_1$, thermal force $f_{\rm th}$ and classical gravity force $f_{G}^{\rm cl}$ via $\chi_q$, which has poles near $\pm \omega_Q$. Depending on the readout angle, $\hat x$ is linearly combined with quadrature fields to yield $\hat b_{\zeta}$, which is measured via homodyne detection (HD). The result of measurement $\xi$, as well as the thermal force $f_{\rm th}$, are used to determine the classical center $x_c$ of the object's self-gravity potential, which is then used to generate $f_G^{\rm cl}$. In delayed measurements, one can insert a time delay before HD. \label{fig:feedback}
    }
\end{figure}

So far, we have been treating $x_c(t)$, the position around which the object's self gravitational potential is centered, as a classical function of time.  Let us now introduce the causal conditional approach of determining the classical variable $x_c(t)$ with a delayed measurement. As shown in Figure~\ref{fig:diagram}, suppose the out-going field is detected at a distance $c\tau/2$ from the output of the cavity, so that for $t'<t-\tau$, the operators $\hat{b}_\zeta(t')$ are measured, and projected to a classical random process $\xi(t')$.  Then we have $x_c(t)$ determined as the conditional expectation of $\hat x(t)$:
\begin{equation}
\label{eqxcdelay}
x_c(t) = E\left[\hat x(t) \big| \{\hat b_\zeta(t') = \xi (t')|t'\le t-\tau\}\right]
\end{equation}
Physically, this embodies the assumption that classical gravity is generated by collecting measurement results made up till this time. Since $\hat b_\zeta$ consists of $\hat b_{\zeta Q}$ the quantum part, and a classical part that is already a function of $\xi$, we can alternatively consider the collapse of $\hat b_{\zeta Q}$ to a classical random  process $z(t)$, during $t' \le t-\tau$, and write
\begin{align}
    \xi=z+\alpha x_{\rm cl}\sin\zeta.
    \label{xi}
\end{align}
The probability for $\hat b_{\zeta Q}$ to collapse to a particular $z$ follows the Born's rule of quantum mechanics, while $x_{\rm cl}$ is determined consistently by taking conditional expectation of Eq.~\eqref{eqxHeis}
\begin{equation}
\begin{split}\label{eqxcdelay2}
    x_{\rm c}(t) =&E[\hat x_Q(t) |\hat b_{\zeta Q}(t')=z(t'),t'\le t-\tau]\\
    &+\chi_Q(M\omega_{\rm SN}^2 x_c+f_{\rm th})
\end{split}
\end{equation}
If we put the information extraction from the experimetalist's measurement result and from the environment on equal fottings, then we can write $x_c(t)$ as conditioned upon both the measurement result  during $t'<t-\tau$ and on the values of $f_{\rm th}(t')$, $t'\le t$ (see Fig~\ref{fig:feedback}), or:
\begin{equation}
    \label{x_c_expectation}
    x_c(t) = E\left[\hat x(t) \big|
    \left\{\hat b_{\zeta}(t'-\tau), f_{\rm th}(t'): t'\le t\right\}\right] 
\end{equation}

\subsubsection{Causal Wiener Filtering}

In order to implement Eq.~\eqref{x_c_expectation}, writing the conditional expectation of $\hat x(t)$ in terms of $z$, the projective measurement result for $\hat{b}_{\zeta Q}$, we first perform a Causal whitening of the field $\hat b_{\zeta Q}$. We first obtain the symmetrized spectrum of $S_{b_{\zeta_Q}b_{\zeta_Q}}$ by using Eq.~\eqref{bzetaq}, and then use
\begin{equation}
    S_{a_1 a_1}(\Omega)=S_{a_1 a_1}(\Omega)=1\,,\quad S_{a_1a_2}=0\,,
\end{equation}
we arrive at 
\begin{equation}
    \label{eq:sbbcausal}
    S_{b_{\zeta{\rm Q}}b_{\zeta{\rm Q}}}(\omega)=\underbrace{\frac{(\omega-\beta)(\omega+\beta^*)}{(\omega-\tilde\omega_Q)(\omega
    +\tilde\omega^*_Q)}}_{\phi_+(\omega)}
    \underbrace{\frac{(\omega-\beta^*)(\omega+\beta)}{(\omega-\tilde\omega^*_Q)(\omega+\tilde\omega_Q)}}_{\phi_-(\omega)}
\end{equation}
with 
\begin{equation}
    \tilde\omega_Q = \sqrt{\omega_Q^2-\gamma^2}-i\gamma\,.
\end{equation}
and 
\begin{align}
    \beta^2&=\omega_Q^2-2\gamma^2+\Lambda^2 \sin\zeta\cos\zeta\notag\\
    &-i\sqrt{4\gamma^2(\omega_Q^2-\gamma^2)+2\gamma^2\Lambda^2 \sin2\zeta+\Lambda^4\sin^4\zeta},
\end{align}
where we have defined
\begin{equation}
    \label{eq:Lambda}
    \Lambda = \sqrt{{\hbar}/{M}} \alpha ={4\mathcal{F}}/{\pi}\sqrt{({2\omega_0 P_{\rm cav}})({Mc^2})}.
\end{equation}
Here $\tilde \omega_Q$ is a complex resonant frequency of the oscillator under both the harmonic potential and its self-gravity, while $\beta$ is a characteristic frequency that depends on the strength of the measurement. The more detailed factoraization is written as 
\begin{align}
    \label{factorSzz}
    &(\omega-\beta)(\omega+\beta^*)(\omega-\beta^*)(\omega+\beta)\nonumber\\
=&(\omega-\tilde\omega_Q)(\omega+\tilde\omega_Q^*)(\omega-\tilde\omega_Q^*)(\omega+\tilde\omega_Q)+\Lambda^4\sin^2\zeta \nonumber\\
-&\Lambda^2\sin\zeta\cos\zeta[(\omega-\tilde\omega_Q)(\omega+\tilde\omega_Q^*)+(\omega-\tilde\omega_Q^*)(\omega+\tilde\omega_Q)]
\end{align}
With $\mathrm{Re}\tilde\omega_Q, \mathrm{Re}\beta>0$ and $\mathrm{Im}\tilde\omega_Q,\mathrm{Im}\beta<0$, Eq.~\eqref{eq:sbbcausal}, is in a {\it causally factored form} of $S_{b_{\zeta_Q}b_{\zeta_Q}}=\phi_+ \phi_-$:
the factor $\phi_+(\omega)$ has poles and zeros on the upper-half complex plane, while $\phi_-(\omega)$ has poles and zeros on the lower-half complex plane. 

\begin{table}[t]
    \centering
    \scalebox{0.9}[0.9]{
    \begin{tabular}{ccccc}
    \hline\hline
    Parameters & Symbol & & Value & \\
    & & Figure~5 & Figure~8 & Figure~9 \\\hline
    Mirror bare frequency & $\omega_m/2\pi$ & $10$~{\rm mHz} & $10$~{\rm mHz} & $100$~{\rm mHz}\\
    SN frequency & $\omega_{\rm SN}/2\pi$ & $57$~{\rm mHz} & $57$~{\rm mHz} & $57$~{\rm mHz}\\
    Quality factor & $Q_m$ & $10^7$ & $3\times10^6$ & $3\times10^7$\\
    Mechanical damping & $2\gamma/2\pi$ & $1$~{\rm nHz} & $3.3$~{\rm nHz} & $3.2$~{\rm nHz}\\ 
    \hline
    Temperature & $T$ & $1$~{\rm mK} & $300$~{\rm K} & $300$~{\rm mK}\\       
\begin{tabular}{c} Thermal occupation \\ number divided by $Q_m$
\end{tabular} &  $n_{\rm th}^c$  & $200$ & $2\times10^6$ & $2000$\\
\hline
    Mirror mass & $M$ & $1$~{\rm mg} & $1$~{\rm mg} & $1$~{\rm g}\\
    Optical wavelength & $\lambda$ & $1064$~{\rm nm} & $1064$~{\rm nm} & $1064$~{\rm nm}\\
    Cavity finesse & $\mathcal{F}$ & $100$ & $275$ & $1000$\\
    Input-cavity power & $P_{\rm cav}$ & $1$~{\rm nW} & $1$~{\rm mW} & $100$~{\rm mW}\\       
    Optomechanical coupling & $\Lambda/2\pi$ & {$57$\,{mHz}} & $350$~{\rm Hz} & $400$~{\rm Hz}\\
    \hline\hline
    \end{tabular}}
    \caption{The parameters of the optomechanical single device expected for tabletop experiments.
    We use the SN frequency for Tungsten \cite{Helou17}.
    If extended to the low-frequency range, the experiment of low-dissipation milligram-scale mirror \cite{matsumoto20} could be promising. }
    \label{tab:my_label}
\end{table}

We are now ready to make a causal whitening of $\hat b_{\zeta_Q}$, defining 
\begin{align}
\hat    w_Q&=\hat b_{\zeta Q}/\phi_+\notag\\
&=\frac{(\omega^2+2i\omega\gamma-\omega_Q^2)(\cos\zeta\hat{a}_1+\sin\zeta\hat{a}_2)-\Lambda^2\sin\zeta\hat{a}_1}{(\omega-\beta)(\omega+\beta^*)}
\end{align}
From Eq.~\eqref{eqxcdelay2}, we have
\begin{equation}
    x_c(t) = \int_{-\infty}^{t-\tau} dt' \langle \hat x_Q (t) \hat w_Q(t')\rangle \hat w_Q(t') +\chi_Q(M\omega_{\rm SN}^2 x_c+f_{\rm th})
    \label{xcdelay}
\end{equation}
This can be rewritten using the Causal Wiener filter (see e.g., Ref.~\cite{Ebhardt09}), leading to 
\begin{align}
    x_c = K_\tau z  +\chi_Q(M\omega_{\rm SN}^2 x_c+f_{\rm th}),\;\;
    K_\tau=
    \big[{S_{xb_{\zeta{\rm Q}}}}/{\phi_-}\big]_\tau/\phi_+.
    \label{K}
\end{align}
Here $S_{x b_{\zeta_Q}}$ is the cross spectrum between $x$ and $b_{\zeta_Q}$, and we have used $[f]_+$ and $[f]_-$ to describe the causal and non-causal parts of $f$. For $f$ whose Fourier transform is a rational function of $\omega$, $[f]_+$ has poles only in the lowe
r-half complex plane, while $[f]_-$ has poles only in the upper half complex plane. The symbol $[...]_\tau$, $\tau>0$, represents inverse Fourier transforming a frequency domain function, taking the part $t\ge \tau$ and then Fourier transform back (see Appendix~\ref{apdx:wiener_delay}). In particular, for a complex number $z$ and function $f$,
\begin{equation}
    f(\omega ) = 1/({\omega - z}),
\end{equation}
we have 
\begin{equation}
\label{defbrackettau}
    [f(\omega )]_\tau = \left\{
    \begin{array}{ll}
    0 & \mathrm{Im} z >0,\\
     \displaystyle {e^{i(\omega-z)\tau}}/{(\omega-z)} & \mathrm{Im} z<0.
    \end{array}
    \right.
\end{equation}
The Wiener filter $K_\tau$ can be obtained analytically as 
\begin{align}\label{eq:Ktau}
    K_\tau  =& \sqrt{\frac{\hbar}{M}}\frac{\Lambda}{(\omega-\beta)(\omega+\beta^*)(\tilde\omega_Q+\tilde\omega_Q^*)}\nonumber\\
    &\qquad\Bigg[ \frac{(\Lambda^2\sin\zeta+4i\gamma\tilde\omega_Q \cos\zeta)(\omega+\tilde\omega_Q^*)e^{i(\omega-\tilde\omega_Q)\tau} }{(\tilde\omega_Q+\beta)(\tilde\omega_Q-\beta^*)} 
    \nonumber\\
    &\quad\;\;-\frac{(\Lambda^2\sin\zeta-4i\gamma\tilde\omega_Q^*\cos\zeta)(\omega-\tilde\omega_Q)e^{i(\omega+\tilde\omega_Q^*)\tau} }{(\tilde\omega_Q^*-\beta)(\tilde\omega_Q^*+\beta^*)} 
    \Bigg]
\end{align}

Solving Eq.~\eqref{xcdelay}, we can write the conditional expectation of mass position $x_c$ and the measurement result $\xi$ as driven by $z$ and $f_{\rm th}$:~\footnote{Recall that $({\chi_c-\chi_Q})/{\chi_Q}=M\omega_{\rm SN}^2$.}
\begin{align}
x_c &=\chi_c f_{\rm th}+({\chi_c}/{\chi_Q})K_\tau z \\
\xi & = \left[1+\alpha M\omega_{\rm SN}^2 \chi_c K_\tau \sin \zeta\right] z +\chi_c f_{\rm th} \alpha\sin\zeta
\end{align}
We recall that the spectrum of $z$ is the same as $S_{b_{\zeta_Q}b_{\zeta_Q}}$ [given by Eq.~\eqref{eq:sbbcausal}] and the spectrum of $f_{\rm th}$ is given by the Fluctuation-Dissipation Theorem
\begin{equation}
    S_{f_{\rm th}f_{\rm th}}(\omega)=4\hbar M\omega\gamma\coth[\hbar\omega/2k_BT]
    \end{equation}
where $T$ is the temperature of the heat bath the mass damps into. 
Here we have defined 
\begin{equation}
\chi_c(\omega)=1/[M(\omega_m^2-2i\gamma\omega-\omega^2)]  
\end{equation}
the susceptibility of the oscillator with its orignal eigenfrequency.   Note that the thermal noise contribution to both $x_c$ and $\hat{b}_\zeta$ are the same as a classical oscillator --- this is very important since this indicates in the classical regime the mass behaves as an oscillator with $\omega_m$, {and is consistent with observations.} The spectrum of $\xi$ for time delay $\tau$ and quadrature $\zeta$ is given by 
\begin{align}
    S_{\xi\xi}^{\rm SN}&=
     \left|1+\alpha\chi_cM\omega_{\rm SN}^2K_\tau\sin\zeta\right|^2S_{zz}
    +\left|\alpha\chi_c\sin\zeta\right|^2S_{f_{\rm th}f_{\rm th}}\,.
    \label{sxixi_gen}
 \end{align}

\begin{figure}
    \includegraphics[width=0.475\textwidth]{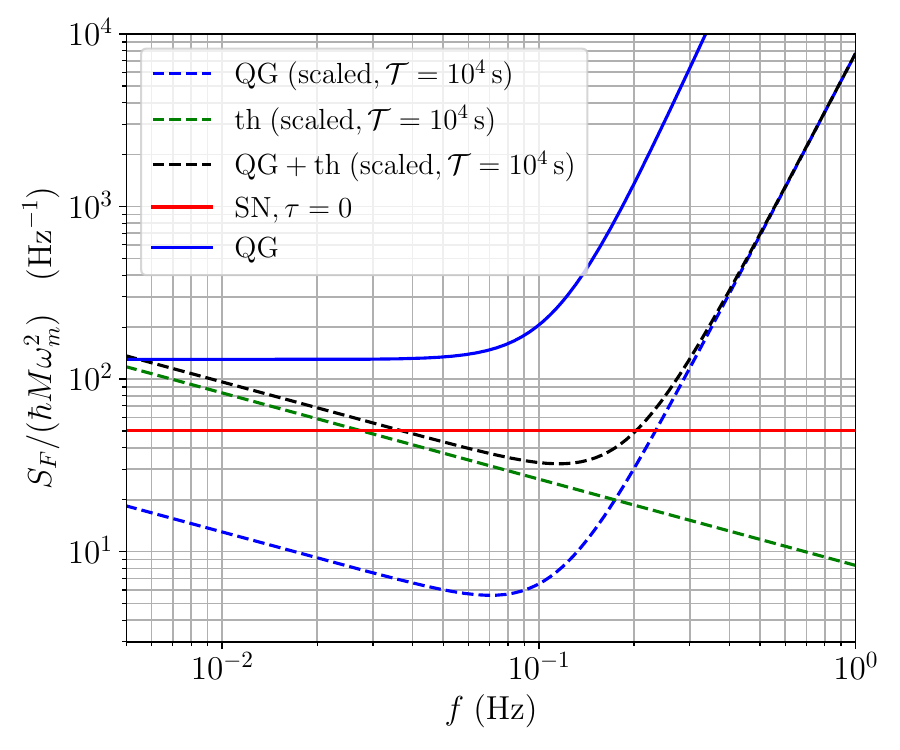}
    \caption{Comparison betweenm $S_F^{\rm SN}/(\hbar M\omega_m^2)$ and the corresponding scaled spectra from QG and thermal contributions (by $\sqrt{2\pi/(\Omega \mathcal{T}
    )}$), their sum, as well as the original QG contribution, for $\omega_m=2\pi\times 10\,$mHz, $T=1\,$mK, $Q_m=10^7$, {$\Lambda=\omega_{\rm SN}=2\pi\times 57\,$mHz} and $\mathcal{T}=10^4$\,sec. Since the SN contribution is above the sum of the QG and thermal contributions, it is detectable from this configuration.   \label{fig:SNbudget}
    }
 \end{figure}

    \subsubsection{Noise Spectra for non-Delayed Measurement}

Before discussing the general case, let us first specialize to $\tau=0$, which has been obtained by Ref.~\cite{Yubao} using Stochastic Schroedinger Equations.  \begin{equation}\label{eq:K}
    K_{\tau=0} = \frac{1}{\alpha \sin\zeta}\left[1-\frac{\omega^2+2i\omega\gamma -\omega_Q^2}{(\omega-\beta)(\omega+\beta^*)}\right]
 \end{equation}
 and the measurement result becomes
 \begin{equation}
   \xi_{\tau=0} = \frac{(\omega-\beta)(\omega+\beta^*)+\omega_{\rm SN}^2}{(\omega-\beta)(\omega+\beta^*)} \frac{\chi_c}{\chi_Q} z+\chi_cf_{\rm th} \alpha\sin\zeta\,.
 \end{equation}
 Note that the $\chi_Q$ on the denominator and the $\chi_c$ on the numerator will shift the frequencies around $\omega_Q$ to around $\omega_c$, erasing the ``additional peak'' in the SN prediction. Indeed, we have the spectrum of the measured optical field as 
 \begin{align}
    S_{\xi\xi}^{\rm SN}=\frac{\left|(\omega+\beta)(\omega-\beta^*)+\omega_{\rm SN}^2\right|^2}{\left|\omega_m^2-2i\gamma\omega-\omega^2\right|^2}
     +\left|\alpha\chi_c\sin\zeta\right|^2S_{f_{\rm th}f_{\rm th}}
    \label{sxixi_nodelay}
 \end{align}
which returns to standard quantum mechanics when $\omega_{\rm SN}=0$. Even in the case $\omega_{\rm SN}>0$, the thermal-noise contribution (the second term) always agrees with standard quantum mechanics. By contrast to the {\it pre-selection} model, this spectrum only has a peak around the {\it classical mechanical resonance frequency}, making it less distinct from standard quantum mechanics.   These results are consistent with that of the previous work\,\cite{Yubao}.

 \begin{figure}
    \includegraphics[width=0.475\textwidth]{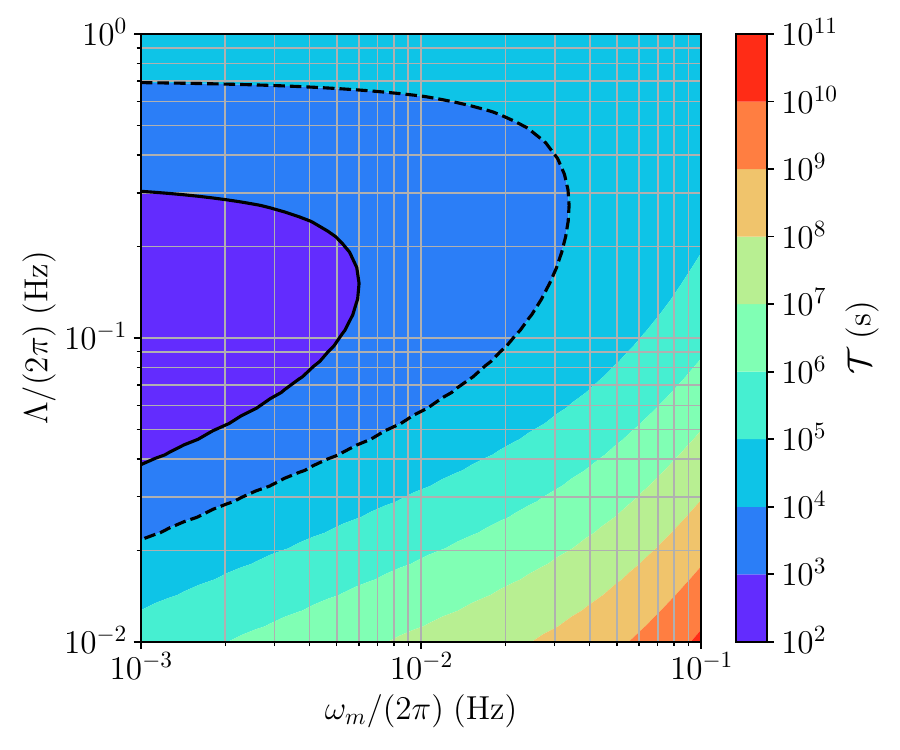}
    \caption{Contour plot of the observation time {$\mathcal{T}$} required in order to achieve Kullback-Leibler SNR $\rho_{\rm SN}^2=1$, fixing $T=1\,$mK, $Q=10^7$.
    The solid and dashed  contours indicate ${\mathcal{T}}=10^3\,$s and $10^4\,$s respectively. \label{fig:nodelaycontour}}
    \end{figure}

 To be more specific, let us focus on the out-going phase quadrature ($\zeta=\pi/2$), which contains displacement of the mirror. Let us also consider $F \equiv \xi/(\alpha\chi_c)$, which is referred to the force acting on the mass, which has a spectrum of 
 \begin{equation}
    S_F =S_F^{\rm QG} + \Delta S_F^{\rm SN} +S_F^{\rm th}
 \end{equation}
 with
 \begin{align}
S_F^{\rm QG}/(\hbar  M\omega_m^2)&=\frac{(\omega^2-\omega_m^2)^2}{\Lambda^2\omega_m^2}+\frac{\Lambda^2}{\omega_m^2}\,, \\
S_F^{\rm th}/(\hbar  M\omega_m^2)&= 4n_{\rm th}^c\,, \\
\Delta S_F^{\rm SN}/(\hbar  M\omega_m^2)&=-\frac{2\Lambda^2\omega_{\rm SN}^2/\omega_m^2}{\omega_Q^2+\sqrt{\Lambda^4+\omega_Q^4}}\,.
 \end{align}
 Here we have defined $Q_m = \omega_m/(2\gamma)$ and $n_{\rm th}^c$ is the thermal occupation number of the mechanical oscillator divided by $Q_m$, 
 \begin{equation}
    \label{eq:nth}
n_{\rm th}^c\equiv     k_B T/(\hbar\omega_m Q_m) 
 \end{equation}
 which characterizes the quantumness of the oscillator. 
 The two terms in the force-referred noise $S_F^{\rm QG}$ corresponds to shot noise and radiation-pressure noise, the thermal noise $S_F^{\rm th}$ is constant, while the SN correction $\Delta S_F^{\rm SN}$ appears as a constant reduction to radiation-pressure noise, which is also constant.

The SN correction term $\Delta S_F^{\rm SN}$'s  lack of spectral features makes this correction hard to identify in practice.  Nevertheless, in principle, the $\Lambda$ dependence of this term might allow it to be distinguished from the two other terms.  We can consider the regime where thermal noise dominates over quantum noise (which is inevitable for the low mechanical frequency band we will consider), and $\Lambda$ greater than several times of $\omega_Q$, one will have to discern a change in spectrum of $2\omega_{\rm SN}^2/\omega_m^2$ from the thermal background.  For an observation bandwidth of $\Gamma\sim\omega_m$ and observation time of $\mathcal{T}$, the error one can discern from thermal noise is 
 {\begin{equation}
     \frac{\Delta S_F^{\rm th}}{\hbar M\omega_m^2}\sim \frac{S_F^{\rm th}}{\hbar M\omega_m^2\sqrt{\Gamma\mathcal{T}}}\sim \frac{S_F^{\rm th}}{\hbar M\omega_m^2\sqrt{\omega_m\mathcal{T}}}
 \end{equation}}
therefore we require {$\Delta S_F^{\rm SN}/(\hbar  M\omega_m^2)\stackrel{>}{_\sim} \Delta S_F^{\rm th}/(\hbar  M\omega_m^2)$, }
or
\begin{equation}
\frac{2 k_B T}{\hbar\omega_{\rm SN}Q_m} \stackrel{<}{_\sim} \frac{\omega_{\rm SN}}{\omega_m}\sqrt{\omega_m\mathcal{T}}
\end{equation}
We can also phrase this in the minimum observation time required to discern the difference, despite the fact that it is difficult to keep experimental parameters stationary over a long intergration time.  The expression is given by 
\begin{align}
    \mathcal{T}_{\rm min} &\approx \frac{\omega_m}{\omega_{\rm SN}^2}\left(\frac{2 k_B T}{\hbar\omega_{\rm SN}Q_m}\right)^2 \nonumber\\
    &=3\times 10^3\,{\rm s}\left(\frac{\omega_m/(2\pi)}{ 10\,{\rm mHz}}\right)\left(\frac{1\,{\rm mK}}{T}\right)^2\left(\frac{10^7}{Q_m}\right)^2\left(\frac{57\,{\rm mHz}}{\omega_{\rm SN}/(2\pi)}\right)^2
\end{align} 
Note that we may be required to use oscillators with $\omega_m$  less than $\omega_{\rm SN}$ in order to meet the requirement. 
We should also check quantum noise, with 
 \begin{equation}
    \Delta S_F^{\rm QG} \sim \frac{\Lambda^2}{\omega_m^2}(\omega_m \mathcal
    {T})^{-1/2}
    \end{equation}
which requires
\begin{equation}
    \mathcal{T}_{\rm min} \approx \left(\frac{\Lambda}{\omega_{\rm SN}}\right)^4\frac{1}{\omega_m}
\end{equation}
This condition can be met if $\Lambda$ is a few times $\omega_{\rm SN}$ and the observation time is substantially longer than the classical period of the oscillator.

\begin{figure*}
    \includegraphics[width=\textwidth]{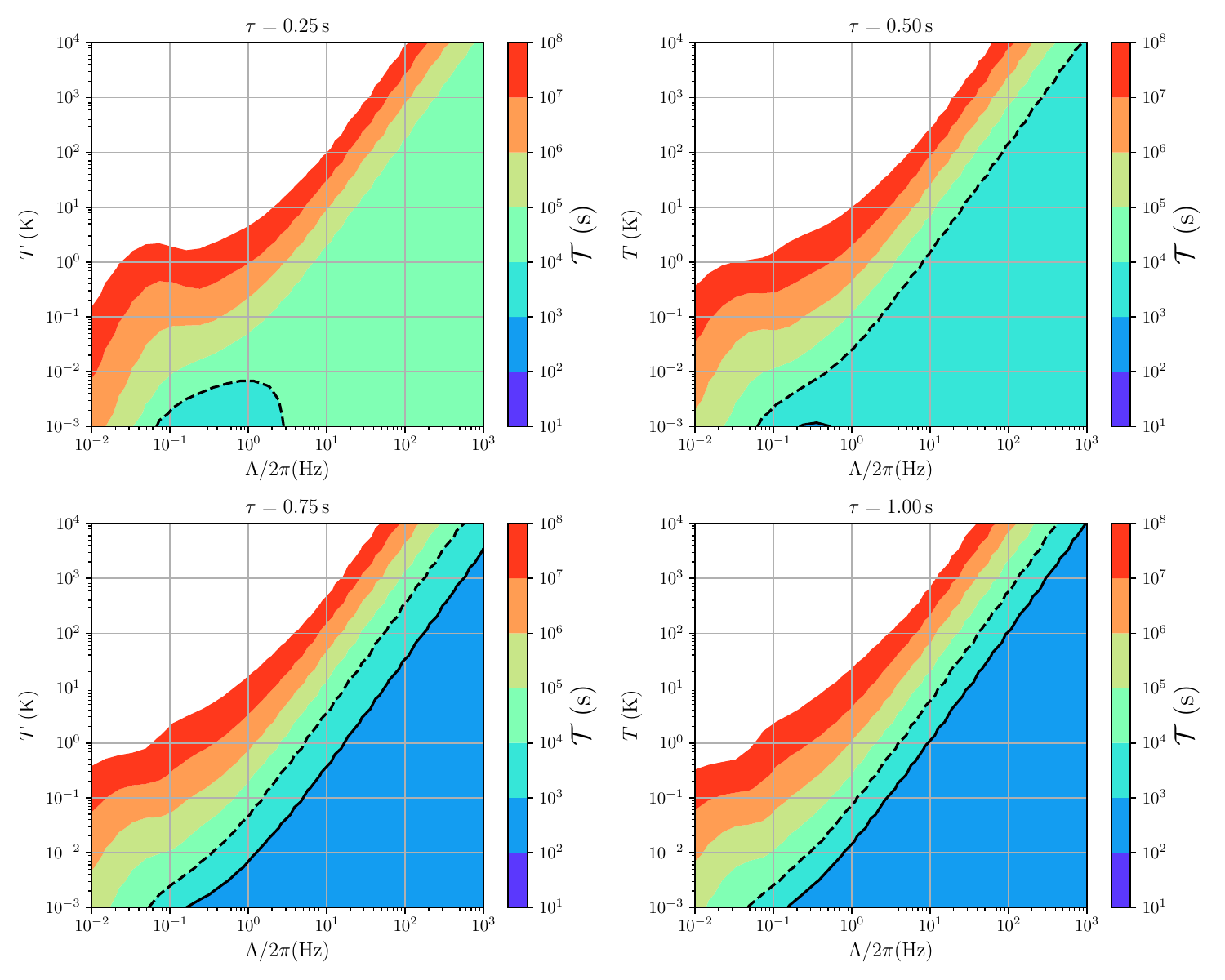}
    \caption{Required observation time {$\mathcal{T}$} for delayed measurements with $\omega_m=2\pi\,10$\,mHz, $Q_m=10^7$ in order to reach $\rho_{\rm SN}^2$=1\,.  Further increasing integration time will improve $\rho_{\rm SN}^2$ by a factor of $\sqrt{\mathcal{T}}$.  The solid and dashed contours indicate ${\mathcal{T}}=10^3$\,s and $10^4$\,s respectively.  \label{fig:delaycontour}}
 \end{figure*}

We can also perform a more detailed analysis, by using the Kullback-Leibler (KL) divergence rate between the standard quantum mechanics spectrum and the one with CCSN. The KL SNR is given by 
\begin{align}
   \rho^2_{\rm SN}&= \mathcal{T} \int \frac{d\Omega}{2\pi}\left(\frac{\Delta S_F^{\rm SN}}{S_F^{\rm th}+S_F^{\rm QG}}\right)^2
   \nonumber\\
   &=\int \frac{d\Omega}{\Omega}\left(\frac{\Delta S_F^{\rm SN}}{\sqrt{\frac{2\pi}{\Omega \mathcal{T}}} (S_F^{\rm th}+S_F^{\rm QG})}\right)^2
\end{align} 
This motivates us to compare $\Delta S_F^{\rm SN}$ with a characteristic QG and thermal spectrum of $\sqrt{2\pi/(\Omega \mathcal{T})}S_F^{\rm QG}(\Omega)$  and 
$\sqrt{2\pi/(\Omega \mathcal{T})}S_F^{\rm th}(\Omega)$ and the effect of SN is detectable if the SN curve dips below the sum of the QG and thermal curves  by a decade of frequency in a log plot. For parameters $\omega_m=2\pi\times 10\,$mHz, $T=1\,$mK, $Q_m=10^7$ and $\mathcal{T}=10^4$\,sec, we plot $\Delta S_F^{\rm SN}/(\hbar M\omega_m^2)$ and the scaled version of thermal and QG noise in Figure~\ref{fig:SNbudget}.  We can see that our sensitivity to the SN correction is limited by thermal noise at low-frequencies, and QG noise at high frequencies. Nevertheless, 
the SN correction is detectable in this configuration.  For a larger parameter space, fixing $T$, $Q_m$ we show the contour plot of the figure of merit $\rho_{\rm SN}^2$ as a function of $\omega_m$ and $\Lambda/\omega_m$ in Figure~\ref{fig:nodelaycontour}.  There exist a parameter space in which CCSN is detectable; we note that the $\omega_m$ tends to be lower than $\omega_{\rm SN}$, and that power should be chosen at an optimal value rather than at the highest possible value.

\begin{figure}
    \includegraphics[width=0.475\textwidth]{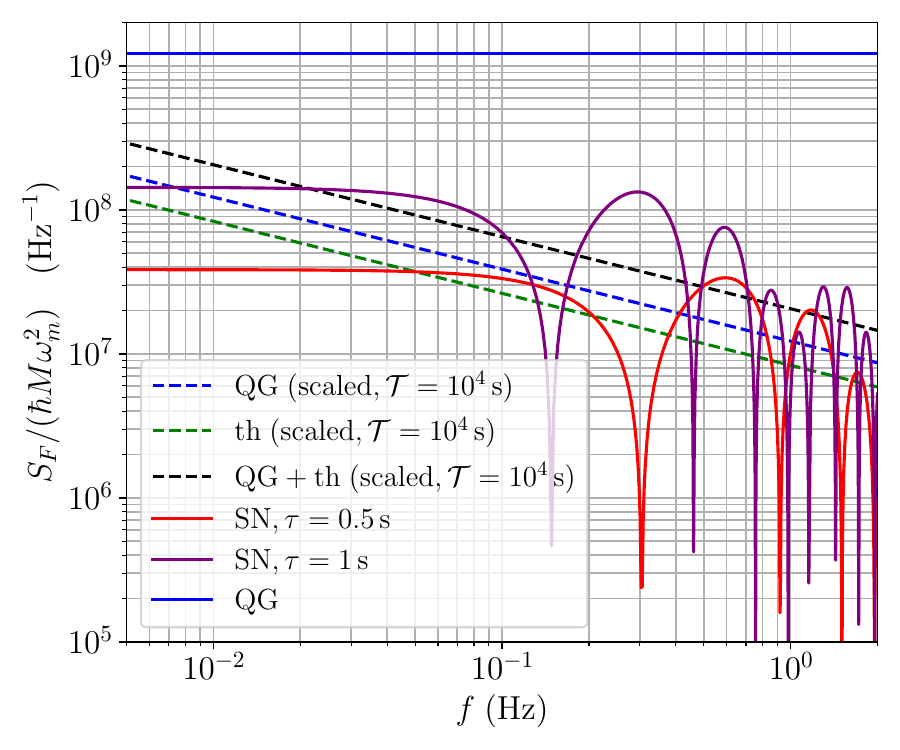}
    \caption{Comparison between $\Delta S_F^{\rm SN}/(\hbar M\omega_m^2)$ and the corresponding scaled spectra from QG and thermal contributions (by $\sqrt{2\pi/(\Omega {\mathcal{T}})}$), their sum, as well as the original QG contribution, for $\omega_m=2\pi\times 10\,$mHz, $T=300\,$K, $Q_m=3\times 10^6$ and $\mathcal{T}=10^4$\,sec. We have used delays of 0.5\,s and 1\,s, and assumed a strong measurement with $\Lambda=2\pi\times 350$\,Hz.  Both configurations the SN contributions are detectable with $\rho_{rm SN}^2 =1.8$ and 13, respectively.     \label{fig:SNbudgetdelay}
    }
 \end{figure}

In summary the CCSN spectrum, although lacking distinct spectral features, does still deviate from the QG spectrum enough to be discerned, if the appropriate calibration can be made.

\subsubsection{Effect of Time Delay}

Turning to the most general case, as it turns out, even with the time delay $\tau$, the zeros in the filter function in from of $S_{zz}$ will cancel poles near $\omega_Q$ and lead to poles around $\omega_m$. Specializing to $\gamma \ll\omega_m$, the form of the spectrum can be written in a compact form. In this case Eq.~\eqref{sxixi_gen} becomes
\begin{equation}
    \label{eq:sxixidelay}
S_{\xi\xi}^{\tau}(\omega) =\left[
    {|\mathcal{Q}(\omega)|^2}+4\Lambda^2 n_{\rm th}^c\omega_m^2 \sin^2\zeta\right]/{\left|\omega_m^2-2i\gamma\omega-\omega^2\right|^2}
\end{equation}
with
\begin{align}
    \mathcal{Q}(\omega)&=(\omega+\beta)(\omega-\beta^*)\nonumber\\
    &+\omega_{\rm SN}^2\Bigg\{
        1+\frac{\Lambda^2\tau 
        e^{i \omega\tau/2}\sin\zeta}{2\omega_q}\Big[
    e^{i\delta}\mathrm{sinc}\frac{(\omega-\omega_q)\tau}{2} \nonumber\\
    &\qquad\qquad\qquad \qquad\qquad\;\;+e^{-i\delta}\mathrm{sinc}\frac{(\omega+\omega_q)\tau}{2} \Big]
    \Bigg\}
\end{align}
where
\begin{equation}
e^{i\delta} ={ie^{i\omega_q\tau/2}\Lambda^2\sin\zeta}/{[(\omega_q-\beta)(\omega_q+\beta^*)]}
\end{equation}
which clearly reduces to Eq.~\eqref{sxixi_nodelay} when $\tau=0$.  We also see that corrections due to time delay $\tau$ does not have peaks  at $\pm\omega_q$. 

In terms of parameter space, we will fix $\omega_m=2\pi\times 10\,$mHz, $Q_m=10^7$ and vary $\Lambda$ and temperature $T$, and plot required ${\mathcal{T}}$ to reach $\rho_{\rm SN}^2=1$, for various values of $\tau$ between 0 and 1\,s.  The results are shown in Figure~\ref{fig:delaycontour}.   It is very interesting to see that as $\tau$ becomes larger than $\sim 0.25$\,sec, required integration time becomes shorter even for higher temperatures --- as one increases measurement strength $\Lambda$. Here we note that the required integration time is the same as long as $QT$ is fixed, while the increase of integration time will lead to SNR $\sim \sqrt{{\mathcal{T}}}$.  This is different from the non-delayed case, where increasing $\Lambda$ beyond a few times $\omega_q$ does not bring benefit. In the non-delayed case, this was presumably because measurement result is promptly used to create gravity, thereby making measurement stronger does not induce more difference from standard quantum gravity.  In this delayed case, the existence of the time delay allows the use of stronger measurement while still maintaining difference between QG and CCSN. In Figure~\ref{fig:SNbudgetdelay}, we consider a room-temperature setup with $\omega_m=2\pi\times 10\,$mHz, $T=300\,$K, $Q_m=3\times 10^6$, and a strong measurement with $\Lambda=2\pi\times 350\,$Hz.  We have used delays of 0.5\,s and 1\,s, and the SN contributions are detectable with $\rho_{\rm SN}^2 =1.8$ and 13, respectively.  This is a significant improvement over the non-delayed case.

Finally, we can examine Eq.~\eqref{eq:sxixidelay} more carefully to understand the above results. In the limit of $\Lambda\gg\omega_q, \omega$,  for $\zeta=\pi/2$ (measuring the out-going phase quadrature which contains displacement signal) we can very roughly write 
\begin{equation}
\frac{S_F}{\hbar M\omega_m^2 }
\approx 
\frac{\Lambda^2 }{\omega_m^2}\left[1+{\omega_{\rm SN}^2\tau^2}g(\omega)\right]+4n_{\rm th}^c
\end{equation} 
where $g(0)=1$ and characterizes the oscillations of the SN contributions shown in Figure~\ref{fig:SNbudgetdelay}. The three terms in the above equation corresponds to standard QG contribution, SN correction, and thermal noise, respectively.  From here, we can see clearly that a high $\Lambda$ will strongly suppress the thermal-noise contribution, thereby allowing the room-temperature operation discussed by Figure~\ref{fig:SNbudgetdelay}.  In order for thermal noise not to dominate over quantum noise, we require
\begin{equation}
    2n_{\rm th}\omega_m/\Lambda \stackrel{<}{_\sim} 1
\end{equation}
which is the familiar result of trapping-cooling experiments.  The threshold of $\tau$, on the other hand, is given by the condition that 
\begin{equation}
\tau \stackrel{>}{_\sim} \frac{1}{\omega_{\rm SN}}(\rho_{\rm SN}^2\Gamma {\mathcal{T}})^{-1/4} \sqrt{1+\left(\frac{2\omega_m}{\Lambda}\right)^2 n^c_{\rm th}}
\end{equation}
Here one might choose $\Gamma$ as the bandwidth where the approximation $g(\omega) \sim 1$ is valid, taking $\Gamma\sim \pi/\tau$.  This leads to the approximate formula of 
\begin{equation}
    \tau \stackrel{>}{_\sim}\frac{1}{\omega_{\rm SN}}\rho_{\rm SN}^{2/3} 
    \left[1+\left(\frac{2\omega_m}{\Lambda}\right)^2 n^c_{\rm th}\right]^{2/3}\left(\pi \omega_{\rm SN} {\mathcal{T}}\right)^{-1/3}
\end{equation}
Note that this formula is only valid for high values of $\Lambda$.  We can see that the delay will have to be at least a fraction of a second in order to take advantage of this regime.

\subsection{Non-stationary measurements}

As we have seen in the previous sections, while a quantum measurement process projects the quantum state of the object continuously, the gravitational potential follows the object, ``erasing'' the peak at $\omega_Q$ predicted by naive classical gravity theories.  The introduction of a time-delayed stationary measurement allowed oscillatory modifications to the spectrum of the out-going field which are also more distinctive from predictions of quantum gravity.  In this subsection, we consider a non-stationary experiment in which the mass was continuously measured during $t<0$, but has measurement turned off at $t=0$.  A follow-up measurement of position --- a state-verification experiment as described by Ref.~\cite{miao2010probing} --- will be performed at a particular moment of time with $t>0$. In this paper, we will only compute the uncertainty of $\hat x$ at $t>0$, without considering the additional sensing and back-action noise of the follow-up measurement.

\begin{figure}[t]
    \includegraphics[width=0.35\textwidth]{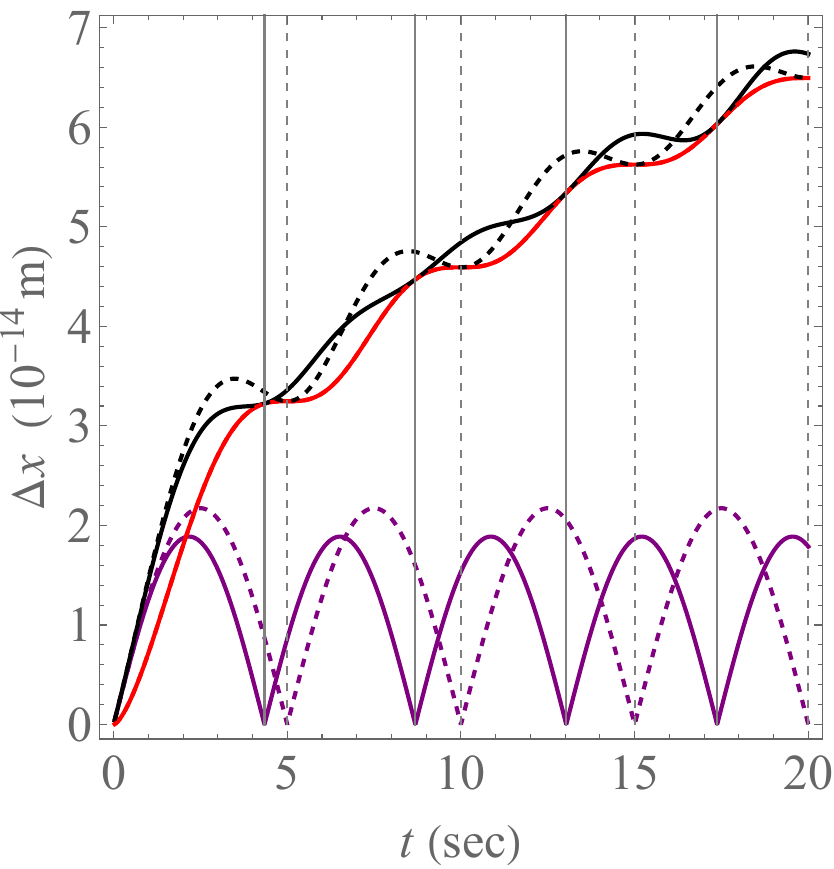}
\caption{Conditional variance of position (i.e., variance after removing best estimate from past measurements) at  $t>0$ after measurement is turned off at $t=0$. The solid purple curve corresponds to contribution from the quantum term, the solid red curve is contribution from classical term, while the black solid curve is the total variance. Solid and dashed vertical grid lines indicate $n\pi/\omega_Q$ and $n\pi/\omega_c$, respectively, with $n=1,2,3,\ldots$. Dashed purple and black curves are quantum and total variances when setting $\omega_{\rm SN}=0$.   \label{fig:nonstationary}}
\end{figure}

The position operator at a time $t>0$ can be given by 
\begin{align}
\hat x(t) &=\int_{-\infty}^{t} \hbar\alpha \chi_Q(t-t')\Theta(-t')\hat a_1(t') dt' \nonumber\\
&+\int_{-\infty}^t \chi_q (t-t')[M\omega_{\rm SN}^2 x_c(t') +f_{\rm th}(t') ]dt'
\end{align}
while the conditioning is only done over $t'<0$, since only during that time $\hat{b}_{\zeta Q}$ has information regarding the test mass's location.  We can also condition over the whitened version $\hat{w}_Q$ of $\hat{b}_{\zeta Q}$. We can then obtain
\begin{equation}
x_c =( {\chi_c}/{\chi_Q} ) E[\hbar\alpha\chi_Q \Theta \hat a_1|\Theta \hat w_Q] +\chi_c f_{\rm th}
\end{equation}
leading to 
\begin{equation}
    \hat x =\hbar\alpha \chi_Q\Theta \hat a_1 +M\omega_{\rm SN}^2\chi_c E [\hbar\alpha\chi_Q \Theta \hat a_1|\Theta \hat w_Q] +\chi_c f_{\rm th}
 \end{equation}
Here we have used the operator $\Theta$ to denote cutting off the $t>0$ part:
\begin{equation}
[\Theta \hat a_1 ](t) =\Theta(-t) \hat a_1 (t)\,.
\end{equation} 
We can further write
\begin{align}
\label{xtotHeisenberg}
    \hat x (t) &= \hbar\alpha\chi_Q\Theta\hat a_1 -
    E[\hbar\alpha\chi_Q\Theta\hat a_1|\Theta \hat{w}_Q]
    \nonumber\\ &+E\left[\hbar\alpha\chi_c\Theta\hat a_1|\Theta \hat{w}_Q\right]+\chi_c f_{\rm th}
\end{align}
We note that the first two terms correspond to a quantum uncertainty that resonates at frequency $\omega_Q$, while the next two terms correspond to classical uncertainty that resonates at $\omega_m$.

Let us now consider an observer who has measured the phase quadrature $\hat{b}_2$ during $t<0$. Note here that the experimental observer measures  the {\it total} $\hat{b}_2$ which is the sum of $\hat{b}_{2Q}$ and the classical contributions. 
\begin{equation}
    \hat{b}_2 = \frac{(\omega-\beta)(\omega+\beta^*)+\omega_{\rm SN}^2}{(\omega-\beta)(\omega+\beta^*)} \frac{\chi_c}{\chi_Q} \hat{b}_{2Q}+\chi_cf_{\rm th} \alpha\sin\zeta\,.
\end{equation}
As it turns out, we can write
\begin{align}
&S_{b_2 b_2}= \phi_+^{\rm tot}\phi_-^{\rm tot}\nonumber\\
\equiv& \frac{(\omega-\beta_{\rm tot})(\omega+\beta_{\rm tot}^*)}{(\omega-\tilde\omega_c+i\gamma)(\omega+\tilde\omega_c+i\gamma)}\cdot 
\frac{(\omega-\beta_{\rm tot}^*)(\omega+\beta_{\rm tot})}{(\omega-\tilde\omega_c-i\gamma)(\omega+\tilde\omega_c-i\gamma)},\\
\end{align}
leading to a whitened output of 
\begin{equation}
    \hat{w}_{\rm tot}={\hat{b}_2}/{\phi_+^{\rm tot}}
\end{equation}
Note that $\hat{w}_{\rm tot}$ only has poles at $(\beta,-\beta^*,\beta_{\rm tot},-\beta_{\rm tot}^*)$.  We can write
\begin{equation}
\hat    w_{\rm tot}=\frac{(\omega-\beta)(\omega+\beta^*)+\omega_{\rm SN}^2}{(\omega-\beta_{\rm tot})(\omega+\beta_{\rm tot}^*)}\hat w_Q-\frac{\alpha f_{\rm th}}{M(\omega-\beta_{\rm tot})(\omega+\beta^*_{\rm tot})}.
\end{equation}

Let us now compute the observer's conditional variance of $\hat x(t)$ with respect to $\hat{w}_{\rm tot}$:
\begin{equation}
V_{xx}^{\rm obs}(\tau)= E\Big[(\hat x(\tau) - E[\hat x(\tau) | \hat w_{\rm tot}(t') : t'<0])^2\Big]
\label{vnonstationary}
\end{equation}
Since $\hat w_{\rm tot}$ at $t<0$ is a linear combination of $\hat w_Q$ at $t<0$ and an additional independent noise, the first two terms of $\hat x(t)$ in Eq.~\eqref{xtotHeisenberg}, which is independent from $\hat w_Q$ and $f_{\rm th}$, will have a vanishing conditional expectation. We show the details of the computation of the third term in Appendix~\ref{apdx:nonstationary}.
\begin{widetext}
Note that the first two terms combined, the third, and fourth terms in $\hat x(\tau)$ are independent from each other, and the combination of the first two terms are independent from $\hat w_{\rm tot}$, we can assemble the conditional variance of $\hat x$ for the observer as 
\begin{equation}
    V_{xx}^{\rm obs}(\tau ) = \frac{1}{2}\int\limits_{-\infty}^{+\infty}\frac{d\Omega}{2\pi}\Bigg[
    \underbrace{\left|[\hbar\alpha\chi_Q]_\tau\right|^2 - \left|\big[[\hbar\alpha\chi_Q]_\tau S_{\hat a_1 \hat w_Q}\big]_\tau\right|^2}_{\mbox{\footnotesize first two terms of Eq.~\eqref{xtotHeisenberg}}}
    +\underbrace{\left|\big[[\hbar\alpha\chi_c]_\tau S_{\hat a_1 \hat w_Q}\big]_\tau\right|^2
    +|\chi_c |^2S_{f_{\rm th} }
    -
    \left|\Big[\big[[\hbar\alpha\chi_c]_\tau S_{\hat a_1 \hat w_Q}\big]_\tau S_{\hat w_Q \hat w_{\rm tot}}\Big]_\tau + \left[\chi_c S_{f_{\rm th} \hat w_{\rm tot}}\right]_\tau 
    \right|^2}_{\mbox{\footnotesize next two terms of Eq.~\eqref{xtotHeisenberg}}}
    \Bigg]\,.
    \label{vxx}
\end{equation}
\end{widetext}

In practice, the first two terms contribute to an oscillatory function that corresponds to the projection of a quantum-noise ellipse.  This ellipse rotates in phase space with frequency $2\omega_Q$  In the case when $\Lambda \gg \omega_Q$, the amplitude of oscillation in $\Delta x$ is given by $\Delta x_Q \sim \sqrt{\hbar \Lambda/(m\omega_Q^2)}$.  The next two terms contribute to both a linearly growing variance (hence $\Delta x  \propto \sqrt{\tau}$) and an oscillation on top of it; with the linear growth term driven by thermal noise and proportional to $\Delta x_{\rm th}\sim \sqrt{n_{\rm th}\tau/m}$. In order for the $\omega_Q$ oscillation to be clearly visible, we will need to impose
\begin{equation}
n_{\rm th} < \Lambda/\omega_Q\,.
\end{equation} 
However, a more precise measurement of the non-stationary evolution of $\Delta x$ over time can reveal a more subtle variation.  In Figure~\ref{fig:nonstationary}, we show $\Delta x(\tau)$ for $\Lambda=2\pi\times400~{\rm Hz}$ and $n_{\rm th}^c = 2000$ (instead of $2\times 10^6$). This requires a 300\,mK temperature for the test mass.  However, increasing the optical power or cavity finesse allows higher-temperature objects to have the same non-stationary signature. 

As we can see from the figure, in SN theory, the quantum contribution to $\Delta x$ (solid purple) indeed differs from standard quantum mechanics (dashed purple curve). In both SN and quantum mechanics, the classical contributions are identical for the parameters we consider, both shown in the solid red curve, and both oscillate at $\omega_c$.  With our parameters, the total conditional variance in the SN case is indeed distinguishable from standard quantum mechanics.

\section{Mutual Gravity Between Two Objects}
\label{sec:Mutual}

In this section, we consider experiments that test whether the 
 mutual gravity between two quantum objects can establish quantum correlations
following protocols proposed in Refs.~\cite{Miao,Datta,Miki}, which can be viewed as optomechanical versions of Refs.~\cite{marletto2017gravitationally,bose2017spin}

These protocols were motivated by the fact that, while a quantum interaction Hamiltonian $\sim \hat x_A \hat x_B$ can entangle two objects,  a classical one  $\sim \langle \hat x_A\rangle \hat x_B +\hat x_A \langle \hat x_B\rangle$ cannot.  Nevertheless, as one performs a test on correlations between masses A and B by measuring out-going light fields, classical information contained in these out-going fields can enter classical gravity field, hence lead to an apparent entanglement. This has indeed been observed by Ref.~\cite{Yubao}, and further elaborated by Ref.~\cite{Liu2024}. Even though such apparent entanglement falls within LOCC correlations, they do make testing of quantum gravity more challenging. 

In this section, we will apply the tools of Wiener filtering to analyze mutual classical gravity between two objects, and confirm that within CCSN they do produce correlations --- as Ref.~\cite{Yubao} found earlier.  In this section, we shall ignore the effect of classical self gravity. 

\begin{figure*}[tbp]
    \centering
    \includegraphics[width=0.65\textwidth]{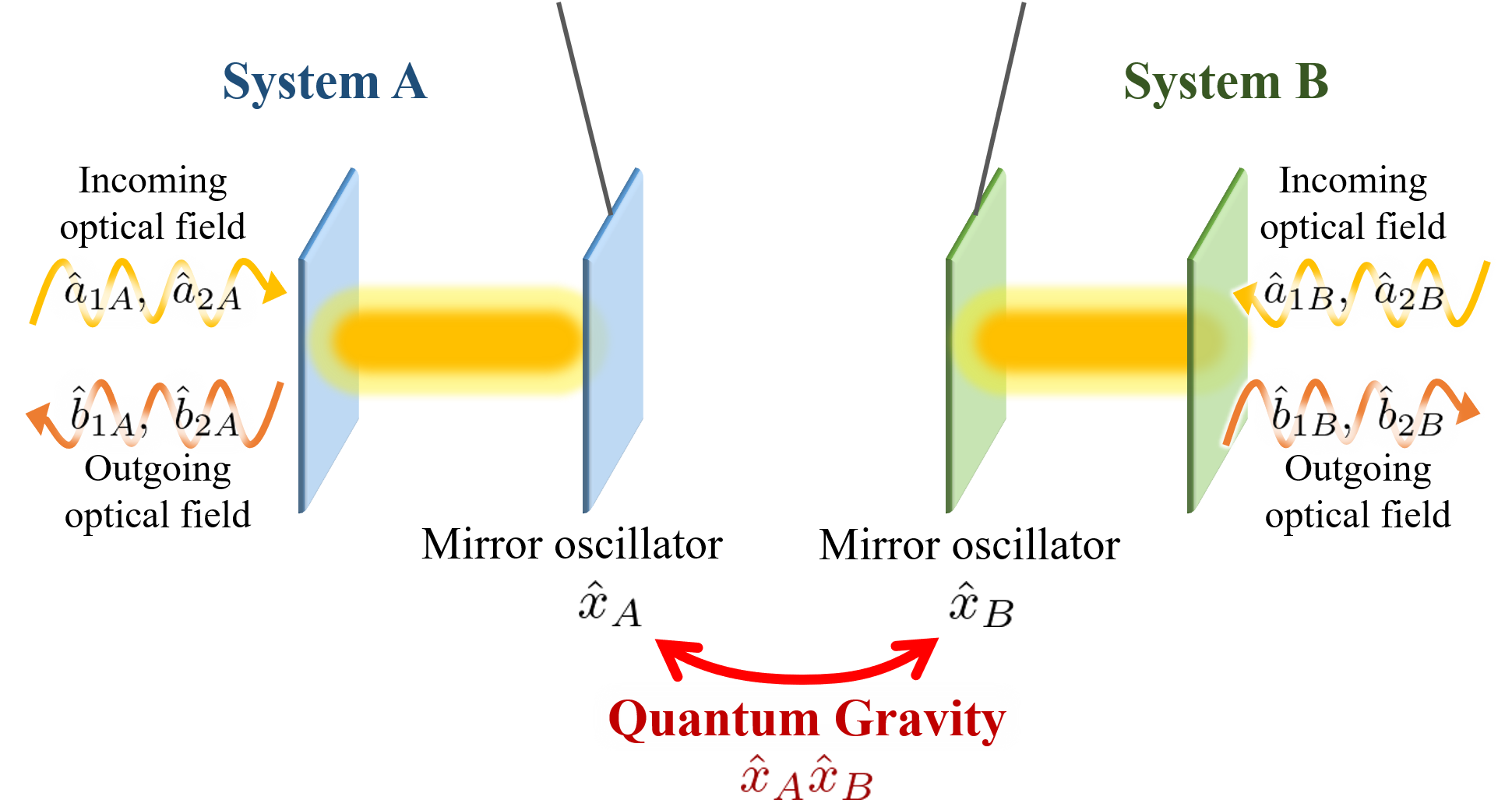}
    \caption{
    Setup for testing mutual quantum gravity between two optomechanical devices, $A$ (left) and $B$ (right). 
    The outgoing field quadrature is measured for each device. Mirror oscillators $A$ and $B$ interact via quantum gravity, described by the coupling term $\hat x_A \hat x_B$.
    }
    \label{fig:setup_mutualQG}
\end{figure*}

\subsection{Setup of the Protocol and Predictions of Quantum gravity}
\label{sec:mutual_gravity_QG}

As shown in Figure~\ref{fig:setup_mutualQG}, we consider two mirrors A and B, with masses  $M_{A/B}$, mechanical eigenfrequencies  $\omega_{A/B}$,  position and momentum operators $(\hat x_{A/B},\hat p_{A/B})$, sensed by incoming optical fields with  amplitude and phase quadratures  $(\hat a_{1A/B},\hat a_{2A/B})$ and  optomechanical couplings $\alpha_{A/B}$.  Mirrors A and B interact with each other through gravity.  The total Hamiltonian is given as
\begin{align}
    \hat H 
    &= \frac{1}{2M_A}\hat p_A^2 + \frac{1}{2}M_A \omega_{mA}^2 \hat x^2 -\hbar\alpha_A \hat x_A \hat a_{1A}
    \nonumber\\
    &
    + \frac{1}{2M_B}\hat p_B^2 + \frac{1}{2}M_B \omega_{mB}^2 \hat x^2 -\hbar\alpha_B \hat x_B \hat a_{1B} 
    +\hat V_{{\rm SN}/{\rm QG}}\,.
\end{align}
We shall measure one quadrature of the out-going optical field of each system, 
$\hat b_{\zeta A}$ and $\hat b_{\zeta B}$. 
%
%
%
The quantum-gravity (QG) interaction Hamiltonian between A and B is given by 
\begin{align}
    \label{eq:VQG}
    &\hat V_{{\rm QG}}
    =-\frac{G M_A M_B}{|\hat x_A-\hat x_B-d_{AB}|}
    \sim -\frac{G M_A M_B}{d_{AB}^3}\left(\hat x_A-\hat x_B\right)^2.
\end{align}
Here $d_{AB}=|x_B^{(0)}-x_A^{(0)}|$ is the distance between the zero-point positions of mirrors.
In the second equality, we performed the Taylor expansion for $\hat x_A-\hat x_B\ll d_{AB}$ and removed a constant term. 
Using this interaction, we obtain the Heisenberg equations of the individual mirrors as follows,
\begin{align}
    \label{eq:mutual_QG}
    \dot{\hat{x}}_A&=
    {\hat{p}_A}/{M_A}
    \notag\\
    \dot{\hat{p}}_A&=
    -M_A(\omega_{QA}^2\hat{x}_A+\omega_{AB}^2\hat x_B)-2\gamma_A\hat{p}_A+f_{\rm thA}+\hbar\alpha_A\hat{a}_{1A}
    \notag\\
    \dot{\hat{x}}_B&=
    {\hat{p}_B}/{M_B}
    \notag\\
    \dot{\hat{p}}_B&=
    -M_B(\omega_{QB}^2\hat{x}_B+\omega_{BA}^2\hat x_A)-2\gamma_B\hat{p}_B+f_{\rm thB}+\hbar\alpha_B\hat{a}_{1B}
\end{align}
Here, the quantum-gravity coupling frequency from B to A is introduced as $\omega_{AB}^2= 2G M_B/d_{AB}^3$, and the one from A to B  is given by $\omega_{BA}^2= 2G M_A/d_{AB}^3$;
$\omega_{QA}^2=\omega_{mA}^2-\omega_{AB}^2$ and $\omega_{QB}^2=\omega_{mB}^2-\omega_{BA}^2$ are the effective frequencies of masses A and B shifted by quantum gravity, respectively. 
Opposite to the self-gravity case given in Eq.~\eqref{eq:selfgrav_effective_freq}, the mututal-gravity contribution to the restoring frequency is negative: the wavefunction of each mass tends to spread out due to the  attractive force from the other mass, whereas in the self-gravity case, the wavefunction tends to be localized due to its self attraction. 
We have also added mechanical dissipations $\gamma_{A/B}$ and the fluctuating thermal forces $f_{\rm thA/B}$.
Note that A and B's Heisenberg equations of motion are coupled via quantum gravity at the level of operators, which will eventually lead to quantum correlations.

Let us simplify the problem by assuming that certain parameters of systems A and B are identical: $\omega_{mA}=\omega_{mB}=\omega_m$, $M_A=M_B=M$, and $\gamma_A=\gamma_B=\gamma$. As a result, the quantum-gravity coupling frequencies are identical, with
$\omega_{AB}=\omega_{BA}=\omega_g$.
Introducing the common and differential displacements of the masses 
$
    \hat x_{\pm}=\hat x_A \pm \hat x_B
$, 
their equations of motion are decoupled,
\begin{align}
    \Ddot{\hat x}_{\pm}
    =-\omega_\pm^2\hat x_{\pm} -2\gamma \dot{\hat x}_{\pm} 
    +\frac{
    \hbar \alpha_A \hat a_{1A}
    + f_{\rm thA} 
    }{M}
    \pm
    \frac{
    \hbar \alpha_B \hat a_{1B}
    + f_{\rm thB}
    }{M}
\end{align}
and can be solved in the Fourier domain as follows:
\begin{align}
    \hat x_{\pm}
    =-\big[
    \hbar \alpha_A \hat a_{1A}
    + f_{\rm thA} 
    \pm
    \left(
    \hbar \alpha_B \hat a_{1B}
    + f_{\rm thB} 
    \right)\big]/(MP_\pm) 
\end{align}
with  $P_{\pm}$ defined as 
\begin{align}\label{eq:mutual_gravity_P}
    &P_{\pm}=\omega^2-\omega_\pm^2+2i\gamma\omega
\end{align}
with $\omega_{\pm}$ the common- and differential-mode eigenfrequencies:
\begin{equation}
    \omega_+ = \omega_m\,,\quad \omega_- = \sqrt{\omega_m^2-2\omega_g^2}\,.
\end{equation}
Using these solutions, we can obtain solutions for the motions of the individual masses A and B:
\begin{align}
    \hat x_A
    &=\chi_Q^{\rm QG}(\hbar\alpha_A \hat a_{1A}+f_{\rm thA})
    +\chi_g(\hbar\alpha_B \hat a_{1B}+f_{\rm thB})
    \label{eq:mutual_QG_xAsol}\\
    \hat x_B
    &=\chi_Q^{\rm QG}(\hbar\alpha_B \hat a_{1B}+f_{\rm thB})
    +\chi_g(\hbar\alpha_A \hat a_{1A}+f_{\rm thA})
    ,\label{eq:mutual_QG_xBsol}
\end{align}
Here we have defined two susceptibilities:
\begin{align}
    \chi_Q^{\rm QG}(\omega)
    &
    =({\omega_m^2-\omega_g^2-2i\gamma\omega-\omega^2})/(MP_+P_-)
    ,
    \\
    \chi_g(\omega)
    &=
    -\omega_g^2
    /
(    MP_+P_-).
\end{align}
From Eq.~\eqref{eq:mutual_QG_xAsol}, we see that each mass responds to forces acting on it with $\chi_Q^{\rm QG}$, while it responds to forces acting on the other mass with $\chi_g$ --- with $\chi_g$ responsible for transferring quantum information. We note that both susceptibilities have poles near both $\omega_m$ and $\omega_g$, while $\chi_g$ is proportional to the square of the coupling frequency, $\omega_g^2$.

As in Sec.~\ref{sec:self_gravity_SN_badcavity}, we introduce the general output quadrature of each cavity light under the bad cavity condition, which is given as follows:
\begin{align}
    \hat{b}_{A\zeta_A}
    &=\hat a_{1A} \cos\zeta_A + (\hat a_{2A}+\alpha_A \hat x_A) \sin\zeta_A 
    ,
    \label{eq:mutual_QG_bzetaA}\\
    \hat{b}_{B\zeta_B}
    &=\hat a_{1B} \cos\zeta_B + (\hat a_{2B}+\alpha_A \hat x_B)\sin\zeta_B 
    .
    \label{eq:mutual_QG_bzetaB}
\end{align}
$\zeta_{A/B}$ is a hand-given measurement parameters that represent the homodyne angle. 
Using Eqs.~\eqref{eq:mutual_QG_xAsol} and \eqref{eq:mutual_QG_xBsol}, we obtain the explicit form of $\hat b_{\zeta_{A/B}}$ as well as the probability distribution of its measured value $\xi_{A/B}$.
Finally, the experimentalist investigate the gravity-induced correlation between systems A and B through the normalized correlation spectrum $\mathcal{C}_{AB}$ defined as follows:
\begin{align}\label{eq:mutual_gravity_CAB}
    \mathcal{C}_{AB}
    :={|S_{\xi_A\xi_B}(\omega)|^2}/[{S_{\xi_A\xi_A}(\omega)S_{\xi_B\xi_B}(\omega)}]
\end{align}
The exact form of $\mathcal{C}_{AB}$ is complicated, and provided by Appendix~\ref{apdx:mutual_gravity}. 
Here, for simplicity, we assume that the parameters for system A and B are identical, except for $\zeta_A$ and $\zeta_B$; specifically, $\alpha_A=\alpha_B=\alpha,~S_{f_{\rm thA}}=S_{f_{\rm thB}}=S_{f_{\rm th}}$. 

In this paper, instead of considering all possible choices of $\zeta_A$ and $\zeta_B$, and computing all correlations between the out-going quadratures --- hence leading to entanglement measures like the logarithmic negativity, we shall restrict ourselves to the   very special homodyne angle setting of $\zeta_A=0,~\zeta_B=\pi/2$, where out-going amplitude quadrature is measured for system A, and out-going phase quadrature is measured for system B.  The former ($\hat b_{A\,0}$) is proportional to the quantum radiation-pressure force acting on mirror A, while the latter ($\hat b_{B\,\pi/2}$) senses the displacement of B.  The quantum fluctuation of $b_{A\,0}$, which is equal to those of the incoming $\hat a_{A\,0}$, drives a ``quantum motion'' of $\hat x_A$.  In quantum gravity, the interaction Hamiltonian term $\hat x_A \hat x_B$ allows the quantum motion of $\hat x_A$ to drive a quantum motion of $\hat x_B$, which in turn gets transduced into $b_{B\,\pi/2}$. This establishes the correlation between  $\hat b_{A\,0}$ and $\hat b_{B\,\pi/2}$
{\it Naively}, one might expect quantum gravity to be indispensible in establishing this correlation, and that classical gravity will not transfer the quantum-radiation-pressure-driven motion of A to B, hence lead to a $\mathcal{C}_{AB}=0$.  In this way, the correlation spectrum $\mathcal{C}_{AB}$ is meant to naively indicate the transfer of quantum information via gravity. We shall explain this naive argument in 
more details in Sec.~\ref{sec:mutual_gravity_naiveSN}, and in Sed.~\ref{sec:mutual_gravity_CCSN} show that this is circumvented if classical gravity can depend on measurement results. 

Specifically, in the QG case, using the solutions Eqs.~\eqref{eq:mutual_QG_xAsol} -- \eqref{eq:mutual_QG_bzetaB}, we obtain the correlation spectrum as follows:
\begin{align}\label{eq:mutual_gravity_CAB_QG_case1}
    \mathcal{C}_{AB}^{\rm QG}  &=\frac{2\Lambda^4\omega_g^4}{2|P_+ P_-|^2+\Lambda^2(\Lambda^2+4n^c_{\rm th}\omega_m^2)(|P_+^2|+|P_-^2|)}
\end{align}
Here $\Lambda$ and $n_{\rm th}^c$ are the same as defined in Eqs.~\eqref{eq:Lambda} and \eqref{eq:nth}.
As long as $\omega_g^2/\omega_m \ll \gamma$, we can simplify the above formula by setting $\gamma\rightarrow 0$, arriving at 
\begin{align}\label{eq:mutual_gravity_CAB_QG_simple}
    \mathcal{C}_{AB}^{\rm QG}  =
\frac{\Lambda^4\omega_g^2}{\Delta^4+\Delta^2(\Lambda_{\rm eff}^4-2\omega_g^4)+\omega_g^4(\Lambda_{\rm eff}^4+\omega_g^4)}
\end{align}
where we have defined
\begin{equation}
    \label{defDelta}
    \Delta =\omega^2-(\omega_m^2-\omega_g^2)\,,\quad \Lambda_{\rm eff}^4 = \Lambda^2(\Lambda^2+4n_{\rm th}^c\omega_m^2)\,.
\end{equation}
We can see that $C_{AB}^{\rm QG}$ is a Lorentzian, which  peaks at $\Delta=0$, or $\omega= \sqrt{\omega_m^2-\omega_g^2}$, with a maximum of 
\begin{equation}
    C_{AB}^{\rm QG\,peak} =( \Lambda/\Lambda_{\rm eff})^4= \Lambda^2/({\Lambda^2+4n_{\rm th}^c\omega_m^2})
\end{equation}
Assuming $\omega_g \ll \omega_m$, the half maxima of this peak are located approximately at $\omega_\pm$, or $\omega_m$ and $\omega_m - \omega_g^2/\omega_m$.  This corresponds to a FWHM of $\omega^2/\omega_m$.  This correlation can be easily observable, as long as $\Lambda \stackrel{>}{_\sim} \sqrt{4n_{\rm th}^c}\omega_m$, and the integration time ${\mathcal{T}}$ is much longer than the inverse of the FWHM to resolve this Lorentzian.  

However, in the next section, we shall consider the role of the measurement process in the mutual gravity case, and show that simply measuring $b_{1A}$ will provide classical information about $A$'s location to classical gravity, hence drive motion of B in a way that has almost the same correlation as the quantum gravity case.  We will also consider the role of time delays in measurements.

\begin{figure*}[t]
    \centering
    \includegraphics[width=0.65\textwidth]{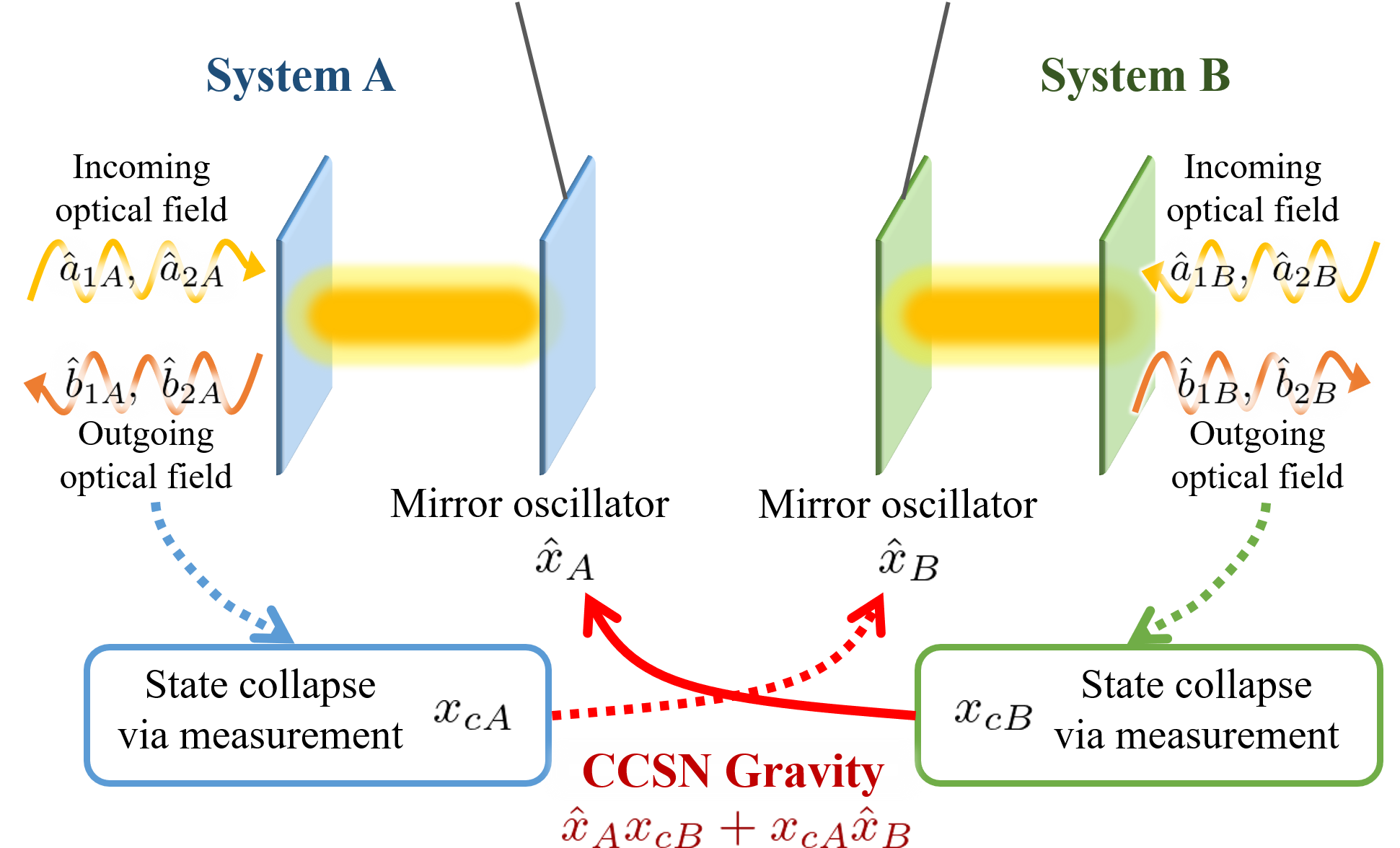}
    \caption{
    Setup for testing mutual CCSN gravity between two optomechanical devices, $A$ (left) and $B$ (right). 
    The outgoing field quadrature is measured for each device, yielding the mirror oscillators to collapse into classical positions, $x_{cA,B}$. CCSN gravity then applies a force based on the measurement result of one system, influencing the other, with the interaction described by the coupling term $\hat{x}A x_{cB} + x_{cA} \hat{x}B$. Through this interaction, the classical positions $x{cA,B}$ can still transfer information between the two devices, creating correlations that mimic those from quantum gravity.
    }
    \label{fig:setup_mutualCCSN}
\end{figure*}  
\subsection{Signatures of Classical Gravity}
\label{sec:mutual_gravity_SN}

Let us now turn to classical gravity interaction between the two masses. According to Sec.~\ref{sec:nonlinear_QM_vs_measurement}, for classical gravity, the quantum interaction Hamiltonian \eqref{eq:VQG} is replaced by a classical one 
\begin{align}
    \hat V_{{\rm SN}}
    &=-\frac{G M_A M_B}{|\hat x_A-x_{cB}-d_{AB}|} -\frac{G M_A M_B}{|x_{cA}-\hat x_B-d_{AB}|} \notag \\
    &\sim -\frac{G M_A M_B}{d_{AB}^3}\left\{
    \left(\hat x_A- x_{cB}\right)^2
    +\left(\hat x_B- x_{cA}\right)^2
    \right\},
\end{align}
with the first term representing the gravity of B acting on A, and  the second term  the gravity of A acting on  B.
The quantities $x_{cA/B}$ are the classical positions of the mirrors, which depend on the particular formulation of quantum gravity, as detailed in Sec.~\ref{sec:auxunify}. In this paper, we shall use conditional expectations computed from the measured quadratures of the out-going fields. 

\subsubsection{Heisenberg Equations and Naive  Classical Gravity}
\label{sec:mutual_gravity_naiveSN}

The Heisenberg equations \eqref{eq:mutual_QG} are replaced by:
\begin{align}
    \dot{\hat{x}}_A&=
    {\hat{p}_A}/{M_A}
    \notag\\
    \dot{\hat{p}}_A&=
    -M_A(\omega_{QA}^2\hat{x}_A+\omega_{AB}^2{\langle \hat x_{B}\rangle})-2\gamma_A\hat{p}_A+f_{\rm thA}+\hbar\alpha_A\hat{a}_{1A}
    \notag\\
    \dot{\hat{x}}_B&=
    {\hat{p}_B}/{M_B}
    \notag\\
    \dot{\hat{p}}_B&=
    -M_B(\omega_{QB}^2\hat{x}_B+\omega_{BA}^2\langle \hat x_{A}\rangle)-2\gamma_B\hat{p}_B+f_{\rm thB}+\hbar\alpha_B\hat{a}_{1B}
    \label{eq:mutual_CCSN_EOM}
\end{align}
with $\omega_{QA,QB,AB,BA}$ defined same as before. Heisenberg equations for the optical fields remain the same as Eqs.~\eqref{eq:mutual_QG_bzetaA} and \eqref{eq:mutual_QG_bzetaB}.




Let us first discuss the naive expectation for classical gravity without considering the impact of measurement on the gravitational interaction.  In doing so, let us ignore thermal noise and only consider quantum fluctuations. If we ignore the fact that quantum measurement collapses the positions of A and B continuously, injecting stochastic classical expectation values, and keep using  $\langle \hat x_A\rangle=\langle \hat x_{B}\rangle=0$ --- as if the system were not being measured --- then the Heisenberg equations  system A is then expressed as
\begin{align}
    \dot{\hat{x}}_A
    =
    {\hat{p}_A}/{M_A}\,,\quad 
    \dot{\hat{p}}_A
=
    -M_A\omega_{QA}^2\hat{x}_A+\hbar\alpha_A\hat{a}_{1A}
\end{align}
and similarly for system B. For simplicity, we neglect mechanical dissipation and the thermal fluctuations.
The output light quadrature of system A is given by
\begin{align}
    \hat{b}_{\zeta A}
    &=\hat a_{1A} \left(\cos\zeta_A
    +\frac{\Lambda_A^2 \sin\zeta_A }{\omega_{QA}^2+2i\gamma\omega-\omega^2}
    \right) + \hat a_{2A} \sin\zeta_A
    \notag
\end{align}
and similarly for system B. The two systems are completely decoupled,  no correlation emerges between $\hat b_{\zeta A}$ and $\hat b_{\zeta B}$, leading to the normalized correlation spectrum of
\begin{align}\label{eq:mutual_gravity_naiveSN_CAB}
    \mathcal{C}_{AB}=0
\end{align}
Therefore, we naively expect significant difference between the quantum gravity and SN gravity scenarios in terms of $\mathcal{C}_{AB}$.  As we take thermal noise into account, $\mathcal{C}_{AB}$ will become non-zero in general, because classical thermal fluctuations of the mirrors can lead to correlations --- yet the dependence of $\mathcal{C}_{AB}$ on $(\zeta_A,\zeta_B)$ can still be used to distinguish between QG and naive classical gravity.  In particular, $\mathcal{C}_{AB}$ for naive classical gravity still vanishes when one of $\zeta_A$ and $\zeta_B$ is set to zero.  For example, $b_{A\,0}$ in naive gravity is still equal to $a_{A\,0}$ the amplitude quadrature of the incoming field, which has an expecation value of zero, and  still does not drive the motion of $\hat x_A$ --- therefore $b_{A\,0}$ is still uncorrelated with any $b_{B\zeta_B}$.  We also note the interesting fact that each oscillator will experience a frequency shift influenced by the presense of the other one, namely from $\omega_A$ to $\omega_{QA}$, and from $\omega_B$ to $\omega_{QB}$. However, unlike the quantum gravity case, here the individual oscillations of $\hat x_A$ and $\hat x_B$ --- instead of a combination  --- are the always the eigenmodes.  The common- and differential-mode split observed in Sec.~\ref{sec:mutual_gravity_QG} for the QG case does not occur here. 

However, once $\hat{b}_{\zeta A/B}$ is measured, the quantum state of the system inevitably collapses.  {If we allow the information of the measurement result to drive the $\langle \hat x_{A/B}\rangle$, which in turn making the gravity force acting on the masses depends on the conditional mean motion $\langle \hat x_{A/B}\rangle\rightarrow  x_{cA/B} $, } this will bring back $\mathcal{C}_{AB}$, as we shall see in the next section.

\subsubsection{CCSN Gravity and Wiener Filtering}
\label{sec:mutual_gravity_CCSN}

We now incorporate the measurement of $\hat{b}_{\zeta A/B}$ and examine how CCSN gravity creates correlations between systems A and B that mimicks quantum gravity.
Figure~\ref{fig:setup_mutualCCSN} shows the setup for testing mutual CCSN gravity, with the same experimental arrangement as in the quantum gravity case.
We shall use the same Wiener filtering technique as in Sec.~\ref{sec:self_gravity_SN_badcavity}, and allow a time delay in the measurement of $\hat{b}_{\zeta A/B}$, as in Sec.~\ref{sec:self_gravity_SN_badcavity}.


To avoid redundancy, we provide explicit expressions only for the physical quantities of system A, as those for system B take a similar form, and can be obtained by exchanging A and B.  The solution of the Heisenberg equations Eqs.~\eqref{eq:mutual_CCSN_EOM} in the Fourier domain are given by
\begin{align}
    \label{eq:mutual_gravity_xAsol}
    \hat x_{A} 
    &= \hat x_{QA} + x_{{\rm cl} A},
\end{align}
where
\begin{align}
    \label{eq:mutual_gravity_xQA_xclA}
    &\hat x_{QA}= \hbar\alpha_A \chi_{QA}^{\rm SN} \hat{a}_{1A}
    ,\;
    x_{{\rm cl} A}= \chi_{QA}^{\rm SN}(
    -M_A\omega_{AB}^2x_{cB}+f_{{\rm th}A}
    ),
\end{align}
and similarly for system B.
Here, the susceptibility $\chi_{QA}^{\rm SN}$ is defined as
\begin{align}
    \chi_{QA}^{\rm SN}(\omega)
    &=\big[{M_{A}\big(\omega_{QA}^2-2i\gamma_{A} \omega-\omega^2\big)}\big]^{-1}.
\end{align}
Eq.~\eqref{eq:mutual_gravity_xAsol} implies that the position of mirror A is modified due to the quantum optomechanical force as denoted by $\hat x_{QA}$ and also displaced due to the SN gravitational interaction and the thermal noise contribution described by $x_{clA}$.

We then suppose the measurement of the out-going light field $\hat{b}_{\zeta A/B}$.
By separating the operator-dependent part from the c-number part, the output light quadrature can be rewritten as follows.
\begin{align}
    \hat{b}_{A \zeta_A}
    &=\hat{b}_{A\zeta_A{\rm Q}}
    +\alpha_A \sin \zeta_A x_{{\rm cl}A},
    \label{eq:mutual_gravity_bzetaA}
    \\
    \hat{b}_{A \zeta_A{\rm Q}}
    &=(\cos\zeta_{A} + \hbar\alpha_{A}^2\chi_{QA}^{\rm SN}\sin\zeta_{A})\hat{a}_{1A}
    +\hat{a}_{2A}\sin\zeta_{A}
    \label{eq:mutual_gravity_bzetaQA}
\end{align}
Similar form holds for system B. 
Let us denote $\xi_{A/B}(t),~z_{A/B}(t)$ as the measured value of $\hat b_{\zeta A/B},~\hat b_{\zeta{\rm Q}A/B}$ respectively, which follows a classical random process. Then, the above quantum-classical separation yields a relation of
\begin{align}
    \label{eq:mutual_gravity_xiA}
    \xi_A 
    &= z_A 
    +\alpha_A \sin\zeta_A \, x_{{\rm cl}A}
    ,
\end{align}
and similarly for system B.

Now, let us consider the time delayed measurement of $\hat{b}_{\zeta A/B}$. Specifically, we suppose that the out-going light field of system A/B is detected at a distance $c\tau_{A/B}$ from the output of the cavity. 
\begin{widetext}
Under this time delayed measurement scheme, the quantity $x_{cA}(t)$ should be conditioned on $\xi_A$ during $t' \le t-\tau_A$, as well as $\xi_B$ during $t' \le t-\tau_B$, while assuming $f_{\rm th A}$ and $f_{\rm th B}$ are classical random processes.  The separation of $\hat x_{A}$, $\hat x_B$, $\hat b_{A\zeta_A}$, and $\hat b_{B\zeta_B}$ into classical and quantum parts allows us to significantly simplify this conditioning process. For $x_{cA}$, it is the sum of the conditional expectation value of $\hat x_{QA}$ and $x_{{\rm cl}A}$, yet the former only depends on $z_A$.  We can write
\begin{align}\label{eq:mutual_gravity_SN_xcA}
    x_{c A}(t)
    &= E\left[
    \hat x_{A}(t)\,\big|\left\{
    \hat b_{A\zeta_A}(t')=\xi_A(t'), t'\leq t -\tau_A\,,\;\;
    \hat b_{B\zeta_B}(t')=\xi_B(t'), t'\leq t -\tau_B
    \right\}\right] \notag \\
    &= E\left[
    \hat x_{QA}(t)\,| \left\{
    \hat{b}_{\zeta A{\rm Q}}(t')=z_A(t') | t'\leq t -\tau_A
    \right\}\right]
    +x_{{\rm cl}A},
\end{align}
and similarly for system B.
\end{widetext}
Following the same approach given in Section~\ref{sec:self_gravity_SN_badcavity}, this can be computed using Causal Wiener filter\cite{Ebhardt09}, resulting in
\begin{align}
    x_{cA}
    &= K_{\tau A}z_A+x_{{\rm cl}A}.
\end{align}
Using Eq.~\eqref{eq:mutual_gravity_xQA_xclA} and solving jointly for systems A and B, we obtain 
\begin{equation}
    x_{cA} =\chi_{AA}^{\rm QG} \frac{K_{\tau A}}{\chi_{QA}}z_A+\chi_{AB}^{\rm QG}  \frac{K_{\tau B}}{\chi_{QB}}z_B + \chi_{AA}^{\rm QG}  f_{\rm thA} +\chi_{AB}^{\rm QG}  f_{\rm thB}
\end{equation}
and similarly for system B.
    Here, $\chi_{AA}^{\rm QG}$, $\chi_{AB}^{\rm QG}$, $\chi_{BA}^{\rm QG}$, and $\chi_{BB}^{\rm QG}$ are the response of mass A and B to driving forces on A and B, respectively, in standard QG. They are given by 
    \begin{equation}
\begin{bmatrix}
    \chi_{AA}^{\rm QG} & \chi_{AB}^{\rm QG} \\
    \chi_{BA}^{\rm QG} & \chi_{BB}^{\rm QG}
\end{bmatrix}
=\frac{
    \begin{bmatrix}
        \chi_{QA} & -\chi_{QA}\chi_{QB}M_A\omega_{AB}^2 \\
        -\chi_{QA}\chi_{QB}M_B\omega_{BA}^2 & \chi_{QB}
    \end{bmatrix}}{1- M_A M_B\chi_{QA}\chi_{QB}\omega_{AB}^2\omega_{BA}^2}
    \end{equation}
We can already see that the classical parts of the motions respond to thermal forces the same way as quantum gravity.  The factors of $\chi_{AA}^{\rm QG}/\chi_{QA}$ and $\chi_{AB}^{\rm QG}/\chi_{QB}$ will also restore poles in the QG case and eliminate those at $\omega_{QA}$ and $\omega_{QB}$ in the Naive classical gravity situation. 
These are indications that correlations in the QG case will be at least partially restored by CCSN.

In a similar way, we can also obtain the expression of $\xi_{A/B}$ in terms of $K_{\tau A/B}$, $z_{A/B}$ and $f_{{\rm th}A/B}$ as follows.
\begin{align}
    \label{eq:mutual_gravity_xiA2}
\xi_A &= \left[
    1+\frac{\chi_{AA}^{\rm QG}-\chi_{QA}}{\chi_{QA}}K_{\tau A} \alpha_A \sin\zeta_A\right]z_A+\frac{\chi_{AB}^{\rm QG}}{\chi_{QB}} K_{\tau B}\alpha_A\sin\zeta_A z_B \nonumber\\
    &+\chi_{AA}^{\rm QG} f_{\rm thA}+\chi_{AB}^{\rm QG} f_{\rm thB}
\end{align}
and similarly for system B.  Similar to before, the Wiener filter $K_{\tau A/B}$ can be obtained by using the Wiener-Hopf method with time delay,
\begin{align}
    K_{\tau A}
&    = \left[ {S_{x_A {b}_{\zeta A{\rm Q}}}}/{\phi_{A-}} \right]_{\tau A}/{\phi_{A+}}\nonumber\\
    &=\frac{1}{\alpha_A\sin\zeta_A}
    \frac{1}{(\omega-\beta_A)(\omega+\beta_A^*)(\tilde{\omega}_{QA}+\tilde{\omega}_{QA}^*)}\notag\\
    &\;\cdot \left[
    (\tilde{\omega}_{QA}-\beta_A)(\tilde{\omega}_{QA}+\beta_A^*)(\omega+\tilde{\omega}_{QA}^*)e^{i(\omega-\tilde{\omega}_{QA})\tau_A}\right.\notag\\
    &   \; \left.-
    (\tilde{\omega}_{QA}^*-\beta_A^*)(\tilde{\omega}_{QA}^*+\beta_A)(\omega-\tilde{\omega}_{QA})e^{i(\omega+\tilde{\omega}_{QA}^*)\tau_A}
    \right]
    \label{eq:KtauA}
\end{align}
Here, $\tilde{\omega}_{QA}=\sqrt{\omega_{QA}^2-\gamma_A^2}-i\gamma_A$ and 
\begin{align}
    \beta_A^2
    &=\omega_{QA}^2-2\gamma_A^2+\Lambda_A^2\sin\zeta_A \cos\zeta_A \notag\\
    &\hspace{5mm}
    -i\sqrt{
    4\gamma_A^2(\omega_{QA}^2-\gamma_A^2)+2\gamma_A^2\Lambda_A^2\sin 2\zeta_A +\Lambda_A^4 \sin^4\zeta_A
    }.
\end{align}
Similar expression applies for system B.  Note that this formula is degenerate for $\zeta_A=0$, in which case we can directly obtain
\begin{equation}
K_{\tau A}=[\alpha_A\chi_{QA}]_\tau
\end{equation}
since $\hat x_{QA}$ simply responds to $\hat a_{1A}$, which is $z_A$in this case.  Substituting the Wiener filter function form into Eq.~\eqref{eq:mutual_gravity_xiA2}, we can obtain the cross spectrum for $\xi_A$ and $\xi_B$.  The exact form of $C_{AB}$ is complicated, which is presented in Appendix~\ref{apdx:mutual_gravity}, where for simplicity, we assume that the parameters for system A and B are identical, except for $\zeta_A$ and $\zeta_B$; specifically, $\omega_{mA}=\omega_{mB}=\omega_m$, $M_A=M_B=M$, $\gamma_A=\gamma_B=\gamma$, $\alpha_A=\alpha_B=\alpha$, $f_{\rm thA}=f_{\rm thB}=f_{\rm th}$, and $\tau_A=\tau_B=\tau$. This results in $\omega_{AB}=\omega_{BA}=\omega_g$.

In the following, we shall discuss two special cases. The first is when $\tau_A=\tau_B=0$, where we will show that at leading order in $\omega_g$, the correlation spectrum is the same as the quantum gravity case. The second is when $(\zeta_A,\zeta_B)=(0,\pi/2)$, which has a particuarly interesting physical meaning --- where we will discuss how the correlation spectrum is modified by the time delay.

\subsubsection{Special case with $\tau=0$}

With $\tau_A=\tau_B=0$, we can simplify the expressions of the Wiener filters, and write (in the $\gamma_A=\gamma_B=0$ limit)  

\begin{align}
    \xi_A = \frac{1}{P_{AB}}&\Big[[(\omega^2-\omega_{QB}^2)(\omega-\beta_A)(\omega+\beta_A^*)-\omega_{BA}^2\omega_{AB}^2]w_A \nonumber\\
    +&\frac{\alpha_A\sin\zeta_A}{\alpha_B\sin\zeta_B}\omega_{AB}^2(\omega_{QB}^2-\beta_B\omega+\beta_B^*\omega-|\beta_B^2|)w_B \nonumber
    \\
    -&\alpha \sin\zeta_A \left(
    \frac{f_{\rm th A}(\omega^2-\omega_{QB})^2}{M_A}
    +\frac{f_{\rm th B}\omega_{AB}^2}{M_B}
    \right)
    \Big] 
\end{align}
\begin{align}
    \xi_B = \frac{1}{P_{AB}}&\Big[[(\omega^2-\omega_{QA}^2)(\omega-\beta_B)(\omega+\beta_B^*)-\omega_{AB}^2\omega_{BA}^2]w_B  \nonumber\\  +&\frac{\alpha_B\sin\zeta_B}{\alpha_A\sin\zeta_A}\omega_{BA}^2(\omega_{QA}^2-\beta_A\omega+\beta_A^*\omega-|\beta_A^2|)w_A \nonumber\\
    -&\alpha \sin\zeta_B \left(
    \frac{f_{\rm th B}(\omega^2-\omega_{QA})^2}{M_B}
    +\frac{f_{\rm th A}\omega_{BA}^2}{M_A}
    \right)
    \Big] 
\end{align}
where $w_A$ and $w_B$ are causally whitened versions of $z_A$ and $z_B$, with $S_{w_A w_A}=S_{w_B w_B}=1$ and $S_{w_A w_B}=S_{w_B w_A}=0$.  The common  denominator is given by $P_{AB}=(\omega^2-\omega_{QA}^2)(\omega^2-\omega_{QB}^2)-\omega_{AB}^2\omega_{BA}^2$, which has poles at the coupled eigenfrequenices.  In this way, the naive expectation of frequency shift in classical gravity does not happen --- classical information contained in the measurement results are able to create the the same coupled eigenfrequencies as quantum gravity predicts.  Classical gravity can only be tested by a quantitative measurement of the auto- and self correlations of the out-going fields. 

Specializing two identical system with arbitrary values of $\zeta_A$ and $\zeta_B$, adopting the same $\Delta$ as in Eq.~\eqref{defDelta} to parametrize $\omega$, and realizing that the regime of interest is when $\Delta$ is at the scale of $\omega_g^2$.  We can perform a Taylor expansion in $\omega_g \sim \sqrt{\Delta}$ --- which amounts to considering frequencies that have $\omega-\omega_Q \sim \omega_g^2/\omega_Q$. In this regime, we have
$    \omega=\omega_Q+{\Delta}/{(2\omega_Q)}
$.
We find that both the auto and the cross correlations of CCSN are identical to the QG at this order.  In other words,
\begin{widetext}
\begin{align}
    S_{AA}^{\rm SN}& =1+
\Big\{\Lambda^2\sin\zeta_A[\Lambda^2(\Delta^2+\omega_g^4)\sin\zeta_A+2\Delta(\omega_g^4-\Delta^2)\cos\zeta_A]+\epsilon_{AA}\Big\}/(\Delta^2-\omega_g^4)^2 \\
    S_{AB}^{\rm SN}& = \Big\{\Lambda^2\omega_g^2(2\Delta\Lambda^2\sin\zeta_A\sin\zeta_B+(\omega_g^4-\Delta^2)\sin(\zeta_A+\zeta_B))+\epsilon_{AB}\Big\}/(\Delta^2-\omega_g^4)^2
\end{align}
\end{widetext}
where $\epsilon_{AA} \sim \omega_g^6$ and $\epsilon_{AB} \sim\omega_g^6$ are corrections due to classical gravity, and are subleading in $\omega_g$. This result significantly deviates from the naive expectation given in Eq.~\eqref{eq:mutual_gravity_naiveSN_CAB}, which predicts no correlation. The correlation induced by CCSN arises from the interplay between measurement process and feedback. In the frequency band of interest, the overall behavior predicted by CCSN is the same as QG.  Quantitative measurements of the auto and cross correlation spectral must be performed with $O(\omega_g^2$) precision to distinguish between classical and quantum gravity. 

\subsubsection{Special case with $(\zeta_A,\zeta_B)=(0,\pi/2)$}

Let us now specialize to the $(\zeta_A,\zeta_B)=(0,\pi/2)$ case, which is the same as the case in Sec.~\ref{sec:mutual_gravity_QG}. We shall for the moment ignore thermal-noise contributions, which are identical for QG and CCSN, let us focus on the quantum configurations, 
\begin{align}
    \xi_A &= z_A \\
    \label{CCSNmutual_xiB}
    \xi_B & =
    \left[
    1+\frac{\chi_{BB}^{\rm QG}-\chi_{QB}}{\chi_{QB}}K_{\tau B} \alpha_B \right]z_B
    +\alpha_B\alpha_A \frac{\chi_{BA}^{\rm QG}}{\chi_{QA}}  [\chi_{QA}]_\tau z_A
\end{align}
Here we can see that the statistics at the output of A is the same as the QG case, with 
\begin{equation} 
S^{\rm SN}_{\xi_A\xi_A}= S^{\rm QG}_{\xi_A\xi_A}\,,\quad(\zeta_A,\zeta_B)=(0,\pi/2)\,.
\end{equation}
  For B, the second term on the RHS of Eq.~\eqref{CCSNmutual_xiB}, leads to the following correlation 
\begin{equation}
S^{{\rm SN}}_{\xi_A\xi_B}=\alpha_A \alpha_B \frac{[\chi_{QA}]_{\tau_A}}{\chi_{QA}}\chi_{BA}^{\rm QG}=\frac{[\chi_{QA}]_{\tau_A}}{\chi_{QA}}S^{\rm QG}_{\xi_A\xi_B}
\end{equation}
which is identical to the QG case if $\tau=0$.  The mechanism behind this CCSN-induced correlation between the two systems lies in the interplay of measurement and feedback.  Specifically, when the experimentalist measures the out-going light field of system A, the mirror position $x_{cA}$ receives measurement feedback, as described in Eq.~\eqref{eq:mutual_gravity_SN_xcA}. This conditioned mirror position $x_{cA}$ then influences system B through the gravitational interaction. Consequently, the measurement outcome of system B $\xi_B$ is affected by the feedback on system A. 
Note that $\xi_A$ remains unaffected by the measurement feedback on system B, as we have set $\zeta_A=0$ for now.
As a result, CCSN produces a correlation between the two systems, which closely resembles that observed in the quantum gravity case. This finding has been previously investigated by Ref.~\cite{Yubao}.

In the limit of low $\gamma$, we have
\begin{equation}
    \label{eq:chiA_timedelay}
    \frac{[\chi_{QA}]_{\tau_A}}{\chi_{QA}} \approx 
    e^{i\omega\tau_A}\left[\cos(\omega_{QA}\tau_A)-\frac{i\omega}{\omega_{QA}}\sin(\omega_{QA}\tau_A)\right]
\end{equation}
This means the correlation near the peak $\omega\sim \omega_{QA}$ can decrease in the time delayed measurement, particularly when $(\omega-\omega_{QA})\tau$ becomes sufficiently large. This suggests that the time delayed measurement mitigates the effects of state collapse, bringing the system behavior closer to the naive expectation of classical gravity, where no correlation between the two gravitating systems is observed. However, it may be experiementally impractical to achieve $(\omega-\omega_Q)\tau \sim 1$.

The first term on the RHS of Eq.~\eqref{CCSNmutual_xiB}, does not have the same spectrum identical to QG even when $\tau =0$. In fact, it is very interesting that when we fix $\zeta_B=\pi/2$, $S_{\xi_B\xi_B}$ actually depends on the value of $\zeta_A$, and here we list two different values, with $\zeta_A=0$ and $\zeta_A=\pi/2$:
\begin{align}
S_{BB}^{{\rm SN}(0,\pi/2)} =1+
\frac{\Lambda^4(\omega_g^4+\Delta^2)+2\omega_g^4\Delta (\sqrt{\Lambda^4+\omega_Q^4}-\omega_Q^2)}{(\Delta^2-\omega_g^4)^2}  \\
S_{BB}^{{\rm SN}(\pi/2,\pi/2)} =1+
\frac{\Lambda^4(\omega_g^4+\Delta^2)+4\omega_g^4\Delta (\sqrt{\Lambda^4+\omega_Q^4}-\omega_Q^2)}{(\Delta^2-\omega_g^4)^2} 
\end{align}
In both equations the second term on the numerators of of the fractions are corrections due to SN. This is demonsrates the interesting effect that merely changing the choice of measurement at A will affect the statistics of results at B.  

At the moment, if we assemble the result of $\mathcal{C}_{AB}^{\rm SN}$ for $\zeta_A=0$ and $\zeta_B=\pi/2$, at $\tau=0$, its difference from $\mathcal{C}_{AB}^{\rm QG}$ only arises from the change in the auto-correlation spectrum,  $S_{BB}$. In the regime of $\omega -\omega_Q\sim \omega_g^2/\omega_Q$, we have
\begin{align}
    \mathcal{C}_{AB}^{\rm SN}&=
    \frac{{S_{BB}^{\rm QG}}}{{S_{BB}^{{\rm SN}(0,\pi/2)}}}
    \mathcal{C}_{AB}^{\rm QG}\nonumber\\
    \nonumber\\
    &=\mathcal{C}_{AB}^{\rm QG}\left[1-\frac{2\omega_g^4\Delta}{(\omega_g^4+\Delta^2)(\omega_Q^2+\sqrt{\Lambda^4+\omega_Q^4})}\right]
\end{align}
The typical fractional change is $\sim (\omega_g/\omega_Q)^2$.  It is also possible to explore how $S_{BB}$ changes with $\tau$, but the resulting change is only significant in the regime where $(\omega-\omega_Q)\tau \sim 1$, which is experimentally very challenging. 

\subsubsection{Detectability of CCSN Mutual Gravity without time delay}

Since the most feasbile observable effect in this section has to do with the correction of $S_{BB}$ at $\tau=0$, let us just insert thermal noise back into $S_{BB}$, and explore whether the shift due to classical gravity is visible.  In fact, we only need to write
\begin{align}
S_{BB}^{{\rm QG}\,\zeta_B=\pi/2}=1+\frac{\Lambda_{\rm eff}^4(\omega_g^4+\Delta^2)}{(\Delta^2-\omega_g^4)^2}
.
\end{align}
Then, we have
\begin{equation}
\frac{S^{\rm SN}_{BB}-S^{\rm QG}_{BB}}{S_{BB}^{\rm QG}}
=\frac{2\Delta\omega_g^4\big(\sqrt{\Lambda^4+\omega_Q^4}-\omega_Q^2\big)}{(\Delta^2-\omega_g^4)^2+\Lambda_{\rm eff}^4(\Delta^2+\omega_g^4)}
.
\end{equation}
This function has a maximum at $\Delta =\omega_g^2$, therefore we can approximately write the deviate rate of 
\begin{align}
    \int \frac{d\omega}{2\pi}\left|\frac{S^{\rm SN}_{BB}-S^{\rm QG}_{BB}}{S_{BB}^{\rm QG}}\right|^2 &\approx \frac{2\omega_g^2}{\omega_m}\left[\frac{\Lambda^2}{\Lambda^2+4n_{\rm th}^c\omega_m^2}\frac{\omega_g^2}{\omega_m^2+\sqrt{\Lambda^4+\omega_m^4}}\right]^2
    \nonumber\\
    &\stackrel{<}{_\sim} ({\omega_g^2}/{\omega_m})({\omega_g^2}/{\omega_m^2 n_{\rm th}^c})^2\,.
\end{align}
Here, the optimal choice for $\Lambda $ is $\omega_m\ll \Lambda \ll \sqrt{4n_{\rm th}^c}\omega_m$.
Therefore, the required observation time to detect the deviation between the two gravity models is given by
\begin{equation}
\label{eq:mutual_Tobs}
    {\mathcal{T}} 
    =\left(\int \frac{d\omega}{2\pi}\left|\frac{S^{\rm SN}_{BB}-S^{\rm QG}_{BB}}{S_{BB}^{\rm QG}}\right|^2\right)^{-1}
    \sim \frac{\omega_m}{\omega_g^2}\left(\frac{n_{\rm th}^c\omega_m^2}{\omega_g^2}\right)^2
\end{equation} 
which is extremely challenging to achieve.  This very large factor arises from the fact that integration time is inversely proportional to the the {\it square} of the change in spectrum we need to detect, and the fact that the change in spectrum is now suppressed by $\propto \omega_g^2$, where the frequency scale for the scaling is $n_{\rm th}^c\omega_m$ and leading to the scaling of  $\omega_g^4/(n_{\rm th}^c\omega_m)^2$.

\subsection{Numerical results and effect of time delay}

\begin{figure}[htbp]
    \centering
    \begin{minipage}[t]{1\linewidth}
    \centering
    \includegraphics[width=0.9\textwidth]{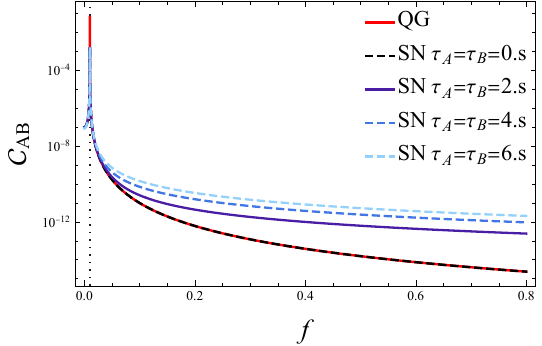}
    \end{minipage}
    \caption{The normalized correlation spectrum $C_{AB}$ for $\zeta_A=0$ and $\zeta_B=\pi/2$. Otherwise, the parameters are set to those listed in Table~\ref{tab:mutual_gravity_parameters}. We have used delays of $0\,\rm{s},~2\,\rm{s},~4\,\rm{s}$ and $6\,\rm{s}$, and assumed a strong measurement with $\Lambda=2\pi\times 350\,\rm{Hz}$. The vertical gray dashed line in the left side shows the spectrum peak at $f=\omega_Q/2\pi$. The effect of time delayed measurement appears in regions away from the peak.}
    \label{fig:CAB}
    \hspace{5mm}
    \begin{minipage}[t]{1\linewidth}
    \centering
    \includegraphics[width=0.9\textwidth]{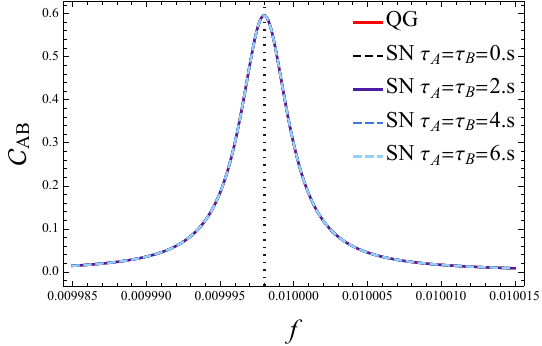}
    \end{minipage}
    \caption{The normalized correlation spectrum $C_{AB}$ around the peak. The parameters are set to be the same as in Figure~\ref{fig:CAB_peak}. The vertical gray dashed line shows the spectrum peak at $f=\omega_Q/2\pi$.}
    \label{fig:CAB_peak}
    \hspace{5mm}
    \begin{minipage}[t]{1\linewidth}
    \centering
    \includegraphics[width=0.98\textwidth]{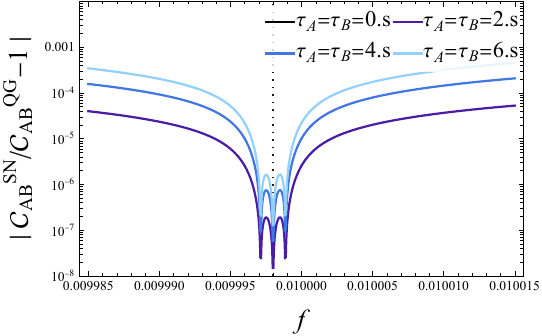}
    \end{minipage}
    \caption{The rate of change between $C_{AB}^{\rm SN}$ and $C_{AB}^{\rm QG}$. The parameters are set to be the same as in Figure~\ref{fig:CAB_peak}. A discrepancy of approximately $0.01\%$ is observed near the spectrum peak $\omega\sim\omega_{\pm}$ for $\tau_A=\tau_B=6\,{\rm s}$.}
    \label{fig:CAB_compareQGSN}
\end{figure}

\begin{figure*}[htbp]
    \centering
    \includegraphics[width=0.9\linewidth]{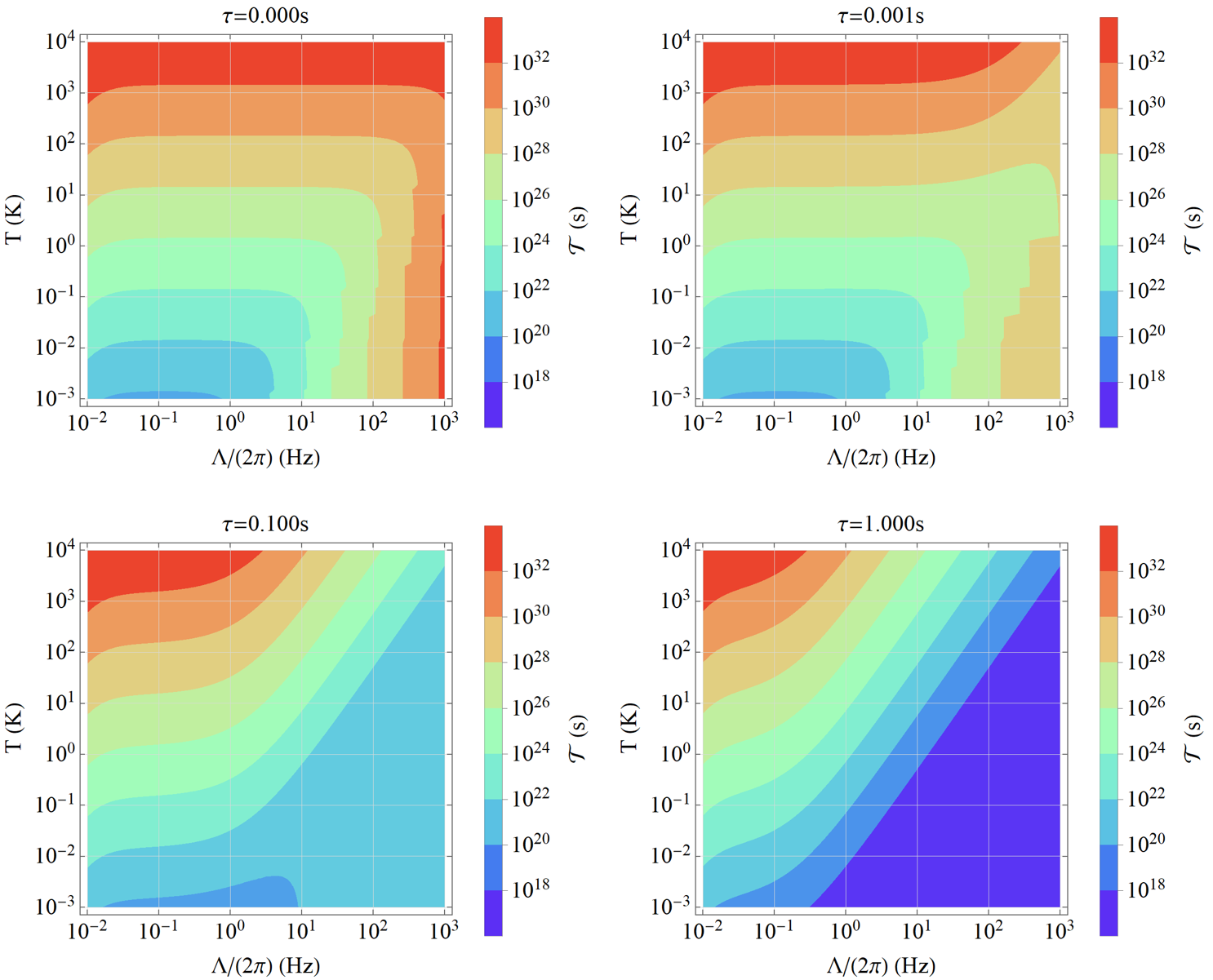}
    \caption{Required observation time ${\mathcal{T}}$. Parameters are set to those listed in Table~\ref{tab:mutual_gravity_parameters}. In each panel, we show different values of the measurement delay time: \(\tau = 0.000\,{\rm s},~0.001\,{\rm s},~0.100\,{\rm s},~1.000\,{\rm s}\).
    }
    \label{fig:Tobs_mutual}
\end{figure*}

\begin{table}[tbp]
    \centering
    \begin{tabular}{ccc}
    \hline\hline
    Parameters & Symbol & Value \\\hline
    Mirror mass & M & $1$~{\rm g}\\
    Mirror bare frequency & $\omega_m/2\pi$ & $10$~{\rm mHz}\\
    SN frequency & $\omega_{\rm g}/2\pi$ & $2\times10^{-4}$~{\rm Hz}\\
    Mechanical damping & $2\gamma/2\pi$ & $1.67\times10^{-8}$~{\rm Hz}\\
    Optical wavelength & $\lambda$ & $1064$~{\rm nm}\\
    Cavity finesse & $\mathcal{F}$ & $4000$\\
    Input-cavity power & $P_{\rm cav}$ & $2000$~{\rm W}\\
    Temperature & $T$ & $300$~{\rm K}\\
    \hline\hline
    \end{tabular}
    \caption{
    The parameters of the optomechanical devices with two gravitating mirrors expected for tabletop experiments~\cite{Yubao}.
    }
    \label{tab:mutual_gravity_parameters}
\end{table}

In this section, we present numerical plots of the normalized correlation spectrum $\mathcal{C}_{AB}$ and examine how the differences between the CCSN and QG cases depend on the delay time of the measurement.
For simplicity, we assume that the parameters for system A and B are identical, except for $\zeta_A$ and $\zeta_B$; Specifically, $\omega_{mA}=\omega_{mB}=\omega_m$, $M_A=M_B=M$, $\gamma_A=\gamma_B=\gamma$, $\alpha_A=\alpha_B=\alpha$, $f_{\rm thA}=f_{\rm thB}=f_{\rm th}$, and $\tau_A=\tau_B=\tau$. Additionally, we take $\zeta_A=0$ and $\zeta_B=\pi/2$.
The numerical plots in the following figures are generated using the parameters listed in Table~\ref{tab:mutual_gravity_parameters} for the mutual gravity case.

Figure~\ref{fig:CAB} shows the normalized correlation spectrum for delay times on the order of $\mathcal{O}(1)$, specifically $\tau_A = \tau_B = \{0,~2,~4,~6\}$. The horizontal axis represents the frequency, $f = \omega/2\pi$. The red line corresponds to the QG case, while the blue lines represent the CCSN gravity case, with varying intensities indicating different delay times $\tau_A = \tau_B = {0,~2,4,6}$, as shown in the legend. When there is no delay time ($\tau_A = \tau_B = 0$), there is no difference between the QG and CCSN gravity spectra, consistent with previous findings\cite{Yubao}. However, as the delay time increases, differences between the two gravity models emerge in the off-resonant region. 
The overall elevation of the spectrum in the off-resonant region due to time-delayed measurement is indicated in Eq.~\eqref{eq:chiA_timedelay}. The ratio of the mutual correlation $S_{\xi A \xi B}$ between QG and CCSN, given in Eq.~\eqref{eq:chiA_timedelay}, generally increases in the off-resonant region ($\omega>\omega_{Q}$) as the delay time increases.

Figure~\ref{fig:CAB_peak} focuses on the details around the peak of the normalized correlation spectrum shown in Figure~\ref{fig:CAB}. Even with the introduction of the measurement time delay, the difference between the two gravity models remains negligible around the peak. This behavior is also consistent with Eq.~\eqref{eq:chiA_timedelay}, which indicates that the impact of time delay become significant only when $(\omega-\omega_Q)\tau\sim1$. 

Figure~\ref{fig:CAB_compareQGSN} shows the rate of change between QG and SN, $|C_{AB}^{\rm SN}/C_{AB}^{\rm QG}-1|$. The intensity of the blue lines again varies with the measurement delay times $\tau_A=\tau_B=\{0,~2,~4,~6\}$. 
This figure shows that the differences are minimal, with a maximum value of less than $0.01\%$ around $\omega \sim \omega_{\pm}$, making them far from detectable.
As a result, distinguishing between the two gravity models under the current delay time conditions $\tau_A=\tau_B=\{0,~2,~4,~6\}$ remains a significant challenge.

Finally, we examine how the required observation time ${\mathcal{T}}$, given in Eq.~\eqref{eq:mutual_Tobs}, is affected by the time delayed measurement. Figure~\ref{fig:Tobs_mutual} presents contour plots of ${\mathcal{T}}$ as a function of the optomechanical coupling $\Lambda$ and temperature $T$. Each panel corresponds to a different delay time: $\tau = 0.000\,{\rm s},~0.001\,{\rm s},~0.100\,{\rm s},~1.000\,{\rm s}$. In all cases, the required observation time decreases as the temperature decreases. Additionally, introducing a time delay further shortens the observation time. However, even with a delay of $\tau=1.000\,{\rm s}$, the required observation time remains at least $\mathcal{O}(10^{18})\,{\rm s}$, making practical realization extremely difficult.
{
The reason for such a long observation time can be understood from the approximate expression in Eq.~\eqref{eq:mutual_Tobs}. The required time is inversely proportional to the sixth power of $\omega_g$, leading to an extraordinaryly large value. This constraint may be alleviated if $\omega_m$ is comparable to $\omega_g$ rather than assuming $\omega_m\gg\omega_g$. However, since Eq.~\eqref{eq:mutual_Tobs} is derived under the assumption $\omega_m\gg\omega_g$, a different form would be required for the case $\omega_m\sim\omega_g$.
}

\section{Conclusions}
\label{sec:conclusions}

In this paper, aiming at contributing to the ongoing discussion of how to ``test the quantum nature of gravity'', we have argued that it is crucial to consider the measurement processes' impact on the classical expectation values of physical observables, which in turn must affect the classical gravitational field.  In fact, this argument dates back to the experiment of Page and Geilker~\cite{page1981indirect}; this paper systematically examined various classical gravity models and discussed how measurement results can be incorporated into classical gravitational fields, proposing a unified theoretical framework.  In this framework, classical information is drawn from three sources: the experimentalists's measurement results, the environment, and auxiliary observers that are additionally introduced to ``collapse'' macroscopic quantum superpositions.  We have argued that the Causal-Conditional Schroedinger-Newton (CCSN) theory is a minimum model within this framework.  By filtering classical information from the environment and the experimentalists, the CCSN model allows the gravity field to receive information to reconstruct gravitational interaction at the macroscopic scale without introducing additional decoherence. By contrast, KTM~\cite{kafri2014classical} and Oppenheim and collaborators~\cite{layton2024healthier,oppenheim2023postquantum} models emphasize the roles of additional observers that collect information.  Nevertheless, if we do consider the fact that measuring devices are usually the most macroscopic objects in experiments, and therefore, the auxiliary observers may mainly extract information also from the devices, these models are much less different from CCSN than at first sight.  The CCSN model is the extreme case where information is solely extracted from the measuring device. 

As we compare to previous, naive models of classical gravity, e.g., the naive SN theory with ``pre-selection''~\cite{Yang13,helou2017measurable}, the CCSN theory produces phenomenology that much closer mimics quantum gravity in many situations~\cite{Liu2024,Yubao}.  In this paper, as we have viewed CCSN as a feedback-control scheme within linear quantum mechanics, we have introduced the use of conventional techniques from linear quantum control theory, in particular Wiener filtering.  Its equivalence to the Stochastic Schr\"odinger Equation treatment of Ref.~\cite{Yubao} has been proved in Ref.~\cite{Liu2024}.   Using Wiener filtering, we analyzed the phenomenology of CCSN in several different situations.


First, we examined the predictions of CCSN for  a single test mass under continuous measurement. In naive classical gravity, or the ``pre-selection'' model of SN, assuming that the classical expectation value of the mass's position operator can be set to zero, the spectral peak of the output light would appear at a frequency shifted by the self-gravity SN effect. Since the typical SN characteristic frequency is two orders of magnitude higher than the typical mutual gravity characteristic frequency, this would have made SN gravity much easier to detect than quantum gravity. 

However, in CCSN,  classical information about the mass feeds back into the gravitational field, shifting the spectral peak back to closely resemble the prediction of quantum gravity, consistent with the results in Ref.~\cite{Yubao}. Nevertheless, we quantified the difference between CCSN gravity and quantum gravity, showing that it is in principle observable, see Fig.~\ref{fig:nodelaycontour}.  To better distinguish between CCSN and quantum gravity, we introduced a time delay in the measurement process. The delayed arrival of test-mass information into the gravity field leads to output light spectra that differ more from quantum gravity --- with a difference that also increase with measurement strength (e.g., optical power). In practice, if a delay of around 0.25\,second can be achieved, the distinctions between SN gravity and standard quantum mechanics become much more easily visible, as we show in Fig.~\ref{fig:delaycontour}. 

Next, we discussed stopping continuous measurement as an alternative approach to the time delay. In this scenario, the conditional variance can be computed using a causal Wiener filter, conditioned on measurement outcomes recorded before stopping the measurement. We demonstrated that the time evolution of the conditional variance of mass position for SN gravity deviates from the evolution under standard quantum mechanics --- almost in the same way a predicted by ``pre-selection'' models of Schr\"odinger-Newton theory~\cite{Yang13}. This deviation also becomes more pronounced for stronger measurement strengths (e.g., stronger power). We note that this non-stationary strategy has not been fully analyzed: we still need to quantify the additional error in $\Delta x$ in the subsequent experiment. Nevertheless, such non-stationary strategies, e.g., also proposed for levitated objects, do seem promising in being able to test CCSN~\cite{Michimura2017,Delic2020,Hosseini2022}. 

Finally, we studied two gravitationally interacting test masses each monitored by light field, each with a possible time delay.  In this case, we ignored CCSN self gravity and only studied the effect of mutual gravity. 
Without time-delayed measurement, CCSN's prediction of the correlations between the two out-going light fields strongly mimics that of quantum gravity, making it very hard to distinguish from what is already a very challeging effect to measure. We also showed that a significant time delay on the order of $(\omega_1-\omega_Q)\tau \sim 10^{-2}$ substantially reduces the correlation spectrum. This reduction arises because time-delayed measurements relax the collapse of each system's quantum state. However, realizing such a significant time delay in the experiment is challenging. 
{Moreover, the required observation time is extremely long, as shown in Fig.~\ref{fig:Tobs_mutual}. Therefore, it seems difficult to distinguish the signal of quantum and semiclassical mutual gravity from the time-delayed measurement. However, the possibility may remain using lower-frequency oscillators or techniques that amplify gravitational coupling \cite{pedernales2022,YK2023}.}

In summary, even though CCSN has more subtle features than original, more naive models of SN theories, it still leads to signatures that are much easier to detect than testing the establishment of mutual entanglement between two objects.  In this way, CCSN can be a great stepping stone toward testing the quantum nature of gravity.  

Before ending this paper, we would like to comment on the very recent proposal of L.\ Diosi on testing the causality violation of Schr\"odinger-Newton theory~\cite{diosi2025causality} by examining the difference of a test-mass' evolution according to whether or how out-going light from the mass has been detected.  While the CCSN formulation already restores causality --- which is preferred by most physicists, our work in Sec.~\ref{sec:single} did show that different strategies of measurement do affect the mass' evolution in non-trivial ways that can be experimentally detectable.  Furthermore, our work in Sec.~\ref{sec:Mutual}, which confirms the ``apparent entanglement via classical gravity'' found by Liu et al.~\cite{Liu2024}, also demonstrates a nontrivial effect of the measurement on the quantum system.~\footnote{The ``apparent entanglement'' found by Liu et al.~\cite{Liu2024} can be called a ``fake entanglement'', in the same spirit as the ``fake action-at-a-distance'' in Ref.~\cite{diosi2025causality}.}   In principle, one can also relax the causality requirement on the feedback mechanisms of CCSN, in which case the predicted causality violation can be subject to experimental tests.

\acknowledgments

We thank Dan Carney, Su Direkci, and Tomohiro Fujita for discussions.
D.M. was supported by JSPS KAKENHI (Grant No. JP22J21267).
Y.K. is supported by Grant-in-Aid for JSPS Fellows.
Y.C. is funded by the Simons Foundation (Award No.\ 568762). Y.M.\ and Y.L. are supported by the National Key R$\&$D Program of China ``Gravitational Wave Detection" (Grant No.2023YFC2205801), National Natural Science Foundation of China under Grant No.12474481, No.12441503

\appendix

\section{Wiener filtering with a time delay}
\label{apdx:wiener_delay}
We consider the conditional expectation value with the time-delayed measurement. Considering the measurement records $\hat b(t')$ for $t'<t-\tau$, the expectation value is given by Eq.~\eqref{eqxcdelay}, which leads to Eq.~\eqref{xcdelay}. Using the Fourier transformation, we can compute the expectation value as
\begin{align}
    x_c(t) &= \int_{-\infty}^{t-\tau} dt' \langle \hat x_Q (t) \hat w_Q(t')\rangle \hat w_Q(t') \nonumber\\
&=     \int_{-\infty}^{t-\tau} dt' \int_{-\infty}^{+\infty} \frac{d\Omega}{2\pi}S_{x_Q w_Q}(\Omega) e^{-i\Omega (t-t')} \hat w_Q(t') \nonumber\\
&=     \int_{-\infty}^{+\infty} dt' \int_{-\infty}^{+\infty} \frac{d\Omega}{2\pi}[S_{x_Q w_Q}(\Omega)]_\tau e^{-i\Omega (t-t')} \hat w_Q(t') \nonumber\\
&=     \int_{-\infty}^{+\infty} \frac{d\Omega}{2\pi}[S_{x_Q w_Q}(\Omega)]_\tau e^{-i\Omega t} \hat w_Q(\Omega),
\end{align}
where $[f]_\tau=\int_\tau^\infty dte^{i\Omega t}f(t)$ represents the causal function for taking $t>\tau\ge0$. Therefore, we derive
\begin{equation}
    x_c(\Omega) = [S_{x_Q w_Q}]_\tau \frac{\hat{b}_{\zeta Q}}{\phi_+}= \left[\frac{S_{x_Q b_{\zeta Q}}}{\phi_-}\right]_\tau \frac{\hat{b}_{\zeta Q}}{\phi_+}.
\end{equation}
\begin{widetext}

\section{Variance for the non-stationary measurement}
\label{apdx:nonstationary}
The conditional variance for the non-stationary measurement is given by Eq.~\eqref{vnonstationary}. For the third term of $\hat x(t)$, we will have to use properties of Fourier transform, as well as the ``[...]'' symbol \eqref{defbrackettau} to write it down in terms of frequency-domain quantities:
\begin{align}
    E[(\alpha\chi_c \Theta \hat a_1)(\tau) |\Theta \hat w_Q]=&\alpha \iint\limits_{-\infty}^0 dt' dt'' \chi_c(\tau - t')\langle  \hat a_1(t') \hat w_Q(t'')\rangle \hat w_Q(t'') \nonumber\\
    =& \alpha \iint\limits_{-\infty}^0 dt' dt'' \iint\limits_{-\infty}^{+\infty} \frac{d\Omega d\Omega'}{(2\pi)^2}\chi_c (\Omega)e^{-i\Omega (\tau - t')}  S_{\hat a_1 \hat w_Q} (\Omega')e^{-i\Omega'(t'-t'')} \hat w_Q(t'') \nonumber\\
    =&\int_{-\infty}^0 dt'' \int_{-\infty}^{+\infty} dt' \iint\limits_{-\infty}^{+\infty} \frac{d\Omega d\Omega'}{(2\pi)^2}[\chi_c]_\tau (\Omega)e^{-i\Omega (\tau - t')}  S_{\hat a_1 \hat w_Q} (\Omega')e^{-i\Omega'(t'-t'')} \hat w_Q(t'') \nonumber\\
    =& \int_{-\infty}^0 dt''
\int\limits_{-\infty}^{+\infty} \frac{d\Omega }{2\pi} [\chi_c]_\tau(\Omega) S_{\hat a_1 \hat w_Q}(\Omega) e^{-i\Omega (\tau-t'')}
\hat w_Q(t'') \nonumber\\
 =&   \int_{-\infty}^{+\infty } dt''
\int\limits_{-\infty}^{+\infty} \frac{d\Omega }{2\pi} \big[[\chi_c]_\tau S_{\hat a_1 \hat w_Q}\big]_\tau (\Omega) e^{-i\Omega (\tau-t'')}
\hat w_Q(t'')\nonumber\\
 =& 
\int\limits_{-\infty}^{+\infty} \frac{d\Omega }{2\pi} \big[[\chi_c]_\tau S_{\hat a_1 \hat w_Q}\big]_\tau (\Omega) e^{-i\Omega \tau}
\hat w_Q(\Omega)
\end{align}
We can then obtain the variance of this term:
We then need to compute its conditional expectation with respect to $\hat w_{\rm tot}$:
\begin{align}
E\Big[        E\big[(\alpha\chi_c \Theta \hat a_1)(\tau) |\Theta \hat w_Q\big] \big|\Theta\hat w_{\rm tot}\Big]
 & =  \int_{-\infty}^{0 } dt' \int_{-\infty}^{+\infty } dt''
\int\limits_{-\infty}^{+\infty} \frac{d\Omega }{2\pi} \big[[\chi_c]_\tau S_{\hat a_1 \hat w_Q}\big]_\tau (\Omega) e^{-i\Omega (\tau-t'')}
\langle \hat w_Q(t'')\hat w_{\rm tot}(t') \rangle \hat w_{\rm tot}(t')\nonumber\\
&= \int_{-\infty}^{0 } dt' \int_{-\infty}^{+\infty } dt''
\iint\limits_{-\infty}^{+\infty} \frac{d\Omega d\Omega '}{(2\pi)^2} \big[[\chi_c]_\tau S_{\hat a_1 \hat w_Q}\big]_\tau (\Omega) e^{-i\Omega (\tau-t'')}
S_{\hat w_Q \hat w_{\rm tot}} (\Omega') e^{-i\Omega'(t''-t')}
 \hat w_{\rm tot}(t')\nonumber\\
 &= \int_{-\infty}^{0 } dt'
\int\limits_{-\infty}^{+\infty} \frac{d\Omega }{2\pi} \big[[\chi_c]_\tau S_{\hat a_1 \hat w_Q}\big]_\tau (\Omega) e^{-i\Omega (\tau-t')}
S_{\hat w_Q \hat w_{\rm tot}} (\Omega) 
 \hat w_{\rm tot}(t') \nonumber\\
 &= 
\int\limits_{-\infty}^{+\infty} \frac{d\Omega }{2\pi}
\Big[
\big[[\chi_c]_\tau S_{\hat a_1 \hat w_Q}\big]_\tau S_{\hat w_Q \hat w_{\rm tot}} \Big]_\tau (\Omega) e^{-i\Omega \tau}
 \hat w_{\rm tot}(\Omega)
 \end{align}
We can compute the other terms and correlations in the same way. As a result, we obtain the conditional variance Eq.~\eqref{vxx}.

\section{The spectrum in the mutual gravity case}
\label{apdx:mutual_gravity}

In this section, we show the explicit form of the spectrum in the mutual gravity case.
For simplicity, we assume that the parameters for system A and B are identical, except for $\zeta_A,~\zeta_B$ and $\tau_A,~\tau_B$; specifically, $\omega_{mA}=\omega_{mB}=\omega_m$, $M_A=M_B=M$, $\gamma_A=\gamma_B=\gamma$, $\alpha_A=\alpha_B=\alpha$, $f_{\rm thA}=f_{\rm thB}=f_{\rm th}$.

\subsection{Quantum gravity}

First, we present the results for the quantum gravity case.

The spectral density for system A is expressed as
\begin{align}
    S_{\xi A \xi A}^{{\rm QG}}(\omega)
    =
    \frac{
    \Sigma_A^{\rm QG}(\omega)
    +\Lambda^2\sin^2\zeta_A\Sigma_{th}^{\rm QG}(\omega)
    }{
    \{(\omega^2-\omega_m^2)^2+4\gamma^2\omega^2\}
    \{\omega^2-\omega_m^2+2\omega_g^2)^2+4\gamma^2\omega^2\}
    },
\end{align}
where
\begin{align}
    \Sigma_A^{\rm QG}(\omega)
    &:=\left\{(\omega^2-\omega_m^2)^2+4\gamma^2\omega^2\right\}
    -\Lambda^2\sin2\zeta_A (\omega^2-\omega_m^2+\omega_g^2)\left\{(\omega^2-\omega_m^2)(\omega^2-\omega_m^2+2\omega_g^2)+4\gamma^2\omega^2\right\}\\
    &+\Lambda^4\sin^2\zeta_A \left\{(\omega^2-\omega_m^2)(\omega^2-\omega_m^2+2\omega_g^2)+2\omega_g^4+4\gamma^2\omega^2\right\},\\
    \Sigma_{th}^{\rm QG}(\omega)
    &:=\frac{S_{f_{\rm th} f_{\rm th}}}{\hbar M}\left\{(\omega^2-\omega_m^2)(\omega^2-\omega_m^2+2\omega_g^2)+2\omega_g^4+4\gamma^2\omega^2\right\}.
\end{align}

The correlation spectrum between the two systems A and B is given by
\begin{align}
    S_{\xi A \xi B}^{{\rm QG}}(\omega)
    =
    \frac{
    \omega_g^2\left\{
    \sin(\zeta_A+\zeta_B)\Gamma_{AB}^{\rm QG}(\omega)
    -2\Lambda^2(1+S_{f_{\rm th} f_{\rm th}}/(\hbar M))(\omega^2-\omega_m^2+\omega_g^2)\sin\zeta_A \sin\zeta_B
    \right\}
    }{
    \{(\omega^2-\omega_m^2)^2+4\gamma^2\omega^2\}
    \{\omega^2-\omega_m^2+2\omega_g^2)^2+4\gamma^2\omega^2\}
    },
\end{align}
where
\begin{align}
    \Gamma_{AB}^{\rm QG}(\omega)
    =\Lambda^2\left\{
    (\omega^2-\omega_m^2)(\omega^2-\omega_m^2+2\omega_g^2)-4\gamma^2\omega^2
    \right\}.
\end{align}

Using these formulas, we derive the explicit form of the normalized correlation spectrum $C_{AB}$, as defined in Eq.~\eqref{eq:mutual_gravity_CAB}.

\subsection{Causal-Conditional Schro\"dinger-Newton gravity}

Next, we present the results for the CCSN case.

The spectral density of the system A is given by
\begin{align}
S_{\xi A \xi A}^{{\rm SN}}(\omega)
=
\frac{
\Sigma_A^{\rm SN}(\omega)
+\Lambda^2\sin^2\zeta_A\Sigma_{th}^{\rm SN}(\omega)
}{
\{(\omega^2-\omega_m^2)^2+4\gamma^2\omega^2\}
\{\omega^2-\omega_m^2+\omega_g^2)^2+4\gamma^2\omega^2\}
\{\omega^2-\omega_m^2+2\omega_g^2)^2+4\gamma^2\omega^2\}
}
\end{align}
where
\begin{align}
\Sigma_A^{\rm SN}(\omega)
&:=\left|
(\omega^2-\beta_A^2)\left\{
(\omega^2-\omega_m^2+2i\gamma\omega)
(\omega^2-\omega_m^2+2\omega_g^2+2i\gamma\omega)
+\omega_g^4\alpha\sin\zeta_A K_{\tau A}(\omega)
\right\}
\right|^2\\
&\hspace{20mm}+\omega_g^4\alpha^2\sin^2\zeta_A
\left|(\omega^2-\beta_B^2)K_{\tau B}(\omega)\right|
\left\{
(\omega^2-\omega_m^2+\omega_g^2)^2+4\gamma^2\omega^2
\right\},\\
\Sigma_{th}^{\rm SN}(\omega)
&:=\frac{S_{f_{\rm th} f_{\rm th}}}{\hbar M}
\left\{
(\omega^2-\omega_m^2+\omega_g^2)^2+4\gamma^2\omega^2
\right\}
\left\{
(\omega^2-\omega_m^2+\omega_g^2)^2+\omega_g^4+4\gamma^2\omega^2
\right\}
.
\end{align}
Here, $K_{\tau A/B}$ is the Wiener filter function for the time delayed measurement given in Eq.~\eqref{eq:KtauA}.

The correlation spectrum between two systems A and B is given by
\begin{align}
S_{\xi A \xi B}^{{\rm SN}}(\omega)
=
\frac{
\omega_g^2\left\{
\alpha \sin\zeta_A \Gamma_A^{\rm SN}(\omega)
+\alpha \sin\zeta_B\Gamma_B^{\rm SN}(\omega)^*
+
2\Lambda^2\sin\zeta_A\sin\zeta_B \Gamma_{th}^{\rm SN}(\omega)
\right\}
}{
\{(\omega^2-\omega_m^2)^2+4\gamma^2\omega^2\}
\{\omega^2-\omega_m^2+\omega_g^2)^2+4\gamma^2\omega^2\}
\{\omega^2-\omega_m^2+2\omega_g^2)^2+4\gamma^2\omega^2\}
},
\end{align}
where
\begin{align}
\Gamma_A^{\rm SN}(\omega)
&:=
K_{\tau B}(\omega)|\omega^2-\beta_B^2|^2(\omega^2-\omega_m^2+\omega_g^2+2i\gamma\omega)
\left\{
(\omega^2-\omega_m^2-2i\gamma\omega)(\omega^2-\omega_m^2+2\omega_g^2-2i\gamma\omega)
+\omega_g^4\alpha\sin\zeta_B K_{\tau B}(\omega)^*
\right\},\\
\Gamma_B^{\rm SN}(\omega)
&:=
K_{\tau A}(\omega)|\omega^2-\beta_A^2|^2(\omega^2-\omega_m^2+\omega_g^2+2i\gamma\omega)
\left\{
(\omega^2-\omega_m^2-2i\gamma\omega)(\omega^2-\omega_m^2+2\omega_g^2-2i\gamma\omega)
+\omega_g^4\alpha\sin\zeta_A K_{\tau A}(\omega)^*
\right\},\\
\Gamma_{th}^{\rm SN}(\omega)
&:=
\frac{S_{f_{\rm th} f_{\rm th}}}{\hbar M}
(\omega^2-\omega_m^2+\omega_g^2)
\left\{
(\omega^2-\omega_m^2+\omega_g^2)^2+4\gamma^2\omega^2
\right\}.
\end{align}

Using these formulas, we derive the explicit form of the normalized correlation spectrum $C_{AB}$ for the SN case.

\end{widetext}


\bibliography{references.bib}

\begin{thebibliography}{49}%
\makeatletter
\providecommand \@ifxundefined [1]{%
 \@ifx{#1\undefined}
}%
\providecommand \@ifnum [1]{%
 \ifnum #1\expandafter \@firstoftwo
 \else \expandafter \@secondoftwo
 \fi
}%
\providecommand \@ifx [1]{%
 \ifx #1\expandafter \@firstoftwo
 \else \expandafter \@secondoftwo
 \fi
}%
\providecommand \natexlab [1]{#1}%
\providecommand \enquote  [1]{``#1''}%
\providecommand \bibnamefont  [1]{#1}%
\providecommand \bibfnamefont [1]{#1}%
\providecommand \citenamefont [1]{#1}%
\providecommand \href@noop [0]{\@secondoftwo}%
\providecommand \href [0]{\begingroup \@sanitize@url \@href}%
\providecommand \@href[1]{\@@startlink{#1}\@@href}%
\providecommand \@@href[1]{\endgroup#1\@@endlink}%
\providecommand \@sanitize@url [0]{\catcode `\\12\catcode `\$12\catcode `\&12\catcode `\#12\catcode `\^12\catcode `\_12\catcode `\%12\relax}%
\providecommand \@@startlink[1]{}%
\providecommand \@@endlink[0]{}%
\providecommand \url  [0]{\begingroup\@sanitize@url \@url }%
\providecommand \@url [1]{\endgroup\@href {#1}{\urlprefix }}%
\providecommand \urlprefix  [0]{URL }%
\providecommand \Eprint [0]{\href }%
\providecommand \doibase [0]{http://dx.doi.org/}%
\providecommand \selectlanguage [0]{\@gobble}%
\providecommand \bibinfo  [0]{\@secondoftwo}%
\providecommand \bibfield  [0]{\@secondoftwo}%
\providecommand \translation [1]{[#1]}%
\providecommand \BibitemOpen [0]{}%
\providecommand \bibitemStop [0]{}%
\providecommand \bibitemNoStop [0]{.\EOS\space}%
\providecommand \EOS [0]{\spacefactor3000\relax}%
\providecommand \BibitemShut  [1]{\csname bibitem#1\endcsname}%
\let\auto@bib@innerbib\@empty
\bibitem [{\citenamefont {Carlip}(2008)}]{carlip2008quantum}%
  \BibitemOpen
  \bibfield  {author} {\bibinfo {author} {\bibfnamefont {S.}~\bibnamefont {Carlip}},\ }\href@noop {} {\bibfield  {journal} {\bibinfo  {journal} {Classical and Quantum Gravity}\ }\textbf {\bibinfo {volume} {25}},\ \bibinfo {pages} {154010} (\bibinfo {year} {2008})}\BibitemShut {NoStop}%
\bibitem [{\citenamefont {Carney}\ \emph {et~al.}(2019)\citenamefont {Carney}, \citenamefont {Stamp},\ and\ \citenamefont {Taylor}}]{carney2019tabletop}%
  \BibitemOpen
  \bibfield  {author} {\bibinfo {author} {\bibfnamefont {D.}~\bibnamefont {Carney}}, \bibinfo {author} {\bibfnamefont {P.~C.}\ \bibnamefont {Stamp}}, \ and\ \bibinfo {author} {\bibfnamefont {J.~M.}\ \bibnamefont {Taylor}},\ }\href@noop {} {\bibfield  {journal} {\bibinfo  {journal} {Classical and Quantum Gravity}\ }\textbf {\bibinfo {volume} {36}},\ \bibinfo {pages} {034001} (\bibinfo {year} {2019})}\BibitemShut {NoStop}%
\bibitem [{\citenamefont {Yang}\ \emph {et~al.}(2013)\citenamefont {Yang}, \citenamefont {Miao}, \citenamefont {Lee}, \citenamefont {Helou},\ and\ \citenamefont {Chen}}]{Yang13}%
  \BibitemOpen
  \bibfield  {author} {\bibinfo {author} {\bibfnamefont {H.}~\bibnamefont {Yang}}, \bibinfo {author} {\bibfnamefont {H.}~\bibnamefont {Miao}}, \bibinfo {author} {\bibfnamefont {D.-S.}\ \bibnamefont {Lee}}, \bibinfo {author} {\bibfnamefont {B.}~\bibnamefont {Helou}}, \ and\ \bibinfo {author} {\bibfnamefont {Y.}~\bibnamefont {Chen}},\ }\href@noop {} {\bibfield  {journal} {\bibinfo  {journal} {Physical Review Letter}\ }\textbf {\bibinfo {volume} {110}},\ \bibinfo {pages} {170401} (\bibinfo {year} {2013})}\BibitemShut {NoStop}%
\bibitem [{\citenamefont {Giulini}\ and\ \citenamefont {Gro{\ss}ardt}(2011)}]{giulini2011gravitationally}%
  \BibitemOpen
  \bibfield  {author} {\bibinfo {author} {\bibfnamefont {D.}~\bibnamefont {Giulini}}\ and\ \bibinfo {author} {\bibfnamefont {A.}~\bibnamefont {Gro{\ss}ardt}},\ }\href@noop {} {\bibfield  {journal} {\bibinfo  {journal} {Classical and Quantum Gravity}\ }\textbf {\bibinfo {volume} {28}},\ \bibinfo {pages} {195026} (\bibinfo {year} {2011})}\BibitemShut {NoStop}%
\bibitem [{\citenamefont {Giulini}\ and\ \citenamefont {Gro{\ss}ardt}(2014)}]{giulini2014centre}%
  \BibitemOpen
  \bibfield  {author} {\bibinfo {author} {\bibfnamefont {D.}~\bibnamefont {Giulini}}\ and\ \bibinfo {author} {\bibfnamefont {A.}~\bibnamefont {Gro{\ss}ardt}},\ }\href@noop {} {\bibfield  {journal} {\bibinfo  {journal} {New Journal of Physics}\ }\textbf {\bibinfo {volume} {16}},\ \bibinfo {pages} {075005} (\bibinfo {year} {2014})}\BibitemShut {NoStop}%
\bibitem [{\citenamefont {Helou}\ \emph {et~al.}(2017{\natexlab{a}})\citenamefont {Helou}, \citenamefont {Luo}, \citenamefont {Yeh}, \citenamefont {Shao}, \citenamefont {Slagmolen}, \citenamefont {McClelland},\ and\ \citenamefont {Chen}}]{helou2017measurable}%
  \BibitemOpen
  \bibfield  {author} {\bibinfo {author} {\bibfnamefont {B.}~\bibnamefont {Helou}}, \bibinfo {author} {\bibfnamefont {J.}~\bibnamefont {Luo}}, \bibinfo {author} {\bibfnamefont {H.-C.}\ \bibnamefont {Yeh}}, \bibinfo {author} {\bibfnamefont {C.-g.}\ \bibnamefont {Shao}}, \bibinfo {author} {\bibfnamefont {B.}~\bibnamefont {Slagmolen}}, \bibinfo {author} {\bibfnamefont {D.~E.}\ \bibnamefont {McClelland}}, \ and\ \bibinfo {author} {\bibfnamefont {Y.}~\bibnamefont {Chen}},\ }\href@noop {} {\bibfield  {journal} {\bibinfo  {journal} {Physical Review D}\ }\textbf {\bibinfo {volume} {96}},\ \bibinfo {pages} {044008} (\bibinfo {year} {2017}{\natexlab{a}})}\BibitemShut {NoStop}%
\bibitem [{\citenamefont {Helou}\ and\ \citenamefont {Chen}(2017)}]{helou2017extensions}%
  \BibitemOpen
  \bibfield  {author} {\bibinfo {author} {\bibfnamefont {B.}~\bibnamefont {Helou}}\ and\ \bibinfo {author} {\bibfnamefont {Y.}~\bibnamefont {Chen}},\ }in\ \href@noop {} {\emph {\bibinfo {booktitle} {Journal of Physics: Conference Series}}},\ Vol.\ \bibinfo {volume} {880}\ (\bibinfo {organization} {IOP Publishing},\ \bibinfo {year} {2017})\ p.\ \bibinfo {pages} {012021}\BibitemShut {NoStop}%
\bibitem [{\citenamefont {Scully}(2022)}]{scully2022semiclassical}%
  \BibitemOpen
  \bibfield  {author} {\bibinfo {author} {\bibfnamefont {S.}~\bibnamefont {Scully}},\ }\emph {\bibinfo {title} {Semiclassical Pictures of Gravity: Investigating and Testing Non-Linear Theories of Quantum Gravity}},\ \href@noop {} {Ph.D. thesis},\ \bibinfo  {school} {The Australian National University (Australia)} (\bibinfo {year} {2022})\BibitemShut {NoStop}%
\bibitem [{\citenamefont {Helou}(2019)}]{helou2019testing}%
  \BibitemOpen
  \bibfield  {author} {\bibinfo {author} {\bibfnamefont {B.}~\bibnamefont {Helou}},\ }\href@noop {} {\emph {\bibinfo {title} {Testing alternative theories of Quantum Mechanics with Optomechanics, and effective modes for Gaussian linear Optomechanics}}}\ (\bibinfo  {publisher} {California Institute of Technology},\ \bibinfo {year} {2019})\BibitemShut {NoStop}%
\bibitem [{\citenamefont {Liu}\ \emph {et~al.}(2023)\citenamefont {Liu}, \citenamefont {Miao}, \citenamefont {Chen},\ and\ \citenamefont {Ma}}]{Yubao}%
  \BibitemOpen
  \bibfield  {author} {\bibinfo {author} {\bibfnamefont {Y.}~\bibnamefont {Liu}}, \bibinfo {author} {\bibfnamefont {H.}~\bibnamefont {Miao}}, \bibinfo {author} {\bibfnamefont {Y.}~\bibnamefont {Chen}}, \ and\ \bibinfo {author} {\bibfnamefont {Y.}~\bibnamefont {Ma}},\ }\href@noop {} {\bibfield  {journal} {\bibinfo  {journal} {Physical Review D}\ }\textbf {\bibinfo {volume} {107}},\ \bibinfo {pages} {024004} (\bibinfo {year} {2023})}\BibitemShut {NoStop}%
\bibitem [{\citenamefont {Liu}\ \emph {et~al.}(2025)\citenamefont {Liu}, \citenamefont {Zhong}, \citenamefont {Chen},\ and\ \citenamefont {Ma}}]{Liu2024}%
  \BibitemOpen
  \bibfield  {author} {\bibinfo {author} {\bibfnamefont {Y.}~\bibnamefont {Liu}}, \bibinfo {author} {\bibfnamefont {W.}~\bibnamefont {Zhong}}, \bibinfo {author} {\bibfnamefont {Y.}~\bibnamefont {Chen}}, \ and\ \bibinfo {author} {\bibfnamefont {Y.}~\bibnamefont {Ma}},\ }\href@noop {} {\bibfield  {journal} {\bibinfo  {journal} {Phys. Rev. D}\ }\textbf {\bibinfo {volume} {111}},\ \bibinfo {pages} {062004} (\bibinfo {year} {2025})}\BibitemShut {NoStop}%
\bibitem [{\citenamefont {Howl}\ \emph {et~al.}(2021)\citenamefont {Howl}, \citenamefont {Vedral}, \citenamefont {Naik}, \citenamefont {Christodoulou}, \citenamefont {Rovelli},\ and\ \citenamefont {Iyer}}]{howl2021non}%
  \BibitemOpen
  \bibfield  {author} {\bibinfo {author} {\bibfnamefont {R.}~\bibnamefont {Howl}}, \bibinfo {author} {\bibfnamefont {V.}~\bibnamefont {Vedral}}, \bibinfo {author} {\bibfnamefont {D.}~\bibnamefont {Naik}}, \bibinfo {author} {\bibfnamefont {M.}~\bibnamefont {Christodoulou}}, \bibinfo {author} {\bibfnamefont {C.}~\bibnamefont {Rovelli}}, \ and\ \bibinfo {author} {\bibfnamefont {A.}~\bibnamefont {Iyer}},\ }\href@noop {} {\bibfield  {journal} {\bibinfo  {journal} {PRX Quantum}\ }\textbf {\bibinfo {volume} {2}},\ \bibinfo {pages} {010325} (\bibinfo {year} {2021})}\BibitemShut {NoStop}%
\bibitem [{\citenamefont {Kafri}\ \emph {et~al.}(2014)\citenamefont {Kafri}, \citenamefont {Taylor},\ and\ \citenamefont {Milburn}}]{kafri2014classical}%
  \BibitemOpen
  \bibfield  {author} {\bibinfo {author} {\bibfnamefont {D.}~\bibnamefont {Kafri}}, \bibinfo {author} {\bibfnamefont {J.}~\bibnamefont {Taylor}}, \ and\ \bibinfo {author} {\bibfnamefont {G.}~\bibnamefont {Milburn}},\ }\href@noop {} {\bibfield  {journal} {\bibinfo  {journal} {New Journal of Physics}\ }\textbf {\bibinfo {volume} {16}},\ \bibinfo {pages} {065020} (\bibinfo {year} {2014})}\BibitemShut {NoStop}%
\bibitem [{\citenamefont {Kafri}\ \emph {et~al.}(2015)\citenamefont {Kafri}, \citenamefont {Milburn},\ and\ \citenamefont {Taylor}}]{kafri2015bounds}%
  \BibitemOpen
  \bibfield  {author} {\bibinfo {author} {\bibfnamefont {D.}~\bibnamefont {Kafri}}, \bibinfo {author} {\bibfnamefont {G.}~\bibnamefont {Milburn}}, \ and\ \bibinfo {author} {\bibfnamefont {J.}~\bibnamefont {Taylor}},\ }\href@noop {} {\bibfield  {journal} {\bibinfo  {journal} {New Journal of Physics}\ }\textbf {\bibinfo {volume} {17}},\ \bibinfo {pages} {015006} (\bibinfo {year} {2015})}\BibitemShut {NoStop}%
\bibitem [{\citenamefont {Bose}\ \emph {et~al.}(2017)\citenamefont {Bose}, \citenamefont {Mazumdar}, \citenamefont {Morley}, \citenamefont {Ulbricht}, \citenamefont {Toro{\v{s}}}, \citenamefont {Paternostro}, \citenamefont {Geraci}, \citenamefont {Barker}, \citenamefont {Kim},\ and\ \citenamefont {Milburn}}]{bose2017spin}%
  \BibitemOpen
  \bibfield  {author} {\bibinfo {author} {\bibfnamefont {S.}~\bibnamefont {Bose}}, \bibinfo {author} {\bibfnamefont {A.}~\bibnamefont {Mazumdar}}, \bibinfo {author} {\bibfnamefont {G.~W.}\ \bibnamefont {Morley}}, \bibinfo {author} {\bibfnamefont {H.}~\bibnamefont {Ulbricht}}, \bibinfo {author} {\bibfnamefont {M.}~\bibnamefont {Toro{\v{s}}}}, \bibinfo {author} {\bibfnamefont {M.}~\bibnamefont {Paternostro}}, \bibinfo {author} {\bibfnamefont {A.~A.}\ \bibnamefont {Geraci}}, \bibinfo {author} {\bibfnamefont {P.~F.}\ \bibnamefont {Barker}}, \bibinfo {author} {\bibfnamefont {M.}~\bibnamefont {Kim}}, \ and\ \bibinfo {author} {\bibfnamefont {G.}~\bibnamefont {Milburn}},\ }\href@noop {} {\bibfield  {journal} {\bibinfo  {journal} {Physical review letters}\ }\textbf {\bibinfo {volume} {119}},\ \bibinfo {pages} {240401} (\bibinfo {year} {2017})}\BibitemShut {NoStop}%
\bibitem [{\citenamefont {Marletto}\ and\ \citenamefont {Vedral}(2017)}]{marletto2017gravitationally}%
  \BibitemOpen
  \bibfield  {author} {\bibinfo {author} {\bibfnamefont {C.}~\bibnamefont {Marletto}}\ and\ \bibinfo {author} {\bibfnamefont {V.}~\bibnamefont {Vedral}},\ }\href@noop {} {\bibfield  {journal} {\bibinfo  {journal} {Physical Review Letters}\ }\textbf {\bibinfo {volume} {119}},\ \bibinfo {pages} {240402} (\bibinfo {year} {2017})}\BibitemShut {NoStop}%
\bibitem [{\citenamefont {Miao}\ \emph {et~al.}(2020)\citenamefont {Miao}, \citenamefont {Martynov}, \citenamefont {Yang},\ and\ \citenamefont {Datta}}]{Miao}%
  \BibitemOpen
  \bibfield  {author} {\bibinfo {author} {\bibfnamefont {H.}~\bibnamefont {Miao}}, \bibinfo {author} {\bibfnamefont {D.}~\bibnamefont {Martynov}}, \bibinfo {author} {\bibfnamefont {H.}~\bibnamefont {Yang}}, \ and\ \bibinfo {author} {\bibfnamefont {A.}~\bibnamefont {Datta}},\ }\href@noop {} {\bibfield  {journal} {\bibinfo  {journal} {Physical Review A}\ }\textbf {\bibinfo {volume} {101}},\ \bibinfo {pages} {063804} (\bibinfo {year} {2020})}\BibitemShut {NoStop}%
\bibitem [{\citenamefont {Lami}\ \emph {et~al.}(2024)\citenamefont {Lami}, \citenamefont {Pedernales},\ and\ \citenamefont {Plenio}}]{lami2024testing}%
  \BibitemOpen
  \bibfield  {author} {\bibinfo {author} {\bibfnamefont {L.}~\bibnamefont {Lami}}, \bibinfo {author} {\bibfnamefont {J.~S.}\ \bibnamefont {Pedernales}}, \ and\ \bibinfo {author} {\bibfnamefont {M.~B.}\ \bibnamefont {Plenio}},\ }\href@noop {} {\bibfield  {journal} {\bibinfo  {journal} {Physical Review X}\ }\textbf {\bibinfo {volume} {14}},\ \bibinfo {pages} {021022} (\bibinfo {year} {2024})}\BibitemShut {NoStop}%
\bibitem [{\citenamefont {Diosi}(1987)}]{diosi1987universal}%
  \BibitemOpen
  \bibfield  {author} {\bibinfo {author} {\bibfnamefont {L.}~\bibnamefont {Diosi}},\ }\href@noop {} {\bibfield  {journal} {\bibinfo  {journal} {Physics Letters A}\ }\textbf {\bibinfo {volume} {120}},\ \bibinfo {pages} {377} (\bibinfo {year} {1987})}\BibitemShut {NoStop}%
\bibitem [{\citenamefont {Penrose}(1996)}]{penrose1996gravity}%
  \BibitemOpen
  \bibfield  {author} {\bibinfo {author} {\bibfnamefont {R.}~\bibnamefont {Penrose}},\ }\href@noop {} {\bibfield  {journal} {\bibinfo  {journal} {General relativity and gravitation}\ }\textbf {\bibinfo {volume} {28}},\ \bibinfo {pages} {581} (\bibinfo {year} {1996})}\BibitemShut {NoStop}%
\bibitem [{\citenamefont {Ghirardi}\ \emph {et~al.}(1986)\citenamefont {Ghirardi}, \citenamefont {Rimini},\ and\ \citenamefont {Weber}}]{ghirardi1986unified}%
  \BibitemOpen
  \bibfield  {author} {\bibinfo {author} {\bibfnamefont {G.~C.}\ \bibnamefont {Ghirardi}}, \bibinfo {author} {\bibfnamefont {A.}~\bibnamefont {Rimini}}, \ and\ \bibinfo {author} {\bibfnamefont {T.}~\bibnamefont {Weber}},\ }\href@noop {} {\bibfield  {journal} {\bibinfo  {journal} {Physical Review D}\ }\textbf {\bibinfo {volume} {34}},\ \bibinfo {pages} {470} (\bibinfo {year} {1986})}\BibitemShut {NoStop}%
\bibitem [{\citenamefont {Bassi}\ \emph {et~al.}(2023)\citenamefont {Bassi}, \citenamefont {Dorato},\ and\ \citenamefont {Ulbricht}}]{bassi2023collapse}%
  \BibitemOpen
  \bibfield  {author} {\bibinfo {author} {\bibfnamefont {A.}~\bibnamefont {Bassi}}, \bibinfo {author} {\bibfnamefont {M.}~\bibnamefont {Dorato}}, \ and\ \bibinfo {author} {\bibfnamefont {H.}~\bibnamefont {Ulbricht}},\ }\href@noop {} {\bibfield  {journal} {\bibinfo  {journal} {Entropy}\ }\textbf {\bibinfo {volume} {25}},\ \bibinfo {pages} {645} (\bibinfo {year} {2023})}\BibitemShut {NoStop}%
\bibitem [{\citenamefont {Nimmrichter}\ and\ \citenamefont {Hornberger}(2015)}]{nimmrichter2015stochastic}%
  \BibitemOpen
  \bibfield  {author} {\bibinfo {author} {\bibfnamefont {S.}~\bibnamefont {Nimmrichter}}\ and\ \bibinfo {author} {\bibfnamefont {K.}~\bibnamefont {Hornberger}},\ }\href@noop {} {\bibfield  {journal} {\bibinfo  {journal} {Physical Review D}\ }\textbf {\bibinfo {volume} {91}},\ \bibinfo {pages} {024016} (\bibinfo {year} {2015})}\BibitemShut {NoStop}%
\bibitem [{\citenamefont {Oppenheim}(2023)}]{oppenheim2023postquantum}%
  \BibitemOpen
  \bibfield  {author} {\bibinfo {author} {\bibfnamefont {J.}~\bibnamefont {Oppenheim}},\ }\href@noop {} {\bibfield  {journal} {\bibinfo  {journal} {Physical Review X}\ }\textbf {\bibinfo {volume} {13}},\ \bibinfo {pages} {041040} (\bibinfo {year} {2023})}\BibitemShut {NoStop}%
\bibitem [{\citenamefont {Page}\ and\ \citenamefont {Geilker}(1981)}]{page1981indirect}%
  \BibitemOpen
  \bibfield  {author} {\bibinfo {author} {\bibfnamefont {D.~N.}\ \bibnamefont {Page}}\ and\ \bibinfo {author} {\bibfnamefont {C.}~\bibnamefont {Geilker}},\ }\href@noop {} {\bibfield  {journal} {\bibinfo  {journal} {Physical Review Letters}\ }\textbf {\bibinfo {volume} {47}},\ \bibinfo {pages} {979} (\bibinfo {year} {1981})}\BibitemShut {NoStop}%
\bibitem [{\citenamefont {Helou}\ \emph {et~al.}(2017{\natexlab{b}})\citenamefont {Helou}, \citenamefont {Slagmolen}, \citenamefont {McClelland},\ and\ \citenamefont {Chen}}]{helou2017lisa}%
  \BibitemOpen
  \bibfield  {author} {\bibinfo {author} {\bibfnamefont {B.}~\bibnamefont {Helou}}, \bibinfo {author} {\bibfnamefont {B.}~\bibnamefont {Slagmolen}}, \bibinfo {author} {\bibfnamefont {D.~E.}\ \bibnamefont {McClelland}}, \ and\ \bibinfo {author} {\bibfnamefont {Y.}~\bibnamefont {Chen}},\ }\href@noop {} {\bibfield  {journal} {\bibinfo  {journal} {Physical Review D}\ }\textbf {\bibinfo {volume} {95}},\ \bibinfo {pages} {084054} (\bibinfo {year} {2017}{\natexlab{b}})}\BibitemShut {NoStop}%
\bibitem [{\citenamefont {Carlesso}\ \emph {et~al.}(2016)\citenamefont {Carlesso}, \citenamefont {Bassi}, \citenamefont {Falferi},\ and\ \citenamefont {Vinante}}]{carlesso2016experimental}%
  \BibitemOpen
  \bibfield  {author} {\bibinfo {author} {\bibfnamefont {M.}~\bibnamefont {Carlesso}}, \bibinfo {author} {\bibfnamefont {A.}~\bibnamefont {Bassi}}, \bibinfo {author} {\bibfnamefont {P.}~\bibnamefont {Falferi}}, \ and\ \bibinfo {author} {\bibfnamefont {A.}~\bibnamefont {Vinante}},\ }\href@noop {} {\bibfield  {journal} {\bibinfo  {journal} {Physical Review D}\ }\textbf {\bibinfo {volume} {94}},\ \bibinfo {pages} {124036} (\bibinfo {year} {2016})}\BibitemShut {NoStop}%
\bibitem [{\citenamefont {Vinante}\ \emph {et~al.}(2017)\citenamefont {Vinante}, \citenamefont {Mezzena}, \citenamefont {Falferi}, \citenamefont {Carlesso},\ and\ \citenamefont {Bassi}}]{vinante2017improved}%
  \BibitemOpen
  \bibfield  {author} {\bibinfo {author} {\bibfnamefont {A.}~\bibnamefont {Vinante}}, \bibinfo {author} {\bibfnamefont {R.}~\bibnamefont {Mezzena}}, \bibinfo {author} {\bibfnamefont {P.}~\bibnamefont {Falferi}}, \bibinfo {author} {\bibfnamefont {M.}~\bibnamefont {Carlesso}}, \ and\ \bibinfo {author} {\bibfnamefont {A.}~\bibnamefont {Bassi}},\ }\href@noop {} {\bibfield  {journal} {\bibinfo  {journal} {Physical Review Letters}\ }\textbf {\bibinfo {volume} {119}},\ \bibinfo {pages} {110401} (\bibinfo {year} {2017})}\BibitemShut {NoStop}%
\bibitem [{\citenamefont {Donadi}\ \emph {et~al.}(2021)\citenamefont {Donadi}, \citenamefont {Piscicchia}, \citenamefont {Curceanu}, \citenamefont {Di{\'o}si}, \citenamefont {Laubenstein},\ and\ \citenamefont {Bassi}}]{ddonadi2021underground}%
  \BibitemOpen
  \bibfield  {author} {\bibinfo {author} {\bibfnamefont {S.}~\bibnamefont {Donadi}}, \bibinfo {author} {\bibfnamefont {K.}~\bibnamefont {Piscicchia}}, \bibinfo {author} {\bibfnamefont {C.}~\bibnamefont {Curceanu}}, \bibinfo {author} {\bibfnamefont {L.}~\bibnamefont {Di{\'o}si}}, \bibinfo {author} {\bibfnamefont {M.}~\bibnamefont {Laubenstein}}, \ and\ \bibinfo {author} {\bibfnamefont {A.}~\bibnamefont {Bassi}},\ }\href@noop {} {\bibfield  {journal} {\bibinfo  {journal} {Nature Physics}\ }\textbf {\bibinfo {volume} {17}},\ \bibinfo {pages} {74} (\bibinfo {year} {2021})}\BibitemShut {NoStop}%
\bibitem [{\citenamefont {Wiseman}\ and\ \citenamefont {Milburn}(2009)}]{wiseman2009quantum}%
  \BibitemOpen
  \bibfield  {author} {\bibinfo {author} {\bibfnamefont {H.~M.}\ \bibnamefont {Wiseman}}\ and\ \bibinfo {author} {\bibfnamefont {G.~J.}\ \bibnamefont {Milburn}},\ }\href@noop {} {\emph {\bibinfo {title} {Quantum measurement and control}}}\ (\bibinfo  {publisher} {Cambridge university press},\ \bibinfo {year} {2009})\BibitemShut {NoStop}%
\bibitem [{\citenamefont {M\"{u}ller-Ebhardt}\ \emph {et~al.}(2009)\citenamefont {M\"{u}ller-Ebhardt}, \citenamefont {Rehbein}, \citenamefont {Li}, \citenamefont {Mino}, \citenamefont {Somiya}, \citenamefont {Schnabel}, \citenamefont {Danzmann},\ and\ \citenamefont {Chen}}]{Ebhardt09}%
  \BibitemOpen
  \bibfield  {author} {\bibinfo {author} {\bibfnamefont {H.}~\bibnamefont {M\"{u}ller-Ebhardt}}, \bibinfo {author} {\bibfnamefont {H.}~\bibnamefont {Rehbein}}, \bibinfo {author} {\bibfnamefont {C.}~\bibnamefont {Li}}, \bibinfo {author} {\bibfnamefont {Y.}~\bibnamefont {Mino}}, \bibinfo {author} {\bibfnamefont {K.}~\bibnamefont {Somiya}}, \bibinfo {author} {\bibfnamefont {R.}~\bibnamefont {Schnabel}}, \bibinfo {author} {\bibfnamefont {K.}~\bibnamefont {Danzmann}}, \ and\ \bibinfo {author} {\bibfnamefont {Y.}~\bibnamefont {Chen}},\ }\href@noop {} {\bibfield  {journal} {\bibinfo  {journal} {Physical Review A}\ }\textbf {\bibinfo {volume} {80}},\ \bibinfo {pages} {043802} (\bibinfo {year} {2009})}\BibitemShut {NoStop}%
\bibitem [{\citenamefont {Chen}(2013)}]{chen2013macroscopic}%
  \BibitemOpen
  \bibfield  {author} {\bibinfo {author} {\bibfnamefont {Y.}~\bibnamefont {Chen}},\ }\href@noop {} {\bibfield  {journal} {\bibinfo  {journal} {Journal of Physics B: Atomic, Molecular and Optical Physics}\ }\textbf {\bibinfo {volume} {46}},\ \bibinfo {pages} {104001} (\bibinfo {year} {2013})}\BibitemShut {NoStop}%
\bibitem [{\citenamefont {Helou}\ \emph {et~al.}(2017{\natexlab{c}})\citenamefont {Helou}, \citenamefont {Luo}, \citenamefont {Yeh}, \citenamefont {gang Shao}, \citenamefont {Slagmolen}, \citenamefont {McClelland},\ and\ \citenamefont {Chen}}]{Helou17}%
  \BibitemOpen
  \bibfield  {author} {\bibinfo {author} {\bibfnamefont {B.}~\bibnamefont {Helou}}, \bibinfo {author} {\bibfnamefont {J.}~\bibnamefont {Luo}}, \bibinfo {author} {\bibfnamefont {H.-C.}\ \bibnamefont {Yeh}}, \bibinfo {author} {\bibfnamefont {C.}~\bibnamefont {gang Shao}}, \bibinfo {author} {\bibfnamefont {B.~J.~J.}\ \bibnamefont {Slagmolen}}, \bibinfo {author} {\bibfnamefont {D.~E.}\ \bibnamefont {McClelland}}, \ and\ \bibinfo {author} {\bibfnamefont {Y.}~\bibnamefont {Chen}},\ }\href@noop {} {\bibfield  {journal} {\bibinfo  {journal} {Physical Review D}\ }\textbf {\bibinfo {volume} {96}},\ \bibinfo {pages} {044008} (\bibinfo {year} {2017}{\natexlab{c}})}\BibitemShut {NoStop}%
\bibitem [{\citenamefont {Polchinski}(1991)}]{polchinski1991weinberg}%
  \BibitemOpen
  \bibfield  {author} {\bibinfo {author} {\bibfnamefont {J.}~\bibnamefont {Polchinski}},\ }\href@noop {} {\bibfield  {journal} {\bibinfo  {journal} {Physical Review Letters}\ }\textbf {\bibinfo {volume} {66}},\ \bibinfo {pages} {397} (\bibinfo {year} {1991})}\BibitemShut {NoStop}%
\bibitem [{\citenamefont {Simon}\ \emph {et~al.}(2001)\citenamefont {Simon}, \citenamefont {Bu{\v{z}}ek},\ and\ \citenamefont {Gisin}}]{simon2001no}%
  \BibitemOpen
  \bibfield  {author} {\bibinfo {author} {\bibfnamefont {C.}~\bibnamefont {Simon}}, \bibinfo {author} {\bibfnamefont {V.}~\bibnamefont {Bu{\v{z}}ek}}, \ and\ \bibinfo {author} {\bibfnamefont {N.}~\bibnamefont {Gisin}},\ }\href@noop {} {\bibfield  {journal} {\bibinfo  {journal} {Physical Review Letters}\ }\textbf {\bibinfo {volume} {87}},\ \bibinfo {pages} {170405} (\bibinfo {year} {2001})}\BibitemShut {NoStop}%
\bibitem [{\citenamefont {Di{\'o}si}(2025)}]{diosi2025causality}%
  \BibitemOpen
  \bibfield  {author} {\bibinfo {author} {\bibfnamefont {L.}~\bibnamefont {Di{\'o}si}},\ }\href@noop {} {\bibfield  {journal} {\bibinfo  {journal} {arXiv preprint arXiv:2503.07458}\ } (\bibinfo {year} {2025})}\BibitemShut {NoStop}%
\bibitem [{\citenamefont {Layton}\ \emph {et~al.}(2024)\citenamefont {Layton}, \citenamefont {Oppenheim},\ and\ \citenamefont {Weller-Davies}}]{layton2024healthier}%
  \BibitemOpen
  \bibfield  {author} {\bibinfo {author} {\bibfnamefont {I.}~\bibnamefont {Layton}}, \bibinfo {author} {\bibfnamefont {J.}~\bibnamefont {Oppenheim}}, \ and\ \bibinfo {author} {\bibfnamefont {Z.}~\bibnamefont {Weller-Davies}},\ }\href@noop {} {\bibfield  {journal} {\bibinfo  {journal} {Quantum}\ }\textbf {\bibinfo {volume} {8}},\ \bibinfo {pages} {1565} (\bibinfo {year} {2024})}\BibitemShut {NoStop}%
\bibitem [{\citenamefont {Pang}\ and\ \citenamefont {Chen}(2018)}]{pang2018quantum}%
  \BibitemOpen
  \bibfield  {author} {\bibinfo {author} {\bibfnamefont {B.}~\bibnamefont {Pang}}\ and\ \bibinfo {author} {\bibfnamefont {Y.}~\bibnamefont {Chen}},\ }\href@noop {} {\bibfield  {journal} {\bibinfo  {journal} {Physical Review D}\ }\textbf {\bibinfo {volume} {98}},\ \bibinfo {pages} {124006} (\bibinfo {year} {2018})}\BibitemShut {NoStop}%
\bibitem [{\citenamefont {Carney}\ and\ \citenamefont {Matsumura}(2024)}]{carney2024classical}%
  \BibitemOpen
  \bibfield  {author} {\bibinfo {author} {\bibfnamefont {D.}~\bibnamefont {Carney}}\ and\ \bibinfo {author} {\bibfnamefont {A.}~\bibnamefont {Matsumura}},\ }\href@noop {} {\bibfield  {journal} {\bibinfo  {journal} {arXiv preprint arXiv:2412.04839}\ } (\bibinfo {year} {2024})}\BibitemShut {NoStop}%
\bibitem [{\citenamefont {Kryhin}\ and\ \citenamefont {Sudhir}(2025)}]{kryhin2025distinguishable}%
  \BibitemOpen
  \bibfield  {author} {\bibinfo {author} {\bibfnamefont {S.}~\bibnamefont {Kryhin}}\ and\ \bibinfo {author} {\bibfnamefont {V.}~\bibnamefont {Sudhir}},\ }\href@noop {} {\bibfield  {journal} {\bibinfo  {journal} {Physical Review Letters}\ }\textbf {\bibinfo {volume} {134}},\ \bibinfo {pages} {061501} (\bibinfo {year} {2025})}\BibitemShut {NoStop}%
\bibitem [{\citenamefont {Cata\~{n}o Lopez}\ \emph {et~al.}(2020)\citenamefont {Cata\~{n}o Lopez}, \citenamefont {Santiago-Condori}, \citenamefont {Edamatsu},\ and\ \citenamefont {Nobuyuki}}]{matsumoto20}%
  \BibitemOpen
  \bibfield  {author} {\bibinfo {author} {\bibfnamefont {S.~B.}\ \bibnamefont {Cata\~{n}o Lopez}}, \bibinfo {author} {\bibfnamefont {J.~G.}\ \bibnamefont {Santiago-Condori}}, \bibinfo {author} {\bibfnamefont {K.}~\bibnamefont {Edamatsu}}, \ and\ \bibinfo {author} {\bibfnamefont {M.}~\bibnamefont {Nobuyuki}},\ }\href@noop {} {\bibfield  {journal} {\bibinfo  {journal} {Physical Review Letter}\ }\textbf {\bibinfo {volume} {124}},\ \bibinfo {pages} {221102} (\bibinfo {year} {2020})}\BibitemShut {NoStop}%
\bibitem [{\citenamefont {Miao}\ \emph {et~al.}(2010)\citenamefont {Miao}, \citenamefont {Danilishin}, \citenamefont {M{\"u}ller-Ebhardt}, \citenamefont {Rehbein}, \citenamefont {Somiya},\ and\ \citenamefont {Chen}}]{miao2010probing}%
  \BibitemOpen
  \bibfield  {author} {\bibinfo {author} {\bibfnamefont {H.}~\bibnamefont {Miao}}, \bibinfo {author} {\bibfnamefont {S.}~\bibnamefont {Danilishin}}, \bibinfo {author} {\bibfnamefont {H.}~\bibnamefont {M{\"u}ller-Ebhardt}}, \bibinfo {author} {\bibfnamefont {H.}~\bibnamefont {Rehbein}}, \bibinfo {author} {\bibfnamefont {K.}~\bibnamefont {Somiya}}, \ and\ \bibinfo {author} {\bibfnamefont {Y.}~\bibnamefont {Chen}},\ }\href@noop {} {\bibfield  {journal} {\bibinfo  {journal} {Physical Review A—Atomic, Molecular, and Optical Physics}\ }\textbf {\bibinfo {volume} {81}},\ \bibinfo {pages} {012114} (\bibinfo {year} {2010})}\BibitemShut {NoStop}%
\bibitem [{\citenamefont {Datta}\ and\ \citenamefont {Miao}(2021)}]{Datta}%
  \BibitemOpen
  \bibfield  {author} {\bibinfo {author} {\bibfnamefont {A.}~\bibnamefont {Datta}}\ and\ \bibinfo {author} {\bibfnamefont {H.}~\bibnamefont {Miao}},\ }\href@noop {} {\bibfield  {journal} {\bibinfo  {journal} {Quantum Science and Technology}\ }\textbf {\bibinfo {volume} {6}},\ \bibinfo {pages} {045014} (\bibinfo {year} {2021})}\BibitemShut {NoStop}%
\bibitem [{\citenamefont {Miki}\ \emph {et~al.}(2024)\citenamefont {Miki}, \citenamefont {Matsumura},\ and\ \citenamefont {Yamamoto}}]{Miki}%
  \BibitemOpen
  \bibfield  {author} {\bibinfo {author} {\bibfnamefont {D.}~\bibnamefont {Miki}}, \bibinfo {author} {\bibfnamefont {A.}~\bibnamefont {Matsumura}}, \ and\ \bibinfo {author} {\bibfnamefont {K.}~\bibnamefont {Yamamoto}},\ }\href@noop {} {\bibfield  {journal} {\bibinfo  {journal} {Physical Review D}\ }\textbf {\bibinfo {volume} {109}},\ \bibinfo {pages} {064090} (\bibinfo {year} {2024})}\BibitemShut {NoStop}%
\bibitem [{\citenamefont {Michimura}\ \emph {et~al.}(2017)\citenamefont {Michimura}, \citenamefont {Kuwahara}, \citenamefont {Ushiba}, \citenamefont {Matsumoto},\ and\ \citenamefont {Ando}}]{Michimura2017}%
  \BibitemOpen
  \bibfield  {author} {\bibinfo {author} {\bibfnamefont {Y.}~\bibnamefont {Michimura}}, \bibinfo {author} {\bibfnamefont {Y.}~\bibnamefont {Kuwahara}}, \bibinfo {author} {\bibfnamefont {T.}~\bibnamefont {Ushiba}}, \bibinfo {author} {\bibfnamefont {N.}~\bibnamefont {Matsumoto}}, \ and\ \bibinfo {author} {\bibfnamefont {M.}~\bibnamefont {Ando}},\ }\href@noop {} {\bibfield  {journal} {\bibinfo  {journal} {Opt. Express}\ }\textbf {\bibinfo {volume} {25}},\ \bibinfo {pages} {13799} (\bibinfo {year} {2017})}\BibitemShut {NoStop}%
\bibitem [{\citenamefont {Deli\'{c}}\ \emph {et~al.}(2020)\citenamefont {Deli\'{c}}, \citenamefont {Reisenbauer}, \citenamefont {Dare}, \citenamefont {Grass}, \citenamefont {Vuleti\'{c}}, \citenamefont {Kiesel},\ and\ \citenamefont {Aspelmeyer}}]{Delic2020}%
  \BibitemOpen
  \bibfield  {author} {\bibinfo {author} {\bibfnamefont {U.}~\bibnamefont {Deli\'{c}}}, \bibinfo {author} {\bibfnamefont {M.}~\bibnamefont {Reisenbauer}}, \bibinfo {author} {\bibfnamefont {K.}~\bibnamefont {Dare}}, \bibinfo {author} {\bibfnamefont {D.}~\bibnamefont {Grass}}, \bibinfo {author} {\bibfnamefont {V.}~\bibnamefont {Vuleti\'{c}}}, \bibinfo {author} {\bibfnamefont {N.}~\bibnamefont {Kiesel}}, \ and\ \bibinfo {author} {\bibfnamefont {M.}~\bibnamefont {Aspelmeyer}},\ }\href@noop {} {\bibfield  {journal} {\bibinfo  {journal} {Science}\ }\textbf {\bibinfo {volume} {367}},\ \bibinfo {pages} {892} (\bibinfo {year} {2020})}\BibitemShut {NoStop}%
\bibitem [{\citenamefont {Jiang}\ and\ \citenamefont {Hosseini}(2022)}]{Hosseini2022}%
  \BibitemOpen
  \bibfield  {author} {\bibinfo {author} {\bibfnamefont {X.}~\bibnamefont {Jiang}}\ and\ \bibinfo {author} {\bibfnamefont {M.}~\bibnamefont {Hosseini}},\ }\href@noop {} {\bibfield  {journal} {\bibinfo  {journal} {Phys. Rev. Res.}\ }\textbf {\bibinfo {volume} {4}},\ \bibinfo {pages} {013132} (\bibinfo {year} {2022})}\BibitemShut {NoStop}%
\bibitem [{\citenamefont {Pedernales}\ \emph {et~al.}(2022)\citenamefont {Pedernales}, \citenamefont {Streltsov},\ and\ \citenamefont {Plenio}}]{pedernales2022}%
  \BibitemOpen
  \bibfield  {author} {\bibinfo {author} {\bibfnamefont {J.~S.}\ \bibnamefont {Pedernales}}, \bibinfo {author} {\bibfnamefont {K.}~\bibnamefont {Streltsov}}, \ and\ \bibinfo {author} {\bibfnamefont {M.~B.}\ \bibnamefont {Plenio}},\ }\href@noop {} {\bibfield  {journal} {\bibinfo  {journal} {Physical Review Letter}\ }\textbf {\bibinfo {volume} {128}},\ \bibinfo {pages} {110401} (\bibinfo {year} {2022})}\BibitemShut {NoStop}%
\bibitem [{\citenamefont {Kaku}\ \emph {et~al.}(2023)\citenamefont {Kaku}, \citenamefont {Fujita},\ and\ \citenamefont {Matsumura}}]{YK2023}%
  \BibitemOpen
  \bibfield  {author} {\bibinfo {author} {\bibfnamefont {Y.}~\bibnamefont {Kaku}}, \bibinfo {author} {\bibfnamefont {T.}~\bibnamefont {Fujita}}, \ and\ \bibinfo {author} {\bibfnamefont {A.}~\bibnamefont {Matsumura}},\ }\href@noop {} {\bibfield  {journal} {\bibinfo  {journal} {Physical Review D}\ }\textbf {\bibinfo {volume} {108}},\ \bibinfo {pages} {106014} (\bibinfo {year} {2023})}\BibitemShut {NoStop}%
\end{thebibliography}%

\end{document}